\documentclass[structabstract]{aa} 
\usepackage{booktabs}
\usepackage{graphicx}
\usepackage{txfonts}
\usepackage{natbib}
\usepackage{longtable}
\usepackage{lscape}
\usepackage{multirow}
\usepackage{pdfpages}
\usepackage{ulem}
\usepackage[colorlinks=true, allcolors=blue]{hyperref}

\bibpunct{(}{)}{;}{a}{}{,}

\newcommand{\kms}{km\,s$^{-1}$}

\newcommand{\Msun}{M$_{\odot}$}

\newcommand{\nodata}{...}

\begin{document}
%
\title{New submillimetre HCN lasers in carbon-rich evolved stars}

\author{W.~Yang\inst{1,2}, K.~T.~Wong\inst{3,4}, H.~Wiesemeyer\inst{2}, K.~M.~Menten\inst{2,}\thanks{During the article review stage, we suffered the painful loss of Prof. Dr. Karl Martin Menten. 
This work is one of his legacy, initiated under his guidance, and serves as a continuation of his research from years ago. This work is dedicated to the memory of Prof. Dr. Menten, whose invaluable contributions to radio astronomy and unwavering support to young generations will always be remembered.}, Y.~Gong\inst{5,2}, J. Cernicharo\inst{6}, E. De Beck\inst{7},  B.~Klein\inst{2}, C. A. Dur{\'a}n\inst{8,2} }

\institute{
School of Astronomy \& Space Science, Nanjing University, 163 Xianlin Avenue, Nanjing 210023, People's Republic of China \\
\email{wjyang@nju.edu.cn}
\and
Max-Planck-Institut f{\"u}r Radioastronomie, Auf dem H{\"u}gel 69, D-53121 Bonn, Germany
\and
Theoretical Astrophysics, Department of Physics and Astronomy, Uppsala University, Box 516, 751 20 Uppsala, Sweden \\
\email{katat.wong@physics.uu.se}
\and
Institut de Radioastronomie Millim{\'e}trique, 300 rue de la Piscine, 38406 Saint-Martin-d'H{\`e}res, France
\and
Purple Mountain Observatory and Key Laboratory of Radio Astronomy, Chinese Academy of Sciences, 10 Yuanhua Road, Nanjing 210033, People's Republic of China
\and
Dept. de Astrof{\'i}sica Molecular, Instituto de F{\'i}sica Fundamental (IFF-CSIC), C/ Serrano 121, 28006, Madrid, Spain
\and
Department of Space, Earth and Environment, Chalmers University of Technology, 41296, Gothenburg, Sweden
\and
Instituto de Radioastronom{\'i}a Milim{\'e}trica, Av. Divina Pastora 7, L20, 18012 Granada, Spain
}

\date{Received date ; accepted date}

\abstract
{
Strong laser emission from hydrogen cyanide (HCN) at 805 and 891~GHz has been discovered towards carbon-rich (C-rich) asymptotic giant branch (AGB) stars. Both lines belong to the Coriolis-coupled system between the (1,1$^{\rm 1e}$,0) and (0,4$^0$,0) vibrational states, which has been extensively studied in early molecular spectroscopy in the laboratory. However, the other lines in this system with frequencies above $\sim$900~GHz, challenging to observe with ground-based telescopes, have remained unexplored in astronomical contexts.}
{We aim to (1) search for new HCN transitions that show laser action in the (0,4$^0$,0), $J=10-9$ line at 894~GHz, the (1,1$^{\rm 1e}$,0)--(0,4$^0$,0), $J=11-10$ line at 964~GHz, the (1,1$^{\rm 1e}$,0), $J=11-10$ at 968~GHz, and the (1,1$^{\rm 1e}$,0), $J=12-11$ line at 1055~GHz towards C-rich AGB stars, (2) study  variability of multiple HCN laser lines, including the two known lasers at 805 and 891 GHz, and (3) construct a complete excitation scenario to the Coriolis-coupled system.}
{We conducted SOFIA/4GREAT observations and combined our data with {\it Herschel}/HIFI archival data, to construct a sample of eight C-rich AGB stars, covering six HCN transitions (i.e. 805, 891, 894, 964, 968 and 1055~GHz lines) in the Coriolis-coupled system.}
{We report the discovery of HCN lasers at 964, 968, and 1055~GHz towards C-rich AGB stars. Laser emission in the 805, 891, and 964~GHz HCN lines was detected in seven C-rich stars, while the 968~GHz laser was detected in six stars and the 1055~GHz laser in five stars. Notably, the 894~GHz line emission was not detected in any of the targets. Among the detected lasers, 
the emission of the cross-ladder line at 891~GHz is always the strongest, with typical luminosities of a few 10$^{44}$ photons~s$^{-1}$. The cross vibrational state 964 GHz laser emission, like a twin of the 891 GHz line, is the second strongest. The 1055~GHz emission always has a stronger 968~GHz counterpart.
Towards IRC+10216, all five HCN laser transitions were observed in six to eight epochs and exhibited significant variations in line profiles and intensities. The 891 and 964 GHz lines exhibit similar variations, and their intensity changes do not follow the near-infrared light curve (i.e. non-periodic variations). In contrast, the variations of the 805, 968, and 1055~GHz lines appear to be quasi-periodic, with a phase lag of 0.1 -- 0.2 relative to the near-infrared light curve. A comparative analysis indicates that these HCN lasers may be seen as analogues to vibrationally excited SiO and H$_2$O masers in oxygen-rich stars.
}
{We suggest chemical pumping and radiative pumping could play an important role in the production of the cross-ladder HCN lasers, while the quasi-periodic behaviour of the rotational HCN laser lines may be modulated by additional collisional and radiative pumping driven by periodic shocks and variations in infrared luminosity. }

\keywords{masers --- stars: AGB and post-AGB --- stars: carbon --- circumstellar matter}

\titlerunning{New submillimetre HCN lasers}

\authorrunning{Yang et al.}

\maketitle


\section{Introduction}


Hydrogen cyanide (HCN) is the most abundant molecule after H$_2$ and CO in the innermost circumstellar envelopes of carbon-rich\footnote{C-rich, with a carbon-to-oxygen abundance ratio, [C/O]$>$1. In addition, oxygen-rich (O-rich) AGB stars have [C/O]$<$1, and S-type AGB stars have [C/O]$\approx $1.} asymptotic giant branch (AGB) stars \citep{1964AnTok...9.....T,2013A&A...550A..78S}. 
HCN is produced efficiently in the stellar atmosphere through both the thermodynamic equilibrium chemistry \citep{1964AnTok...9.....T} and non-equilibrium shock-induced chemistry \citep{2006A&A...456.1001C}, making it one of the most important and abundant parent molecules in C-rich AGB stars. Therefore, this molecule serves as an excellent tracer for studying the dynamics and physical conditions near the photospheres of AGB stars as well as in large parts of their circumstellar envelopes \citep[e.g. see details in][]{2024A&A...684A...4U}.

The linear triatomic HCN molecule has three vibrational modes  ($\nu_1$,$\nu_2$,$\nu_3$). These are the doubly degenerate bending mode\footnote{For $\nu_2 \neq$ 0, the degeneracy is lifted, and the level is split in sublevels with different e-f parity \citep{1975JMoSp..55..500B} by rotational-vibrational interactions. The e and f sublevels correspond to the lower and upper split levels \citep{1996JMoSp.180..323M}, respectively.
A new quantum number, the vibrational angular momentum ($l$), is introduced, where $l$ can take values of 0 (even $\nu_2$) or 1 (odd $\nu_2$), and is incremented by 2 up to $l\leq\nu_2$ and $l\leq J$.} ($\nu_2$), and two stretching vibrations, the C-N stretching mode ($\nu_1$) and the C-H stretching mode ($\nu_3$)\footnote{We follow the notation used by previous studies that predicted and first detected laser transitions in the Coriolis-coupled system \citep{1967ApPhL..11...62L,1967PhLA...25..489H,2000ApJ...528L..37S,2003ApJ...583..446S}. It is important to note that in some studies \citep[e.g.][]{2006MNRAS.367..400H,2014MNRAS.437.1828B}, the $\nu_1$ refers to the C-H stretching, and the $\nu_3$ refers to the C-N stretching, which is different from the present work. Besides, the vibrational states of (1,1$^{\rm 1e}$,0) and (0,$4^{0}$,0) are also commonly labelled as $\nu_1+\nu_2^{\rm e}$ and $4\nu_2^0$ (see Appendix~\ref{Sec:appendix-d} as an example) in literature.}. 
In addition to thermal emission from various rotational transitions within different vibrationally excited states at millimetre, submillimetre, and far-infrared wavelengths \citep[e.g.][]{1989A&A...218L..20L,
1996A&A...315L.201C,2011A&A...529L...3C,2013ApJ...778L..25C}, various HCN  lines have been found to be inverted towards several stars. 
HCN masers (and lasers\footnote{In this work, we use the term `laser' for the stimulated HCN emission in the far-infrared and submillimetre regimes, which follows the terminology from the laboratory literature and astronomical discoveries of the HCN lasers at 805 and 891~GHz \citep{2000ApJ...528L..37S,2003ApJ...583..446S}.}) are found to be the most widespread maser species in C-rich stars. 
To date, a total of 75 HCN masers have been identified towards 36 C-rich AGB stars, arising from 10 different vibrational states \citep[e.g.][]{2018A&A...613A..49M,2022A&A...666A..69J}.

Remarkably, two submillimetre HCN transitions that exhibit laser action in the laboratory, the (0,4$^0$,0), $J=9-8$ transition at 805~GHz and the (1,1$^{\rm 1e}$,0)--(0,4$^0$,0), $J=10-9$ transition at 891~GHz, are also reported to show laser emission in astronomical objects (IRC+10216, CIT~6 and Y~CVn,  \citealt{2000ApJ...528L..37S,2003ApJ...583..446S}). First discovered in the 1960s \citep{1964Natur.202..685G, 1965ElL.....1...45M}, both
laser transitions are related to rotation-vibration interactions between the (1,1$^{\rm 1e}$,0) and (0,4$^0$,0) states (i.e. Coriolis coupling), where several rotational levels in both states with identical $J$ are very close in energy \citep{1967ApPhL..11...62L}.
The upper energy levels for both lines are very high ($>$4200~K), well above the temperature of the photosphere or that of the dust formation zone, suggesting that the laser emission originates from the innermost regions of the circumstellar envelope \citep{2000ApJ...528L..37S}. 
In addition, \cite{2003ApJ...583..446S} found that the 891~GHz line is about an order of magnitude stronger than the 805~GHz line, and observations spaced about half a year apart show evidence of variability.

Recently, \cite{2019asrc.confE..55W} performed a pilot survey with the Atacama Large Millimeter/submillimeter Array (ALMA) to search for these two HCN laser lines in Band 10. This survey led to the detection of the 805~GHz laser in four sources (R~For, R~Lep, CQ~Pyx, and V~Hya) and the 891~GHz laser in six sources (R~For, R~Lep, CQ~Pyx, IRC+10216, X~Vel, and V~Hya). They found that the extent of the HCN laser emitting regions was found to be $\sim$11--16~au in V~Hya and $\lesssim$30~au in IRC+10216. 
Note that the diameter of IRC+10216's radio photosphere has been measured to be 10.8 au (at an infrared phase of 0.79), which implies 3.8 au as the diameter of its optical photosphere \citep{2012A&A...543A..73M}.
\cite{2023ApJ...958...86A} imaged the 891 GHz laser in R~Lep with ALMA using baselines up to 16 km. They reported $\sim$1.1$\times$10$^8$~K for the peak brightness temperature at an angular resolution of 5.4~mas $\times$ 4.9~mas, and characterised the distribution of the HCN laser as a ring-like morphology with a diameter of 10--60~au.

\begin{figure}[htbp]
\center
\includegraphics[width=0.44\textwidth]{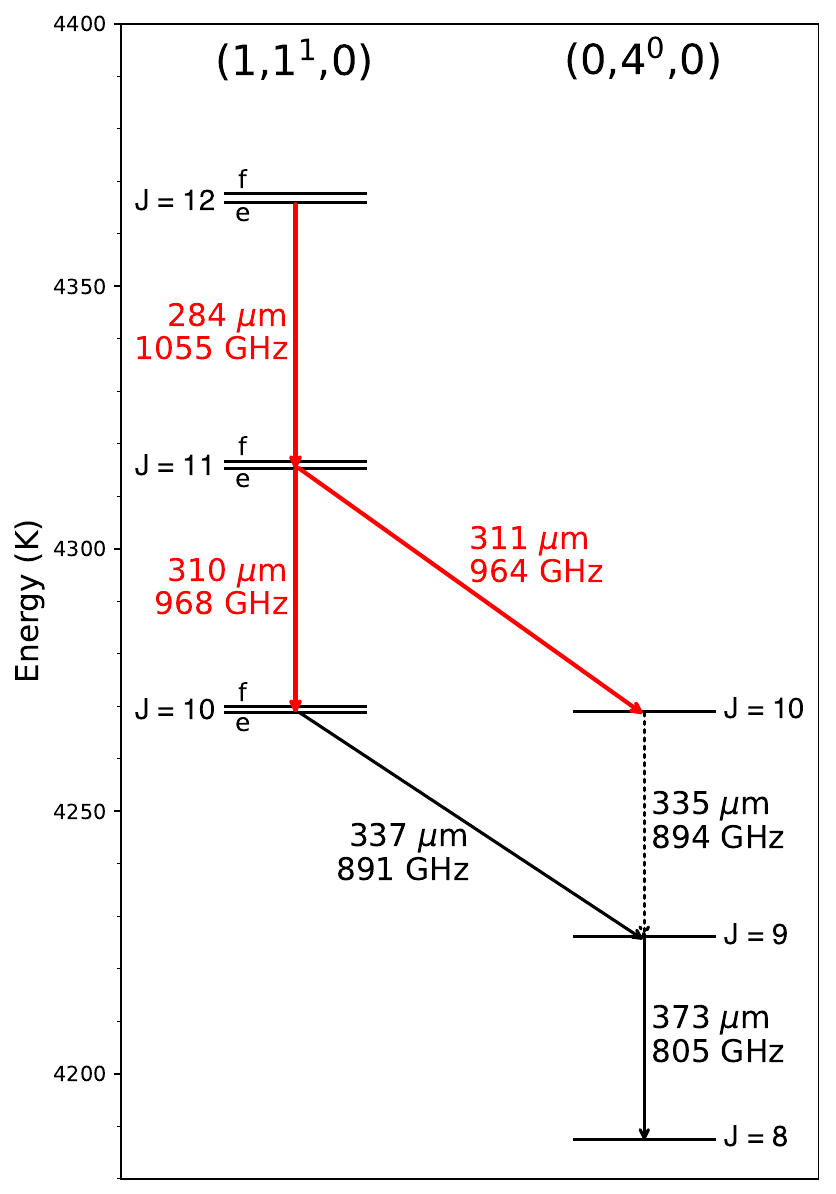}
\caption{Level diagram of the (1,1$^1$,0) and (0,4$^0$,0) vibrational states of HCN near the Coriolis resonance involving the rotational levels from $J=8$ to $J=12$. 
The energy levels are adopted from \cite{2006MNRAS.367..400H}.
The wavelength \citep{1967PhLA...25..489H} and frequency (see Table \ref{Tab:obs_line} for exact numbers) of each transition are labelled.
The red arrows highlight the astronomical laser transitions discovered in the course of this work, while the black solid arrows represent the previously detected laser lines of the HCN transitions towards IRC+10216. The black dotted line indicates the non-detected 894 GHz line \citep{2000ApJ...528L..37S,2003ApJ...583..446S}.
\label{fig:HCN_level}}
\end{figure}

In addition to the two astronomical laser lines, two laser lines at 964 GHz and 968 GHz are also attributed to the Coriolis coupled system by \citet[][see Fig.~\ref{fig:HCN_level}]{1967ApPhL..11...62L}. Furthermore, \citet{1967PhLA...25..489H} observed laser emission near 1055 GHz and 894 GHz, which are pure rotational transitions in the (1,1$^{\rm 1e}$,0) and (0,4$^0$,0) states, respectively, and occur in a cascade simultaneously with the 964~GHz laser line. In their experiments, \citet{1967PhLA...25..489H} managed to measure the precise frequencies for five out of six transitions near the Coriolis resonance except for the line near 1055 GHz due to low intensity. We adopt the rest frequencies measured by \citet{1967PhLA...25..489H} in this work. 
However, these transitions above 900~GHz have never been explored towards astronomical objects and are missing pieces for an understanding of circumstellar HCN laser excitation.



Observations of the highest frequency  submillimetre laser transitions ($\gtrsim$900~GHz) are challenging  or nearly impossible using ground-based telescopes due to the Earth’s atmospheric absorption.
Thanks to the Heterodyne Instrument for the Far-infrared \citep[HIFI;][]{2010A&A...518L...6D} on the {\it Herschel} Space Observatory \citep{2010A&A...518L...1P}, which operated between 2009 and 2013, it has been possible to explore these high-frequency HCN transitions.
Following the end of the {\it Herschel} mission, the Stratospheric Observatory For Infrared Astronomy \citep[SOFIA;][]{2012ApJ...749L..17Y} became the only observatory capable of accessing these transitions (until the end of the end of the SOFIA mission in September 2022).
In particular, 4GREAT \citep{2021ITTST..11..194D}, an extension of the German REceiver for Astronomy at Terahertz Frequencies (GREAT\footnote{GREAT is a development by the MPI f{\"u}r Radioastronomie and the KOSMA/Universit{\"a}t zu K{\"o}ln, in cooperation with the MPI f{\"u}r Sonnensys-temforschung and the DLR Institut f{\"u}r Planetenforschung.}), instrument on SOFIA enabled high-resolution spectroscopy, which is essential for characterising laser emission.


We conducted SOFIA/4GREAT observations, which we combine with {\it Herschel}/HIFI archival data, to 1) search for new HCN laser emissions in C-rich stars and study their physical properties including variability, and 2) shed light on the laser lines' excitation conditions and possible pumping mechanisms. 
This work is organised as follows. Sect.~\ref{Sec:obs} introduces the SOFIA/4GREAT observations and the {\it Herschel}/HIFI archival data used in this work.  
Sect.~\ref{Sec:result} presents the laser detections, the lines' variabilities, and comparisons between laser lines.
In Sect.~\ref{Sec:discuss}, we discuss the laser excitation considerations and possible pumping schemes. A summary of this work and highlighted conclusions are provided in Sect.~\ref{Sec:sum}.

\section{Observations}\label{Sec:obs}

\begin{table*}[hbt]
\caption{Stellar information of the sample of C-rich stars in this work.}\label{Tab:obs_source}
\normalsize
\centering
\setlength{\tabcolsep}{5pt}
\begin{tabular}{lrrcccccc}
\hline \hline 
Source name &   $\alpha$(J2000)   & $\delta$(J2000) & Type & Distance & $V_{\rm LSR}^*$ & $V_{\rm exp}$ & $\dot{M}$ & Period \\
 & (hh:mm:ss) & (dd:mm:ss) &  & (pc) & (\kms) & (\kms)  & (10$^{-7}$~\Msun~yr$^{-1}$) & (day) \\ 
\hline
IRC+10216 (CW~Leo) \tablefootmark{(a)} & 09:47:57.40 & 13:16:43.5 & M & 140 & $-$26.5 & 18 & 200--400 & 630 \\
CIT~6 (RW~LMi) \tablefootmark{(a)}  & 10:16:02.28 & 30:34:18.9 & M & 460 & $-$2 & 15 & 26--140 & 617\\
Y~CVn \tablefootmark{(b)}     & 12:45:07.82 & 45:26:24.9 & SRb & 220 & +22 & 7 & 1.5 & 268\\
S~Cep \tablefootmark{(b)}     & 21:35:12.82 & 78:37:28.2 & M & 380 & $-$15.3 & 22.5 & 12 & 484\\
IRC+50096 (V~384~Per) \tablefootmark{(b)} & 03:26:29.51 & 47:31:48.6 & M & 560 & $-$16.8 & 15.5 & 23 & 535 \\ 
V~Cyg \tablefootmark{(b)}  & 20:41:18.27 & 48:08:28.8 & M & 366 & +13.5 & 12 & 16 & 421\\ 
CRL~3068 (LL~Peg) \tablefootmark{(b)} & 23:19:12.39 & 17:11:35.4 & M & 1300 & $-$31.5 & 14.5 & 250 & 696 \\
II~Lup (IRAS~15194$-$5115) \tablefootmark{(a)} & 15:23:05.07 & $-$51:25:58.7 & M & 640 & $-$15 & 23 & 100 & 576\\
\hline
\hline
\end{tabular}
\normalsize
\tablefoot{Columns 1--3 list the names and coordinates of sources (the top four sources with SOFIA observations). 
Column 4 list the variability type: M (Mira) and SR (Semi-regular), as adopted from \cite{2014A&A...566A.145R}, with the type for S~Cep taken from the American Association of Variable Star Observers (AAVSO; \url{https://www.aavso.org/}).
Columns 5--8 provide the distance, the stellar LSR velocity ($V_{\rm LSR}^*$), the terminal expansion velocity of the envelope ($V_{\rm exp}$), the mass-loss rate ($\dot{M}$) and the stellar pulsation period, respectively.
\tablefoottext{a}{Stellar parameters for IRC+10216, CIT~6, and II~Lup are adopted from \cite{2018A&A...613A..49M}, and the distances derived from the Mira period-luminosity relation are used in this work. We caution that the adopted $V_{\rm LSR}^*$ may be slightly overestimated as the contribution from microturbulence has not been subtracted.
}
\tablefoottext{b}{Stellar parameters for Y~CVn, S~Cep, IRC+50096, V~Cyg and CRL~3068 are adopted from \cite{2018A&A...611A..29M}, except for their stellar pulsation periods that are derived from the AAVSO.}
}
\end{table*}

\begin{table*}[hbt]
\caption{Spectroscopic properties of HCN transitions.}\label{Tab:obs_line} 
\small
\centering
\setlength{\tabcolsep}{3pt}
\begin{tabular}{ccccccccccccccc}
\hline \hline 
\multicolumn{2}{c}{HCN transition}  & Rest frequency\tablefootmark{(a)} & $E_{\rm up}$\tablefootmark{(b)} & $A_{\rm ul}$\tablefootmark{(b)} & \multicolumn{5}{c}{SOFIA/4GREAT\tablefootmark{(c)}} & \multicolumn{4}{c}{{\it Herschel}/HIFI\tablefootmark{(d)}}  \\
\cmidrule(lr){1-2} \cmidrule(lr){6-10} \cmidrule(lr){11-14}
($\nu_1$,$\nu_2$,$\nu_3$) & $J$ &  &  &  & Channel & HPBW  & $\eta_{\rm mb}$ & Factor\tablefootmark{(e)}& Obs.\tablefootmark{(f)} & Band & HPBW  & $\eta_{\rm mb}$ &  Factor\tablefootmark{(e)} \\
 &  & (GHz) & (K) & (s$^{-1}$) & & (\arcsec)  &  & (Jy K$^{-1}$) & & & (\arcsec)  &  & (Jy K$^{-1}$)\\
\hline
(0,4$^0$,0)                     & 9--8   & 804.7509       & 4226.1  & 2.3$\times$10$^{-2}$ & ---       & ---  & ---  & --- & ---     & 3a    &  26.1    & 0.63   & 355 \\ 
(0,4$^0$,0)                     & 10--9   & 894.4142       & 4269.0 & 3.1$\times$10$^{-2}$ & ---       & ---  & ---  & --- & ---     & 3b    &  23.6    & 0.63  & 353 \\ 
(1,1$^{\rm 1e}$,0)--(0,4$^0$,0) & 10--9  & 890.7607       & 4268.8 & ... & 4G2 (LSB) & 31.4 & 0.56 & 609 & 1       & 3b    & 23.6 & 0.63 & 353 \\ 
(1,1$^{\rm 1e}$,0)--(0,4$^0$,0) & 11--10 & 964.3134       & 4315.3 & ... & 4G2 (USB) & 28.8 & 0.56 & 599 & 1,2,3,4 & 4a    & 21.8 & 0.64 & 359 \\ 
(1,1$^{\rm 1e}$,0)              & 11--10 & 967.9658       & 4315.3 & 4.3$\times$10$^{-2}$ & 4G2 (USB) & 28.8 & 0.56 & 598 & 1,2,3,4 & 4a    & 21.7 & 0.64 & 359 \\ 
(1,1$^{\rm 1e}$,0)              & 12--11 & 1055.54$^\dagger$ & 4365.9 & 5.6$\times$10$^{-2}$ & 4G2 (LSB) & 26.2 & 0.56 & 588 & 1       & 4a/4b & 20.0 & 0.64 & 378 \\ 
\hline
\hline
\end{tabular}
\normalsize
\tablefoot{The HCN transitions at 805 and 894~GHz were not observed by SOFIA, as marked by ``---" in the corresponding rows.
\tablefoottext{a}{The frequencies of HCN transitions are adopted from \cite{1967PhLA...25..489H}, with the exception of that of the 1055~GHz line whose frequency is estimated from our detection, labelled by a dagger.}
\tablefoottext{b}{The energy of upper level and the Einstein A coefficient of each transition are adopted from \cite{2006MNRAS.367..400H} and \cite{2014MNRAS.437.1828B}, while the Einstein A coefficients for the 891 and 964~GHz lines are not given, labelled as ``...". }
\tablefoottext{c}{Reference: \cite{2021ITTST..11..194D}.} 
\tablefoottext{d}{Reference: \cite{2017A&A...608A..49S}.}
\tablefoottext{e}{Scaling factors to convert main-beam temperature to flux density. The factors for {\it Herschel}/HIFI data are the averages of H and V polarisations.}
\tablefoottext{f}{Observed sources at the corresponding transition: (1) IRC+10216; (2) CIT~6; (3) Y~CVn; (4) S~Cep.}
}
\end{table*}

\subsection{SOFIA/4GREAT observations}\label{Sec:4great_obs}


We searched for HCN emission from the (1,1$^{\rm 1e}$,0)--(0,4$^0$,0), $J=11-10$ cross-ladder transition at 964~GHz, and from the (1,1$^{\rm 1e}$,0), $J=11-10$ transition at 968~GHz towards IRC+10216, CIT~6, Y~CVn, and S~Cep. 
We also searched for HCN emission of the (1,1$^{\rm 1e}$,0), $J=12-11$ line at 1055~GHz and re-visited laser emission of the (1,1$^{\rm 1e}$,0)--(0,4$^0$,0), $J=11-10$ line at 891~GHz towards IRC+10216.
Table~\ref{Tab:obs_source} lists the stellar information for these four targets. 


Table~\ref{Tab:obs_line} provides the details for the observed HCN transitions and lists the sources observed in each transition. The observations were performed at altitudes between 12.2 and 13.2~km on SOFIA flight \#540 on 2018 December 17 (Cycle 6), under the guaranteed-time project 83\_0625 (PI: Karl M. Menten). The SIS mixer of  band-2  of the 4GREAT receiver \citep{2021ITTST..11..194D} on board was used with four separate tunings. Background signals from hardware contributions, thermal emissions from the telescope, and the atmosphere were removed using double-beam switching between the target and reference positions at $\Delta \alpha = \pm60\arcsec$ from the target. Because the laser emissions are spatially unresolved in these observations, the telescope was pointed toward the stellar positions. The raw data streams from 4GREAT were converted to 4~GHz wide astronomical spectra by fast Fourier-transform spectrometers \citep{2012A&A...542L...3K}. Typical single-sideband system temperatures range from 1000 -- 1100~K and 1000 -- 1300~K in the 964 and 968~GHz lines, respectively, and are around 2600 and 650~K for the 891 and 1055~GHz lines, respectively. 

The raw SOFIA data was corrected for the atmospheric transmission employing calibration-load scans preceding each on-off scan, applying the following procedure provided by the \textit{kalibrate} software implemented in the KOSMA package \citep{2012A&A...542L...4G}: The total power count rates from the sky and the loads at ambient and cold temperatures were used to fit the \textit{am atmospheric model} \citep{2022zndo...6774378P} to the measured atmospheric emission (calibrated to Rayleigh-Jeans equivalent forward-beam brightness temperatures with a forward efficiency of 0.97). In the least-square fitting, both the wet and dry contents of the atmosphere were kept as free parameters, resulting in precipitable water-vapour columns of typically 6 -- 10~$\mu$m and dry-constituent concentrations at most $\pm 20\%$ from the model predictions. The inferred pressure-dependent opacity coefficients were converted to transmission corrections subsequently applied to the data. The remaining data reduction steps were performed with the CLASS software which is part of the GILDAS\footnote{\url{https://www.iram.fr/IRAMFR/GILDAS}} software \citep{2005sf2a.conf..721P}. 
First-order baselines were removed for the calibrated spectra. To enhance the signal-to-noise ratio and to allow for better comparison with the HIFI data, the spectra were boxcar-smoothed to $\sim$0.5~MHz and $\sim$1.0~MHz resolutions for IRC+10216 and the other three targets, respectively (except for the 968~GHz line towards S~Cep, for which the spectra were smoothed to $\sim$2.0~MHz), 
Finally, the resulting spectra were converted from antenna temperature to flux density using main beam efficiencies and scaling factors (listed in Table~\ref{Tab:obs_line}).

\subsection{{\it Herschel}/HIFI archival data}\label{Sec:HIFI_data}

To investigate the behaviours of the multiple HCN laser transitions in the Coriolis-coupled system, we collected from the {\it Herschel} Science Archive all related observations of C-rich stars with the HIFI instrument that cover all six lines (see Fig.~\ref{fig:HCN_level} and Table~\ref{Tab:obs_line}). A total of eight C-rich stars had such data (see Table~\ref{Tab:obs_source}), and towards four of them follow-up SOFIA/4GREAT observations had been conducted. In Appendix~\ref{sec:appendix-a}, Table~\ref{Tab:hifi} summarises the HIFI spectra used in this study. 
Each of the eight AGB stars is known to have at least one HCN maser or laser detection \citep[e.g.][]{2022A&A...666A..69J}.

For IRC+10216, the HIFI data (Project ids: GT1\_jcernich\_4, DDT\_jcernich\_7, OT2\_jcernich\_9, DDT\_jcernich\_10, PI: Jos{\'e} Cernicharo) were gathered from a comprehensive line survey which used all HIFI bands between 480 and 1907~GHz, along with monitoring observations within bands from 1a to 5b \citep{2010A&A...521L...8C,
2014ApJ...796L..21C}. 
Among the {\it Herschel}/HIFI archival data, the HCN laser observations across six epochs were uniquely conducted towards this source, which enables us to study laser variability.
For CIT~6, Y~CVn, S~Cep, IRC+50096, V~Cyg, and CRL~3068, the HCN HIFI data come from a systematic survey for HCN masers in C-rich evolved stars, which comprises all rotational lines ranging from $J=6-5$ up to $J=13-12$ from within multiple vibrational states (project id: OT2\_jcernich\_8, PI: Jos{\'e} Cernicharo).
For II~Lup, the HIFI observations of the HCN transitions are covered by a spectral line survey towards this star (project id: OT2\_edebeck\_2, PI: Elvire De Beck).

We utilised the Level 2.5 data products from the {\it Herschel} Science Archive\footnote{\url{http://archives.esac.esa.int/hsa/whsa/}}, and these products have been further processed with the {\it Herschel} Standard Product Generation (SPG) pipeline (v14.1.0).
These data were obtained from the observations that were performed in dual 
beam switching mode, using the wideband spectrometer (WBS) to record the signal with a spectral resolution of 1.1~MHz. 
We note that the Level 2.5 data products have been regridded in frequency with a uniform spacing of 0.5~MHz \citep{2017A&A...608A..49S}. The forward efficiency is 0.96 for all HIFI bands \citep{2017A&A...608A..49S}, and the HPBW and main beam efficiency for each frequency are listed in Table~\ref{Tab:obs_line}.

To enhance the S/N ratios, we averaged the signals from the horizontal (H) and vertical (V) polarisation channels. 
We note that the 891~GHz laser spectra of IRC+10216 that were observed in single point mode exhibited contamination from $^{13}$CO ($J=8-7$) emission at 881.272808~GHz from the other sideband. Details on the removal of contamination from the 891~GHz spectra are provided in Appendix~\ref{sec:appendix-b}. Additional data reduction was performed with the GILDAS/CLASS
software package. A linear baseline was subtracted from all observed spectra. 

We note that the intensities of the $^{13}$CO line shown in Fig.~\ref{fig:irc10216_891_contamination} exhibit variations of less than 10\%.  
This suggests that the accuracy of the flux density calibration is within 10\%.

\section{Results} \label{Sec:result}

\begin{figure*}[htbp]
\center
\includegraphics[width=0.188\textwidth]{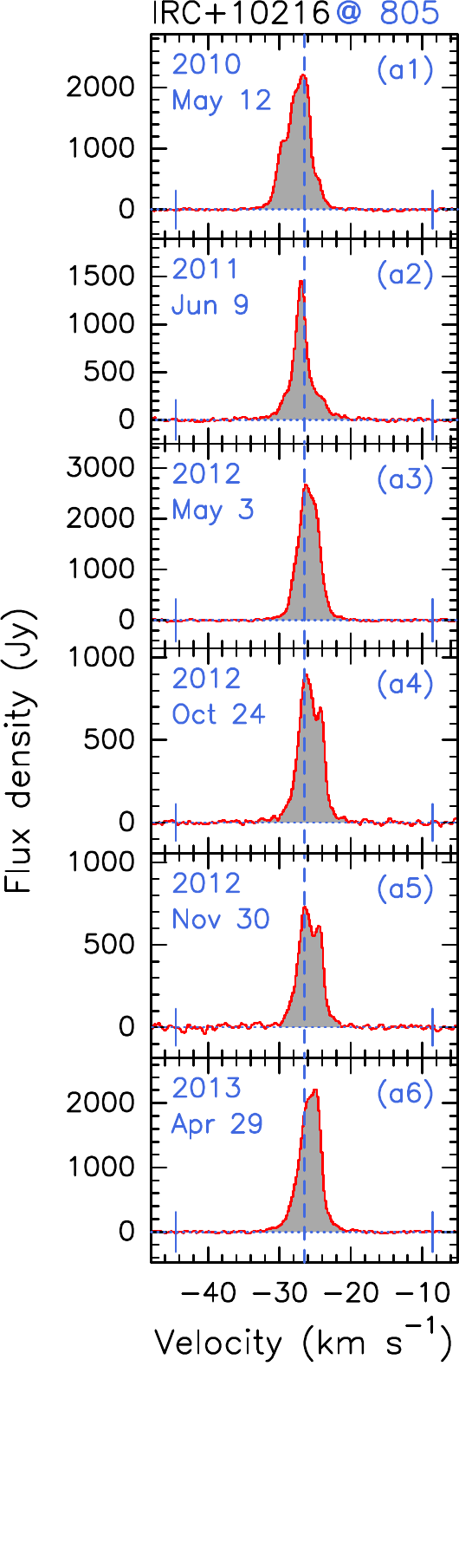}
\includegraphics[width=0.188\textwidth]{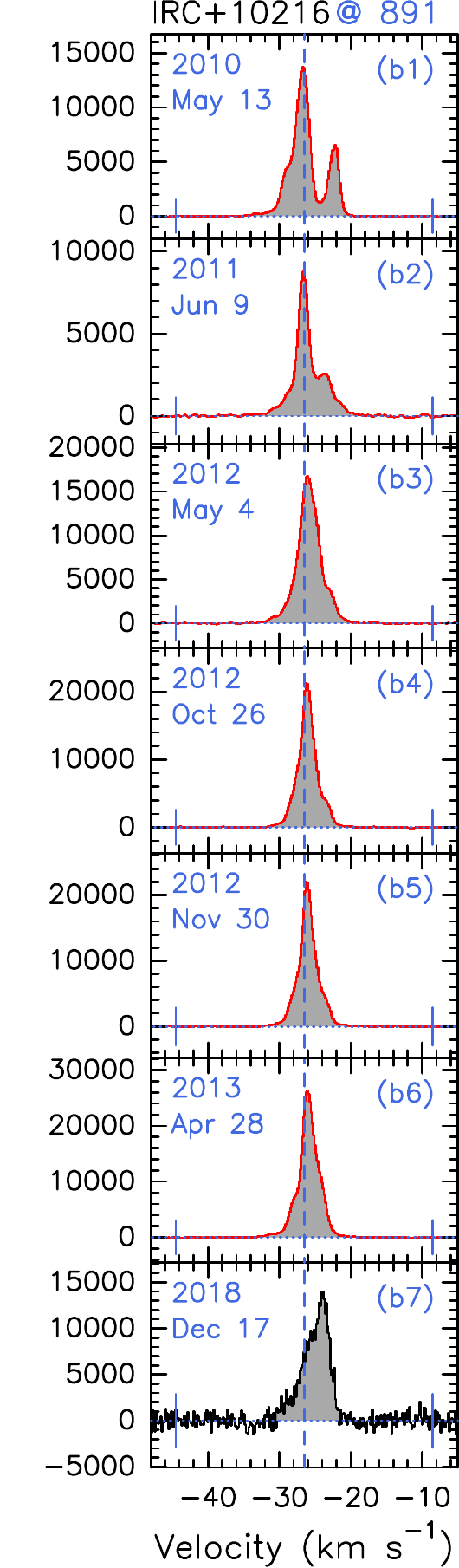}
\includegraphics[width=0.188\textwidth]{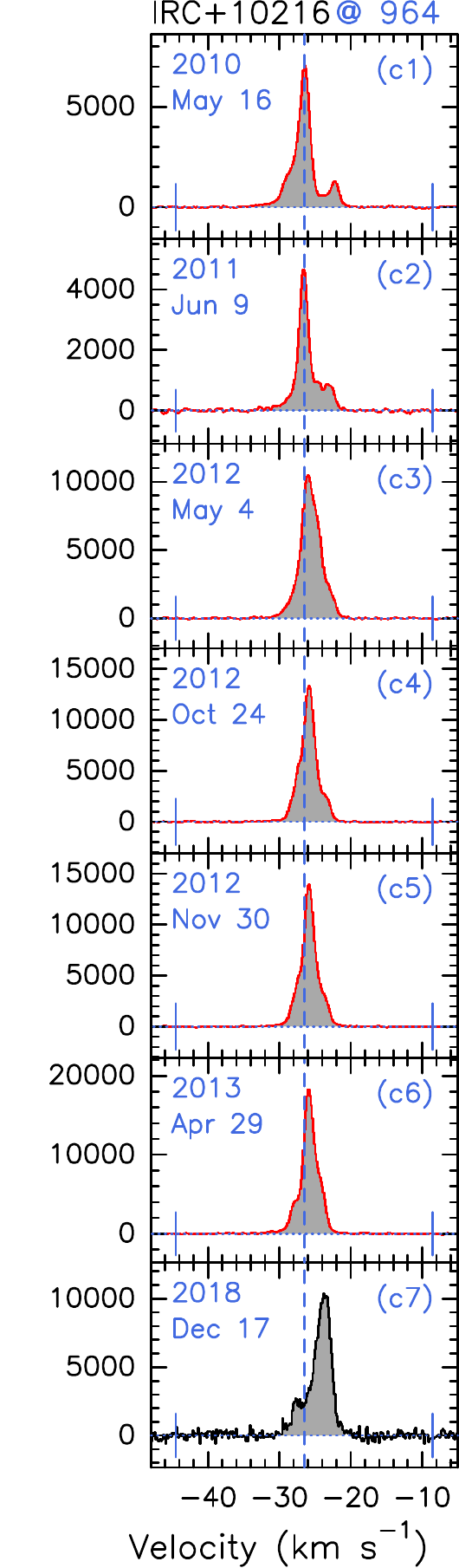}
\includegraphics[width=0.188\textwidth]{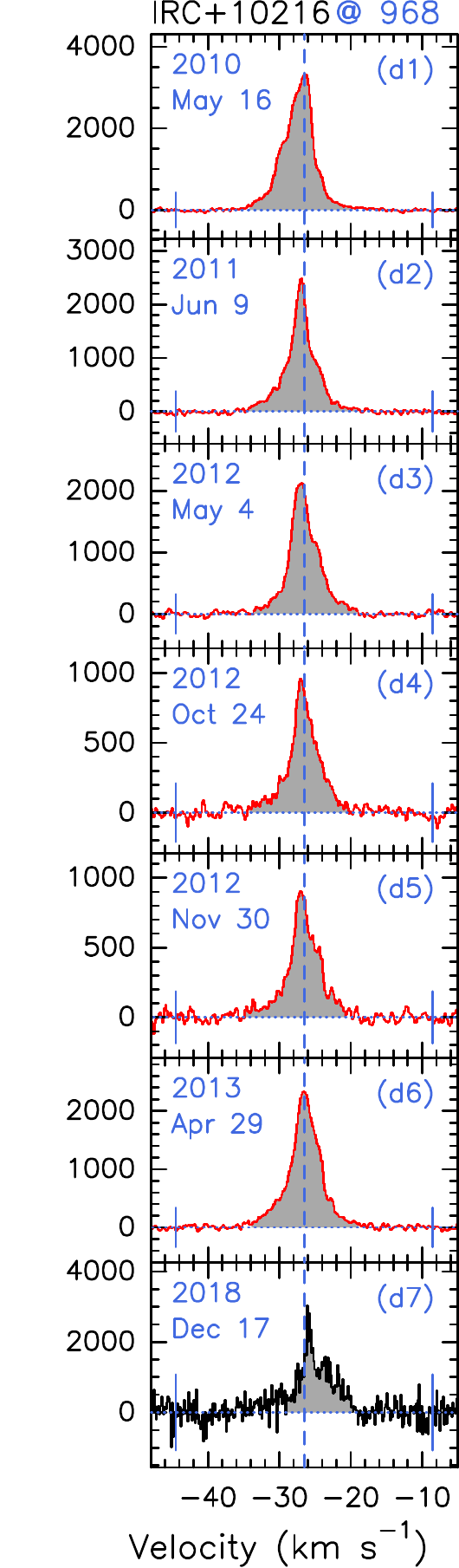}
\includegraphics[width=0.188\textwidth]{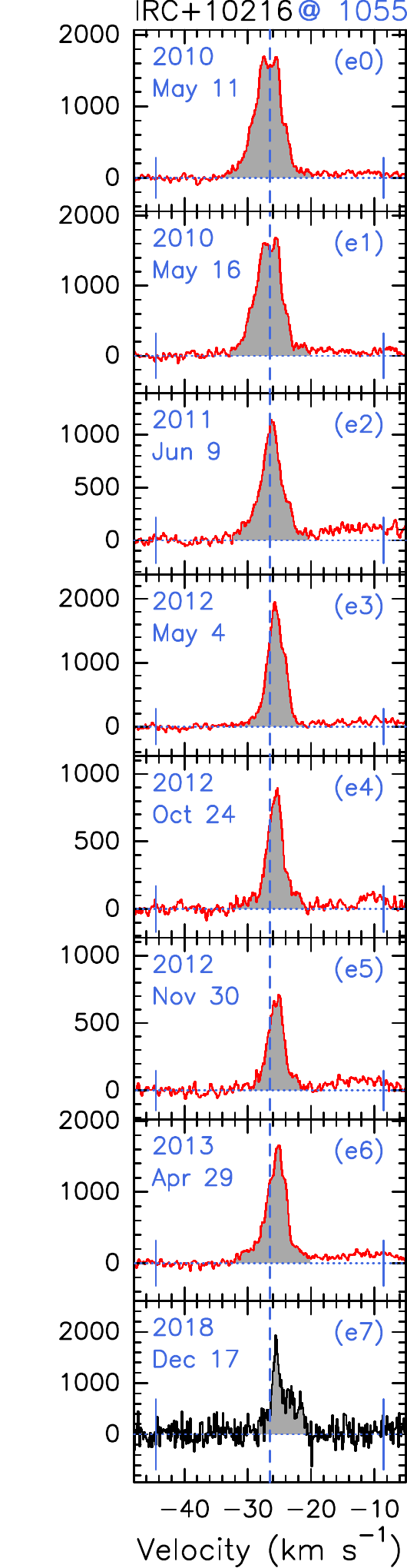}
\caption{Spectra of submillimetre HCN laser transitions within the Coriolis-coupled system observed in six to eight epochs towards IRC+10216. 
Spectra are presented from left to right for the following transitions: (a) the (0,4$^0$,0), $J=9-8$ line at 805~GHz; (b) the (1,1$^{\rm 1e}$,0)--(0,4$^0$,0), $J=10-9$ line at 891~GHz; (c) the (1,1$^{\rm 1e}$,0)--(0,4$^0$,0), $J=11-10$ line at 964~GHz; (d) the (1,1$^{\rm 1e}$,0), $J=11-10$ at 968~GHz; and (e) the (1,1$^{\rm 1e}$,0), $J=12-11$ line at 1055 GHz. 
The {\it Herschel}/HIFI spectra are plotted in red, while the SOFIA/4GREAT spectra, smoothed to a channel spacing of $\sim$0.15~\kms, are shown in black. The observing dates are labelled in their respective panels.
In each panel, the vertical blue dashed line indicates the stellar velocity of $-$26.5~\kms, the vertical blue bars mark the terminal velocity of $\pm$18~\kms, 
and the horizontal dotted blue line represents the baseline.
The grey-shaded regions indicate the velocity ranges used to determine the integrated intensities.
\label{fig:IRC10216_5laser}}
\end{figure*}

\begin{figure*}[htbp]
\center
\includegraphics[width=0.24\textwidth]{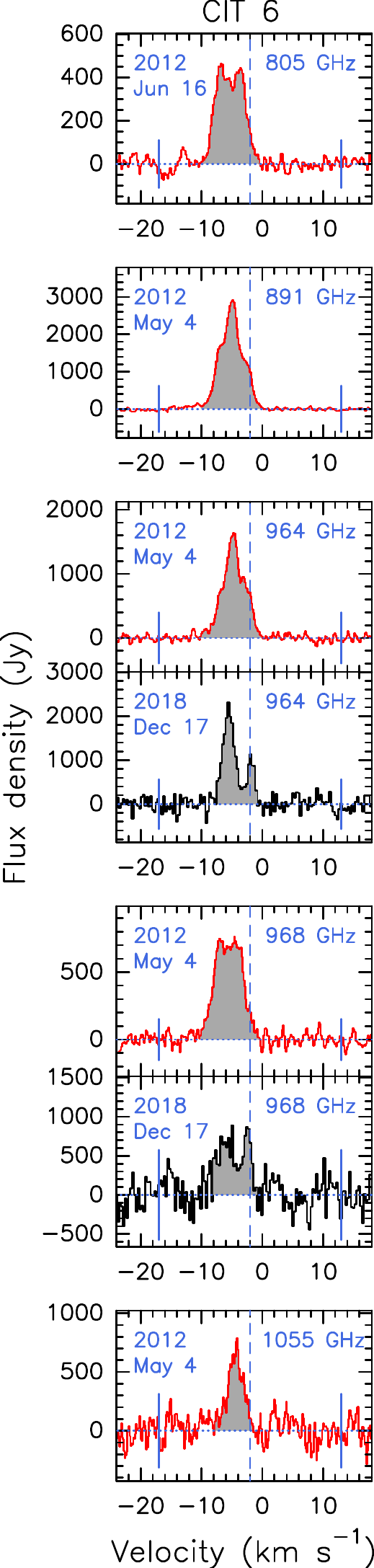}
\includegraphics[width=0.24\textwidth]{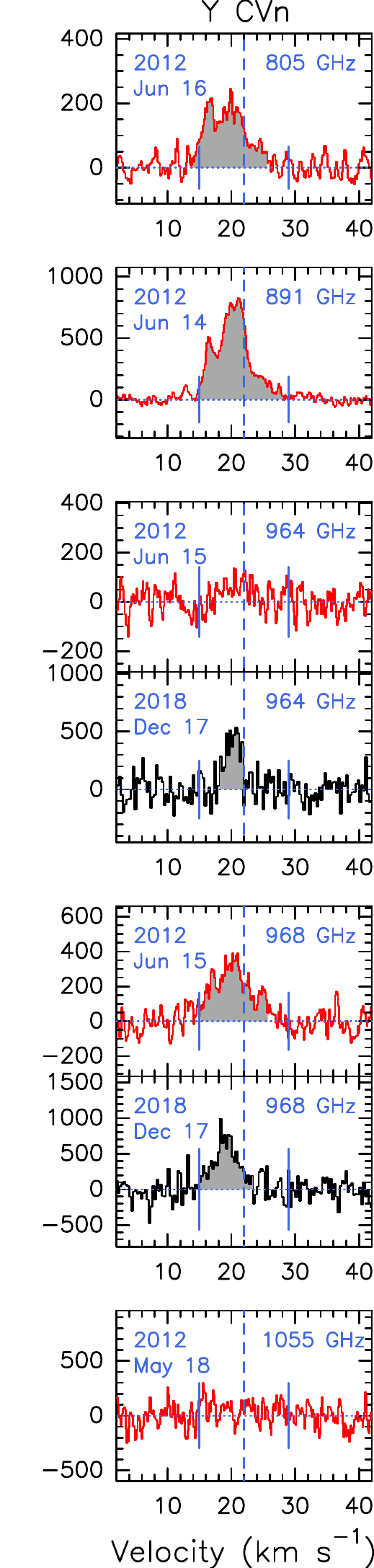}
\includegraphics[width=0.24\textwidth]{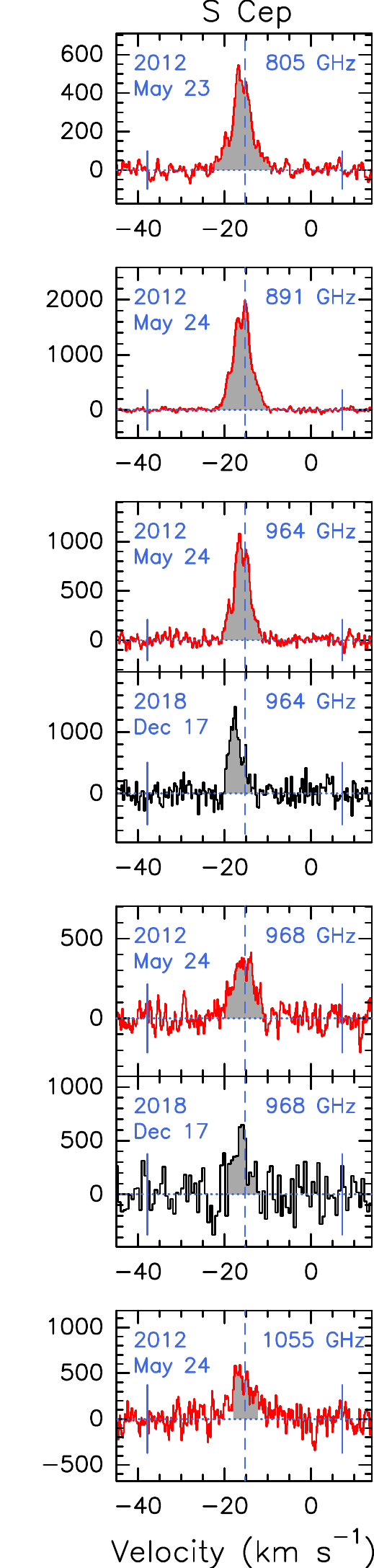}
\caption{Spectra of submillimetre HCN laser transitions towards CIT~6 (left), Y~CVn (middle), and S~Cep (right).
The {\it Herschel}/HIFI spectra are plotted in red, while the SOFIA/4GREAT spectra, smoothed to a channel spacing of $\sim$0.3~\kms\,(except for the tentative detection the 968~GHz emission towards S~Cep, smoothed to a channel spacing of 0.6~\kms ), are plotted in black.
The observing dates are labelled in the perspective panels. 
The vertical blue dashed line indicates the stellar velocity, the vertical blue bars mark the terminal velocity (listed in Table~\ref{Tab:obs_source}), the horizontal dotted blue line represents the baseline, and the grey-shaded regions indicate the velocity ranges used to determine the integrated intensities.
\label{fig:3stars_laser}}
\end{figure*}

\begin{figure*}[htbp]
\center
\includegraphics[width=0.24\textwidth]{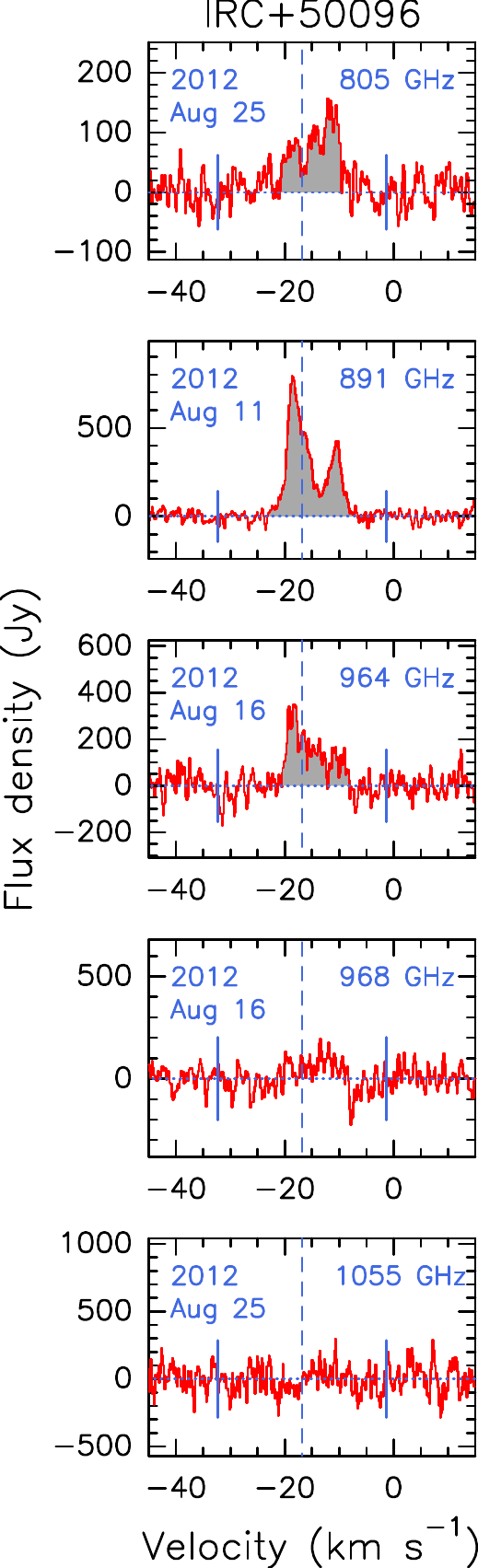}
\includegraphics[width=0.24\textwidth]{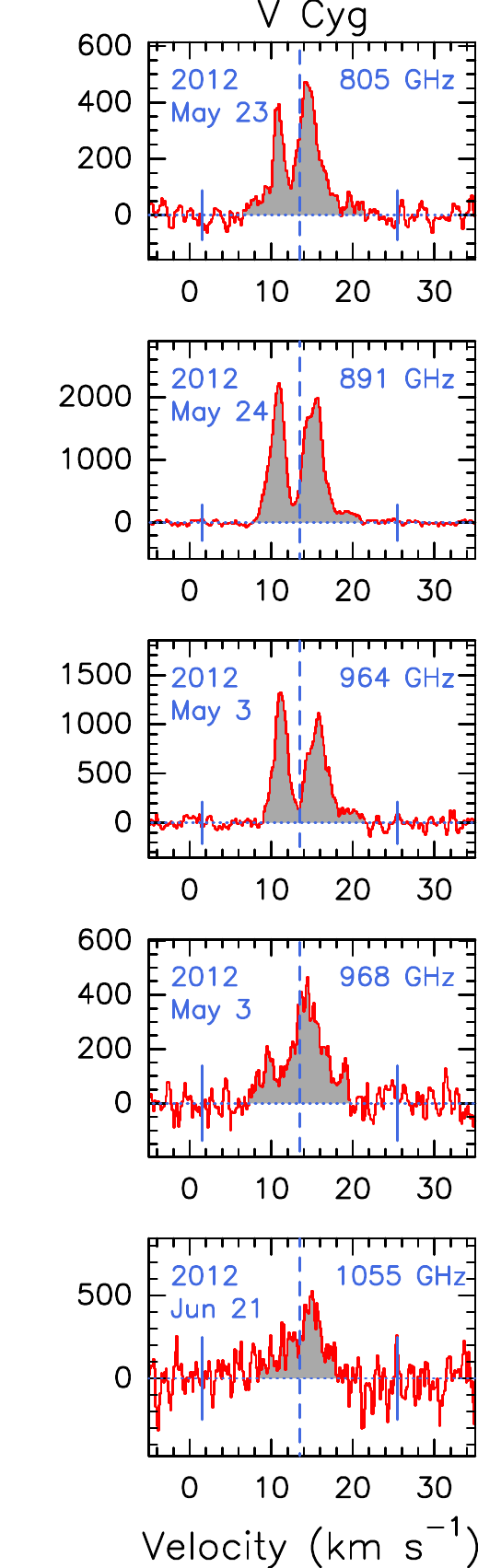}
\includegraphics[width=0.24\textwidth]{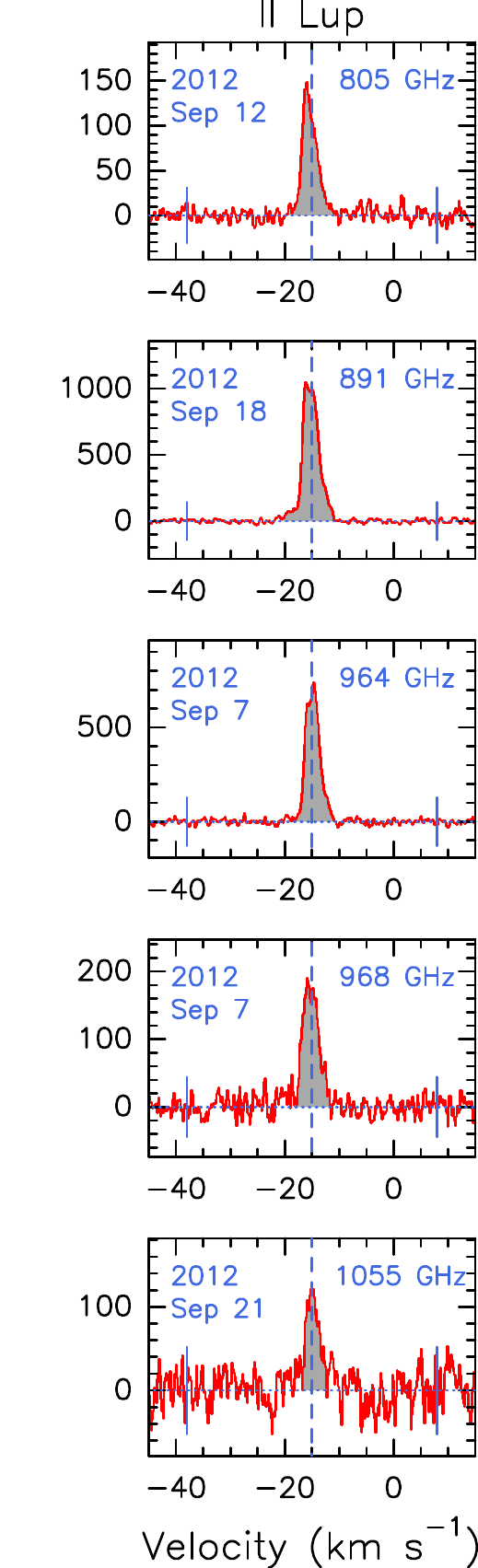}
\includegraphics[width=0.24\textwidth]{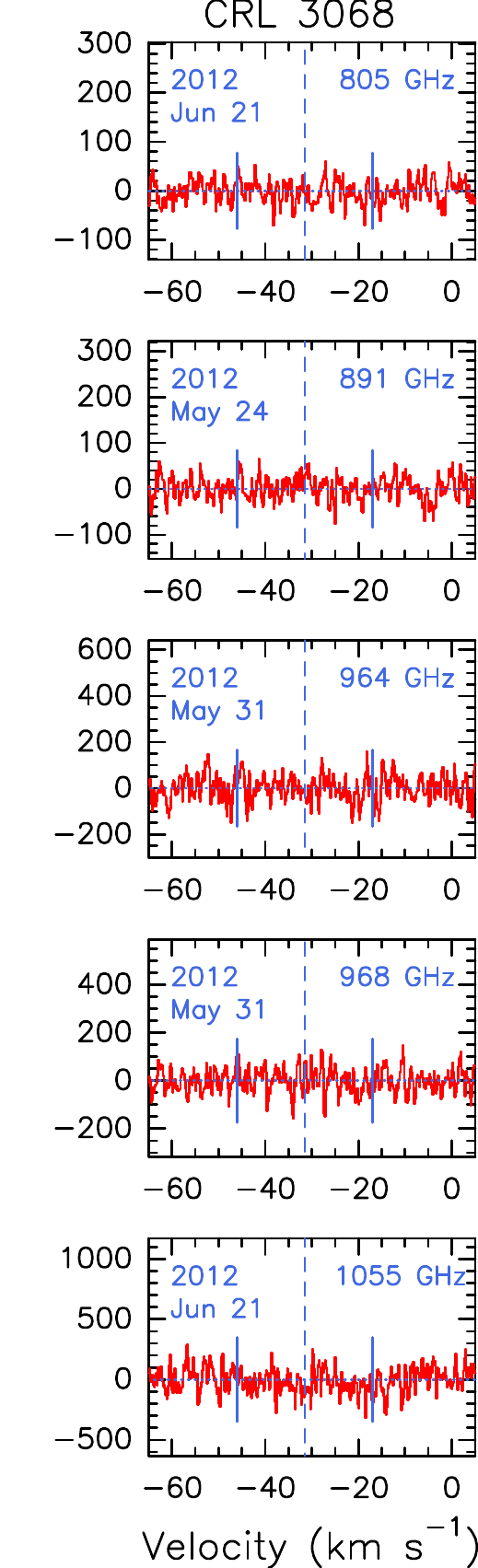}
\caption{Same as Fig.~\ref{fig:3stars_laser}, but for the {\it Herschel}/HIFI spectra of IRC+50096, V~Cyg, II~Lup, and CRL~3068 (from left to right), respectively. \label{fig:3stars_laser_2}}
\end{figure*}

\subsection{Do the detected HCN lines represent laser action?} \label{Sec:laser_identify}

From the {\it Herschel}/HIFI archival data and our SOFIA/4GREAT observations, the (0,4$^0$,0), $J=9-8$ line at 805~GHz, the (1,1$^{\rm 1e}$,0)--(0,4$^0$,0), $J=10-9$ line at 891~GHz, and the (1,1$^{\rm 1e}$,0)--(0,4$^0$,0), $J=11-10$ line at 964~GHz were detected towards seven sources, IRC+10216, CIT~6, Y~CVn, S~Cep, IRC+50096, V~Cyg and II~Lup. 
Among these stars, the (1,1$^{\rm 1e}$,0), $J=11-10$ line at 968~GHz was detected towards six sources (excluding IRC+50096), and the (1,1$^{\rm 1e}$,0), $J=12-11$ line at 1055~GHz was detected towards five sources (except for Y~CVn and IRC+50096). 
No emission from any of the aforementioned HCN transitions was detected towards CRL~3068.
The spectra of these HCN transitions in each source are shown in Figs.~\ref{fig:IRC10216_5laser}, \ref{fig:3stars_laser} and \ref{fig:3stars_laser_2}. 
Our results are summarised in Tables~\ref{Tab:IRC+10216_laser} and \ref{Tab:6stars_laser}.

The (0,4$^0$,0), $J=10-9$ line at 894~GHz was not detected towards any of our targets (with a typical 1$\sigma$ noise level of 20--35~Jy at a channel width of 0.17~\kms). 
Two decades ago, \citet{2003ApJ...583..446S} also failed to detect this line towards IRC+10216. 


We identify the HCN laser emission based on these criteria: 
Firstly, the line profiles are asymmetric and present typical laser characteristics, including the strong intensities, narrow features, and a limited velocity coverage that is significantly smaller than the full width at zero power (FWZP) of the symmetric (parabola-like) spectra of low-excitation lines, for example from HCN. 
Secondly, transitions observed over multiple epochs show variability in peak intensity, the number of narrow features, and the radial velocity of the strongest component (see Sect.~\ref{Sec:variation_description} for details). 
In addition, the non-detections of the 894~GHz emission, which has an upper energy and Einstein $A$ coefficient comparable to those of the other five HCN submillimetre lines, also supports that these detected features are not of thermal origins (see details in Sect.~\ref{Sec:exciatation}).
Our detected spectra meet all these criteria, thus confirming that these detections can be classified as laser emission in this work.

Based on the classification, we therefore report that the detection rates for the 805, 891 and 964~GHz HCN laser emissions are 88\%, although the 964~GHz HCN laser emission was not detected in one observational epoch towards Y~CVn. In contrast, the detection rates for the 968 and 1055~GHz HCN laser emission are 75\% and 63\%, respectively. The high detection rates indicate that these five lasers are quite common in C-rich stars. 
Comments on individual stars (including CRL~3068) are presented in Appendix~\ref{Sec:individual}.

\citet{1967PhLA...25..489H} measured the precise frequencies of the lines, with an error of $\sim$ 1~MHz, except for the 1055~GHz line, for which no precise measurement was provided. In the current work, the frequency of the 1055~GHz line is estimated assuming its peak emission in IRC+10216 is close to the systemic velocity, as seen in the other HCN lines within the Coriolis-coupled system in this star. Since different HCN lines show different line profiles, it is uncertain whether the peak of the 1055~GHz line should align with other lines. We report the estimated frequency of this line only up to two decimal places in GHz (Table~\ref{Tab:obs_line}). The calculated rest frequency of this line in the ExoMol\footnote{\url{https://exomol.com/}; \citet{2024JQSRT.32609083T}} line list is 35.20883~cm$^{-1}$ or 1055.5342~GHz \citep{2006MNRAS.367..400H,2014MNRAS.437.1828B}. The discrepancy between the estimated and calculated frequencies corresponds to a velocity of ${<}2$~\kms, which is consistent with most other HCN lines in the Coriolis-coupled system.

\begin{figure}[htbp]
\center
\includegraphics[width=0.38\textwidth]{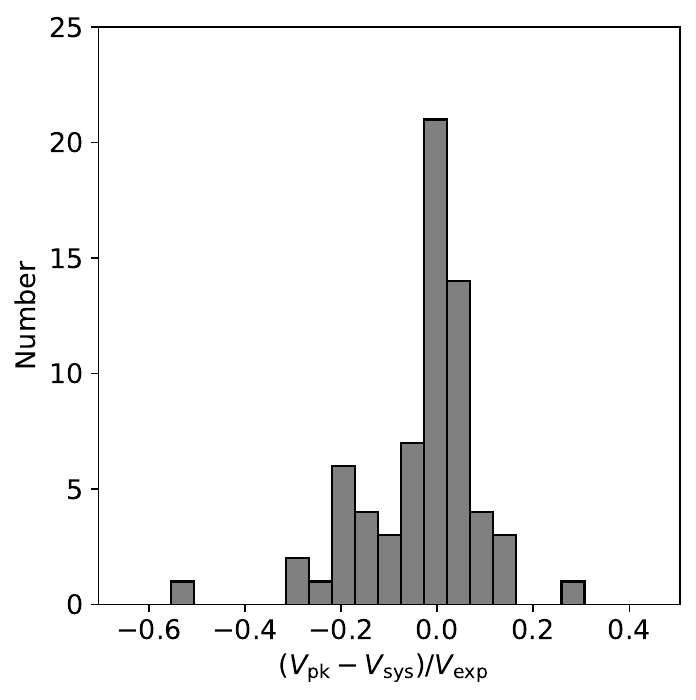}
\caption{Distribution of the velocity differences between the strongest HCN laser features and their stellar systemic velocities versus the terminal expansion velocities for all detected emissions in this work.
\label{fig:velocity_hist}}
\end{figure}

The detected HCN laser emissions from different transitions towards the same source exhibit similar velocity ranges, typically much less than twice their corresponding terminal expansion velocities. 
Figure~\ref{fig:velocity_hist} shows the distribution of the velocity differences between the strongest HCN laser features and their stellar systemic velocities, normalised by the terminal expansion velocities (i.e. ($V_{\rm pk}-V_{\rm sys})/V_{\rm exp}$), for all detected transitions towards C-rich stars in this work.
All ratios are below 0.52, with the majority lower than 0.2, indicates that these lasers do not reach terminal velocities. 
ALMA observations \citep{2019asrc.confE..55W,2023ApJ...958...86A} have revealed that these HCN lasers originate from regions close to the stellar photosphere. 
Therefore, the observed small velocity differences ($<$5~\kms) between the strongest HCN laser features and the stellar systemic velocities indicate that the lasers locate in the wind acceleration zones. 
Assuming the velocity model of IRC+10216 \citep{2012A&A...543A..48A}, these HCN lasers could be confined within 5$R_{\star}$, where $R_{\star}$ is the stellar radius.

We estimate the isotropic luminosity, or ``photon rate", of these 
HCN lasers using the following equation \citep[e.g.][]{2024ApJ...961..190Y}: 
\begin{equation}
L_{\rm HCN} ({\rm photons\,s}^{-1}) = 6.04\times10^{41}~D^2({\rm kpc}) \int S\mathrm{dv} ({\rm Jy\,km~s}^{-1}),
\end{equation}
where $D$ is the distance, $\int S\mathrm{dv}$ is the velocity-integrated flux density of the laser emission.
The HCN laser luminosity of each transition for the individual targets in each observed epoch is listed in Tables~\ref{Tab:IRC+10216_laser} and \ref{Tab:6stars_laser}. The luminosities range from $2.7\times 10^{43}$ to $1.6\times 10^{45}$~${\rm photons\,s}^{-1}$, generally surpassing those of HCN masers at lower frequencies \citep[e.g. $\sim10^{41-44}$~${\rm photons\,s}^{-1}$; see Table~8 in][]{2022A&A...666A..69J}. This makes the HCN lasers in the Coriolis-coupled system the brightest beacons in C-rich stars.


\subsection{Laser variability} \label{Sec:variation_description}

\cite{2003ApJ...583..446S} reported that the 891~GHz ((1,1$^{\rm 1e}$,0)--(0,4$^0$,0), $J=10-9$) laser varied over an interval of four to eight months towards IRC+10216 and CIT~6, while the 805~GHz laser exhibited variability over an interval of about two years towards IRC+10216.
On the other hand, the thermal emission lines of several molecules (e.g. CCH, SiS, HCN, HNC, CN) towards IRC+10216 are also known to be variable \citep{2014ApJ...796L..21C,2018A&A...615L...4P,2019ApJ...883..165H}. This variability is primarily observed as fluctuations in peak intensity, while the line shape, typically parabolic or double-horned, remains unchanged.
We aim to investigate whether the laser emissions show similar variability pattern. 
In particular, the spectra of the five HCN laser transitions observed towards IRC+10216 in six to eight epochs (see Fig.~\ref{fig:IRC10216_5laser}), with intervals spanning nearly a week, a month and half a year, provide an opportunity to examine the laser variability on different timescales.

The spectra of the 1055~GHz ((1,1$^{\rm 1e}$,0), $J=12-11$) 
line towards IRC+10216 on 2010 May 11 and 16 (see Figs.~\ref{fig:IRC10216_5laser}e0 and e1) appear to be identical. 
Both spectra show two peaks. 
The peak intensity of the $-$25.6~\kms\,component remained constant, whereas the peak intensity of the $-$27.4~\kms\,component slightly decreased by 5($\pm$3)\%. Since the subtle intensity change was less than three times the noise level of the observed spectra, we infer that the HCN lasing gas was stable over five days. 
This stability facilitates comparisons of HCN laser emissions observed on  dates that are close to each other (see further discussions in Sect.~\ref{Sec:spectral}).

The observational epochs of 2012 October and November allow us to explore the laser variability over a period of about one month.
No significant changes in line shapes were observed for any of the five laser transitions (see Figs.~\ref{fig:IRC10216_5laser}a4, a5, b4, b5 ... e4 and e5), but the laser peak and integrated intensities varied to different amounts for different transitions.
The laser emissions at 805~GHz and 1055~GHz exhibited the most significant changes, with peak intensities decreasing by 19($\pm$2)\% and 22($\pm$5)\%, respectively.


Over a longer timescale ($\geq$ 5 months), the profiles of all five laser transitions towards IRC+10216 changed dramatically, affecting the number of laser features, line widths of individual features, peak velocities and intensities, and total integrated intensities. This suggests that the variability pattern of these laser lines differs from that of thermal emissions, which supports our identification of laser emission. 
Comparing the spectra obtained in 2018 December ($\phi_{\rm IR}$= 0.22) with those obtained in 2010 May ($\phi_{\rm IR}$= 0.23), we found, even with similar stellar phases, that the laser line profiles change remarkably over several stellar cycles. Taking the 891~GHz laser as an example (see Figs.~\ref{fig:IRC10216_5laser}b1 and b7), the most significant changes in the line profile were a decrease in the number of distinct laser components from two to one, and a shift in the peak velocity from the systemic velocity to the redshifted side, at approximately 2.5~\kms.
Similar changes in the line profile were also observed for SiO masers in O-rich AGB stars \citep{Alcolea1999,2004A&A...424..145P}.

\begin{figure*}[htbp]
\center
\mbox{
\begin{minipage}[b]{9.5cm}
\includegraphics[width=0.95\textwidth]{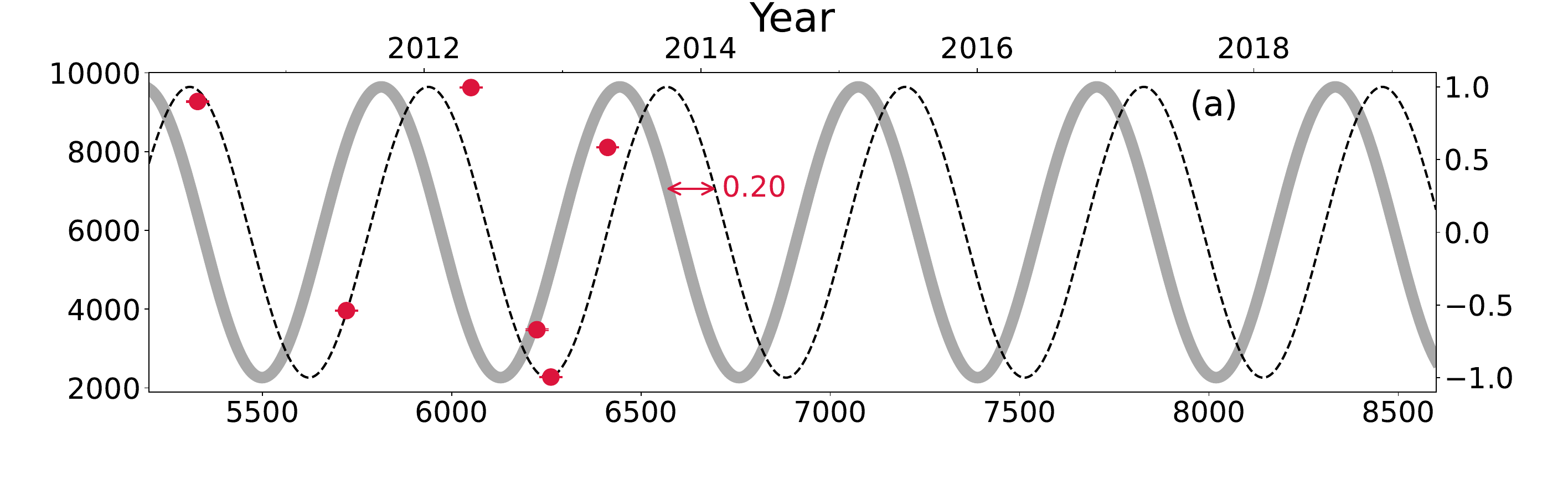}
\vspace{-3mm}
\end{minipage}
\hspace{-4mm}
\begin{minipage}[b]{9.5cm}
\includegraphics[width=0.95\textwidth]{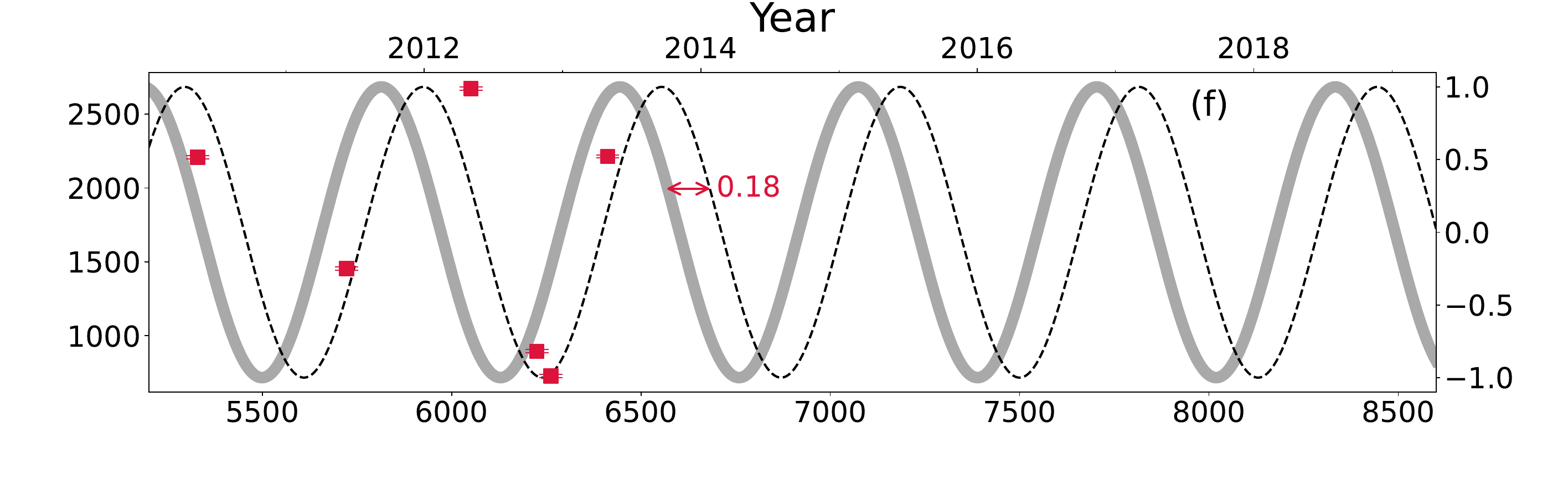}
\vspace{-3mm}
\end{minipage}
}
\mbox{
\begin{minipage}[b]{9.5cm}
\includegraphics[width=0.95\textwidth]{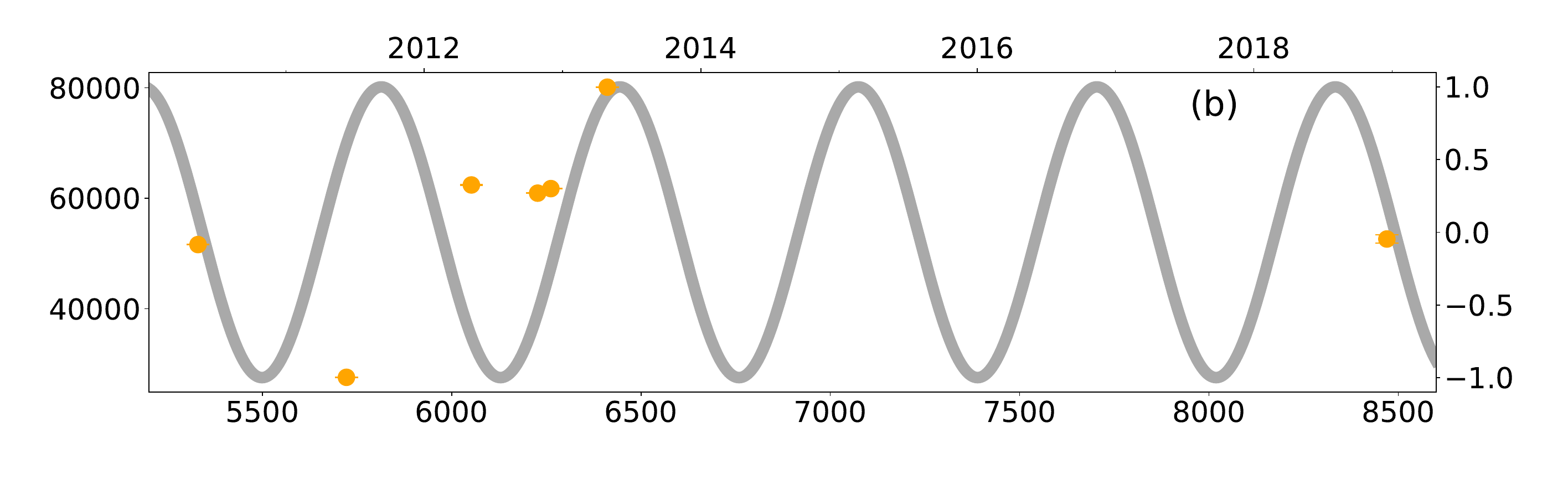}
\vspace{-3mm}
\end{minipage}
\hspace{-4mm}
\begin{minipage}[b]{9.5cm}
\includegraphics[width=0.95\textwidth]{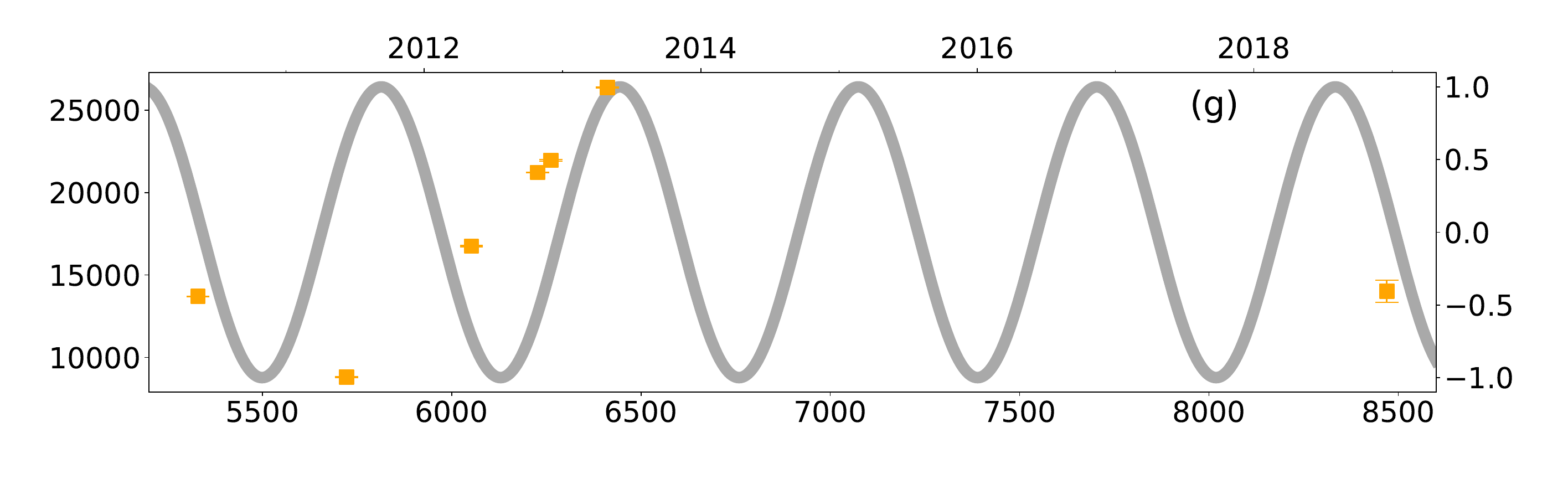}
\vspace{-3mm}
\end{minipage}
}
\mbox{
\begin{minipage}[b]{9.5cm}
\includegraphics[width=0.95\textwidth]{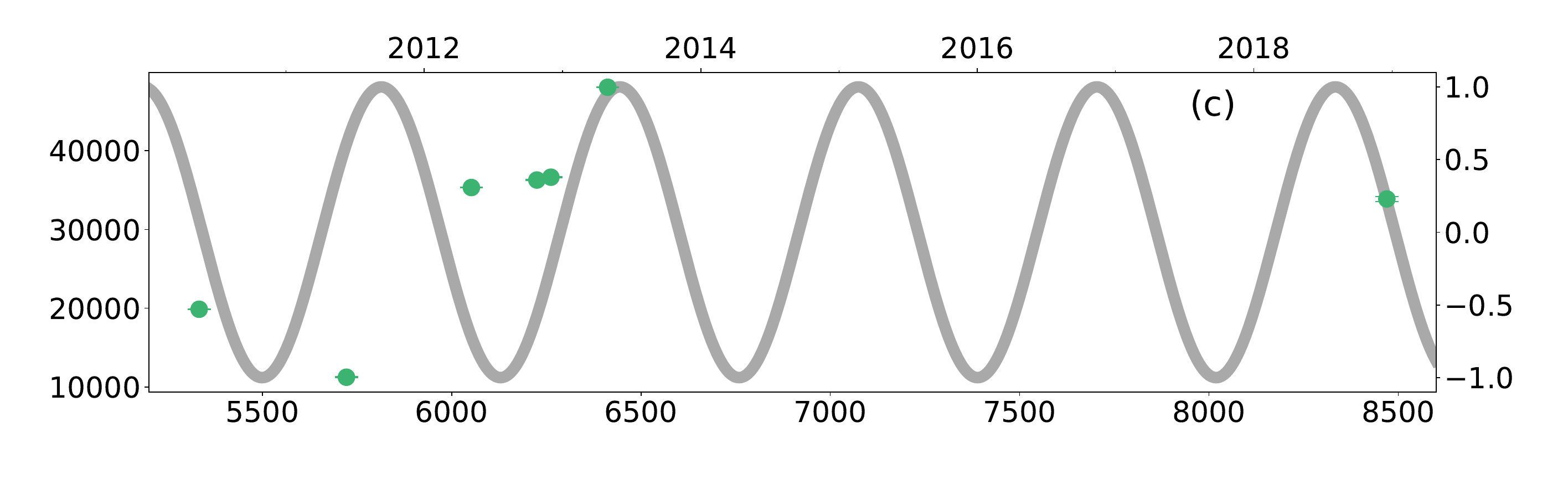}
\vspace{-3mm}
\end{minipage}
\hspace{-4mm}
\begin{minipage}[b]{9.5cm}
\includegraphics[width=0.95\textwidth]{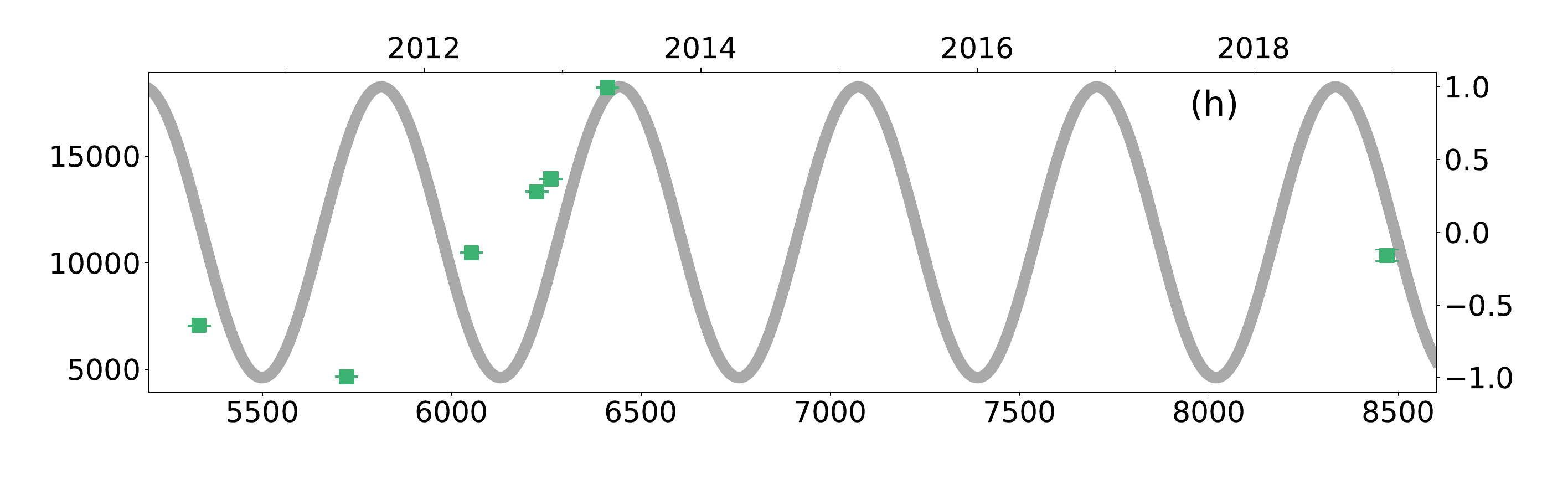}
\vspace{-3mm}
\end{minipage}
}
\mbox{
\begin{minipage}[b]{9.5cm}
\includegraphics[width=0.95\textwidth]{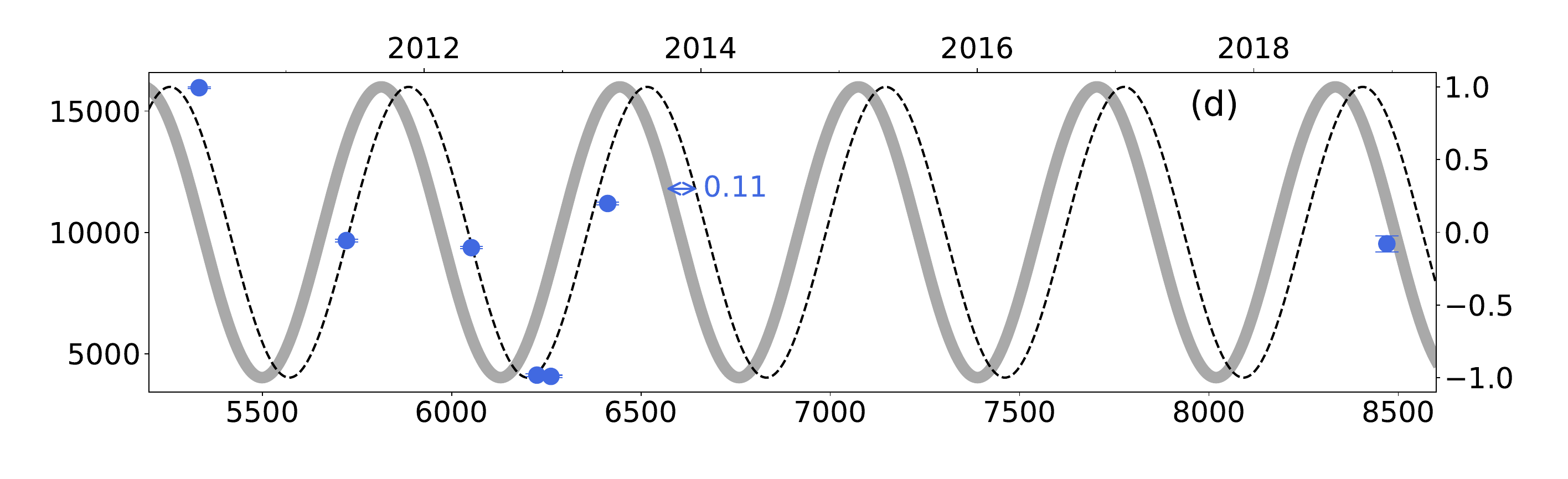}
\vspace{-3mm}
\end{minipage}
\hspace{-4mm}
\begin{minipage}[b]{9.5cm}
\includegraphics[width=0.95\textwidth]{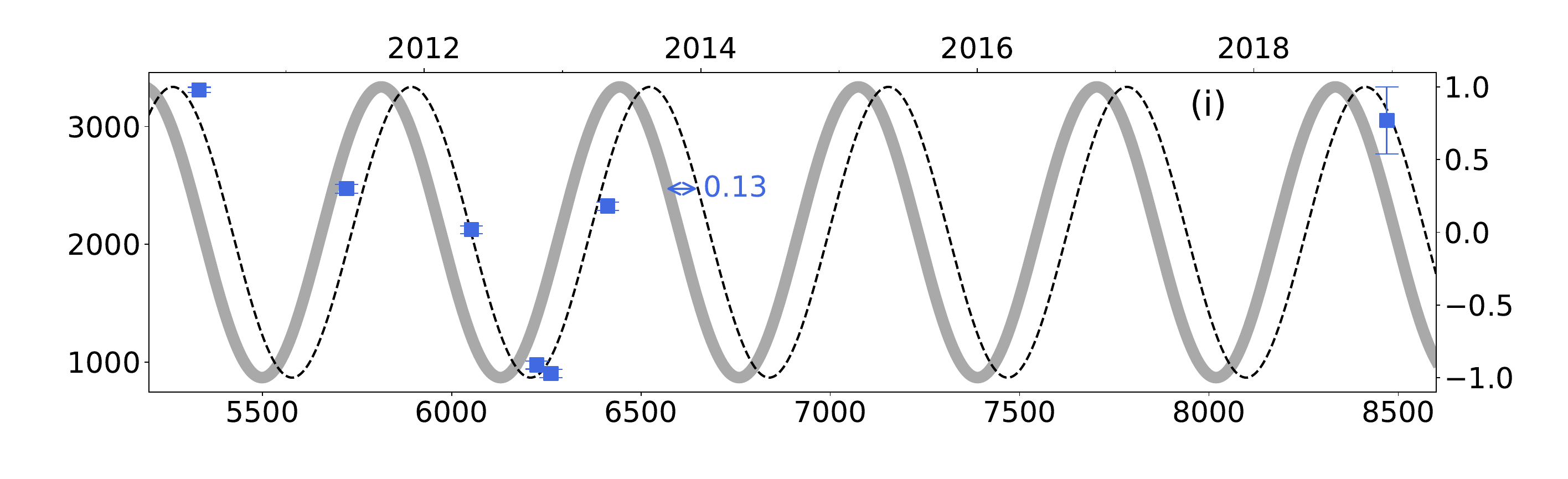}
\vspace{-3mm}
\end{minipage}
}
\mbox{
\begin{minipage}[b]{9.5cm}
\includegraphics[width=0.95\textwidth]{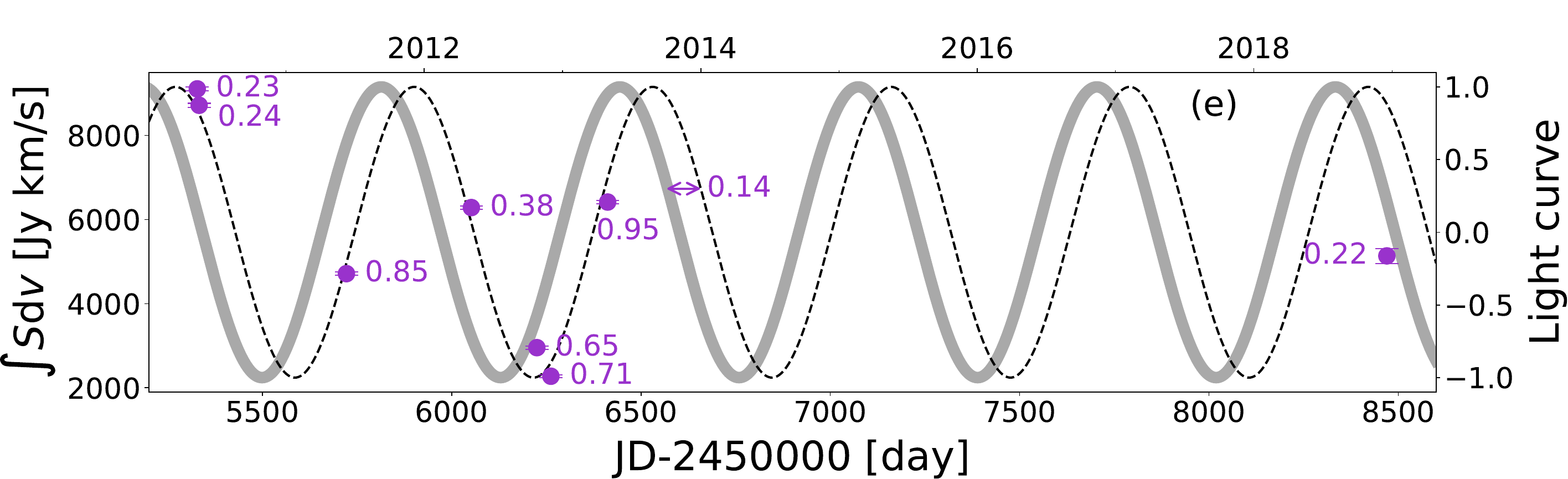}
\end{minipage}
\hspace{-4mm}
\begin{minipage}[b]{9.5cm}
\includegraphics[width=0.95\textwidth]{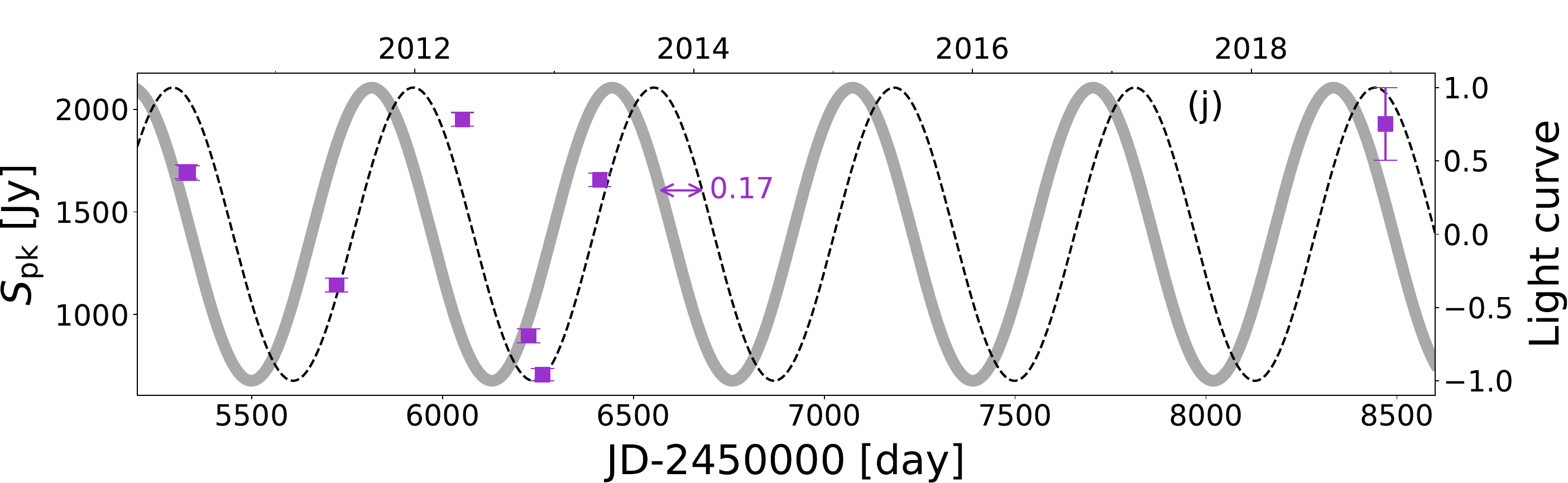}
\end{minipage}
}
\caption{Integrated intensities (left panel, dots) and peak intensities (right panel, squares) of HCN laser emission in each epoch with NIR light curve of IRC+10216. From top to bottom, the red, orange, green, blue and purple symbols represent the HCN laser transitions at 805, 891, 964, 968 and 1055~GHz, respectively. The grey curve indicates the NIR light curve, with a period of 630 days and the epoch of the maximum ($\phi_{\rm IR}$= 0), 2454554, in Julian Day format \citep{2012A&A...543A..73M}. 
The stellar phases for observed dates are labelled in the bottom left panel.
In the panels showing HCN laser emissions at 805, 968, and 1055~GHz, the black dashed curve depicts a shift from the NIR light curve, with the phase lag derived from the best fit and its value labelled in each panel.
\label{fig:IRC10216_variation_laser}}
\end{figure*}

Figure~\ref{fig:IRC10216_variation_laser} shows the substantial variations in the integrated and peak intensities of each HCN laser transition towards IRC+10216 across the observed epochs.
During these epochs, each line reached its maximum at different times. 
The integrated and peak intensities of the 891~GHz ((1,1$^{\rm 1e}$,0)--(0,4$^0$,0), $J=10-9$) and 964~GHz ((1,1$^{\rm 1e}$,0)--(0,4$^0$,0), $J=11-10$) cross-ladder lines show similar variability trends (see Figs.~\ref{fig:IRC10216_variation_laser}b,c,g,h), but they do not follow the near-infrared (NIR) light curve and differ from the trends observed in the other three 
laser lines. 
In contrast, the variations in the integrated and peak intensities of the 805~GHz ((0,4$^0$,0), $J=9-8$), 968~GHz ((1,1$^{\rm 1e}$,0), $J=11-10$), and 1055~GHz ((1,1$^{\rm 1e}$,0), $J=12-11$) lasers appear to be quasi-periodic (see Figs.~\ref{fig:IRC10216_variation_laser}a,d,e,f,i,j), similar to the NIR light curve. 
Assuming that the emissions from these three rotational transitions vary with time periodically at a constant amplitude, and further assuming that the lines share the same period as the NIR light curve but with a phase lag, we determined the phase lag for each line using least-square fitting. We found that the phase lags for the three lines range from 0.1 -- 0.2, resembling the SiS-NIR phase lag for the SiS masers in IRC+10216 \citep{2018ApJ...860..162F}, and the SiO-optical phase lag for the SiO masers in O-rich AGB stars \citep[e.g.][]{2004A&A...424..145P}.
It is worth to note that the variability behaviours of the three lines differ from the HCN $J=3-2$ lines in the $\nu_2=$ 0, 1, 2, and 3 states \citep{2019ApJ...883..165H}, as the latter vary in phase with the NIR light, with no phase lags being reported.

Our 805 and 891~GHz laser spectra towards IRC+10216 exhibit notable differences in line profiles and intensities compared to the spectra observed in 1998--2000 \citep{2000ApJ...528L..37S,2003ApJ...583..446S}. Despite these variations, the presence of these lasers seems to be sustained over a long timescale. Based on the timeline of these observations, we infer that the activities of both the 805 GHz and 891 GHz lasers may have persisted for at least 15 years.

Such notable variability in laser line profiles is not only found towards IRC+10216, but also for other C-rich stars. As shown in Fig.~\ref{fig:3stars_laser}, the HCN laser emission in the 964 and 968~GHz lines towards CIT~6, Y~CVn, and S~Cep exhibited significant variations over an observed interval of about six years. Additionally, our HCN laser spectra at 805 and 891~GHz towards CIT~6 and Y~CVn are also dramatically different from those obtained 12 years ago \citep{2003ApJ...583..446S}.

\subsection{Comparison between the lines in the Coriolis-coupled system} \label{Sec:spectral}

From Sect.~\ref{Sec:variation_description}, we observe that the spectra of IRC+10216 remain essentially unchanged over a short timescale of order a few days.
This facilitates comparisons between laser emission from different transitions observed at nearby dates. Hence, the spectra shown in each row of Fig.~\ref{fig:IRC10216_5laser} can be meaningfully compared. 

Among the six epochs during which all five laser transitions were observed by {\it Herschel}/HIFI (see Figs.~\ref{fig:IRC10216_5laser}a1--e1), the line profiles of the 891 GHz ((1,1$^{\rm 1e}$,0)--(0,4$^0$,0), $J=10-9$) and 964~GHz ((1,1$^{\rm 1e}$,0)--(0,4$^0$,0), $J=11-10$) lines show notable similarities, in contrast to the profiles of the other three laser lines.
Moreover, both the integrated and peak intensities of the 891~GHz line exhibit a variability trend (see Fig.~\ref{fig:IRC10216_variation_laser}) that is similar to that of the 964~GHz line, contrasting with the trends observed in the other three laser lines.
Y~CVn could be an exception, as it showed the strong 891~GHz laser emission on 2012 June 14, but the 964~GHz emission failed to be detected one day later. The non-detection was unlikely due to instrumental issues because strong rotational line emission at 968~GHz was detected in the same setup (Table~\ref{Tab:6stars_laser}).
For the other five C-rich stars, while the duration of stability for their laser line profiles remains undetermined, the observed line profiles at 891 and 964 GHz, which were observed within a short period of time, ranging from the same day to within three weeks, are consistently similar. Notably, as can be seen in the second and third rows of Figs.~\ref{fig:3stars_laser}--\ref{fig:3stars_laser_2}, both lines display a comparable number of discernible emission components, which are aligned in velocity.
These facts indicate that the 891 and 964~GHz HCN lasers are closely related, 
with their pumping mechanisms being strongly linked.

Among all HCN laser transitions in the Coriolis-coupled system, the 891~GHz laser emission emerges as the most luminous laser line with a photon rate of a few 10$^{44-45}$ photons~s$^{-1}$. 
Apart from Y~CVn, the 964~GHz line has the second strongest and second brightest (a few 10$^{43-44}$ photons~s$^{-1}$) laser emission. 
The 968~GHz ((1,1$^{\rm 1e}$,0), $J=11-10$) line usually shows the third brightest laser emission, followed by the weaker laser emissions at 805~GHz ((0,4$^0$,0), $J=9-8$) and 1055~GHz ((1,1$^{\rm 1e}$,0), $J=12-11$) laser lines.
The 968 and 1055~GHz lines were not detected in IRC+50096, and 1055~GHz emission was not detected in Y~CVn, whereas both sources exhibited 805~GHz laser emission. The non-detections may be attributed to inadequate sensitivity and potential variability.


For IRC+10216, the peak intensity ratio ($S_{\rm pk,891}/S_{\rm pk,964}$) of the 891 and 964~GHz laser emissions decreased from 1.9 to 1.4 over the observed epochs. 
For the other C-rich stars, the $S_{\rm pk,891}/S_{\rm pk,964}$ ratios range from 1.4 (II~Lup) to 2.3 (IRC+50096), similar to the values observed in IRC+10216.
We note that the observations of these two lines were not always conducted on the same day (with a maximum separation of 21 days for V~Cyg).
The 964~GHz laser emission was not detected towards Y~CVn on 2012 June 15 with a 1$\sigma$ noise level of 51~Jy at 0.16~\kms, leading to $S_{\rm pk,891}/S_{\rm pk,964}>$5.4. This indicates that Y~CVn exhibits a different behaviour compared to the other C-rich stars. Note that it has a much smaller mass-loss rate than any of the others and belongs to a different variability type, as it is an SRb variable, while the rest are Miras.



\section{Discussion} \label{Sec:discuss}
\subsection{Laser excitation}\label{Sec:exciatation}
Laser emission occurs due to population inversion between corresponding energy levels. 
In this section, we explore circumstellar HCN laser excitation in the Coriolis-coupled system (see Fig.~\ref{fig:HCN_level}). 
In Appendix~\ref{Sec:appendix-d}, we present additional HCN lines to have a set of spectra fully covering the rotational levels from $J=7$ to $J=13$, within and between two vibrational states, for a more comprehensive understanding for laser excitation. 

Based on the SOFIA/4GREAT and {\it Herschel}/HIFI spectra (more details given in Sect.~\ref{Sec:spectral}), our analysis reveals key observational findings: (1) the 891~GHz ((1,1$^{\rm 1e}$,0)--(0,4$^0$,0), $J=10-9$) laser emerges as the strongest; (2) the 964~GHz ((1,1$^{\rm 1e}$,0)--(0,4$^0$,0), $J=11-10$) laser is closely related to the 891~GHz laser, ranking as the second strongest; (3) the 1055~GHz ((1,1$^{\rm 1e}$,0), $J=12-11$)  emission, if detected, always has a 968~GHz ((1,1$^{\rm 1e}$,0), $J=11-10$) counterpart, with the latter being stronger; (4) the 805~GHz ((0,4$^0$,0), $J=9-8$) line co-exists with the 891~GHz line, 
and (5) the 894~GHz ((0,4$^0$,0), $J=10-9$) line is not detected towards any observed targets.

Observational findings (1) and (2) match well the scenario revealed in early laboratory studies 
\citep[e.g.][]{1964JMoSp..12...45M,1967ApPhL..11...62L}, between the vibrational states (1,1$^{\rm 1e}$,0) and (0,4$^0$,0) where significant mixing of the vibrational states at nearly degenerate levels at $J=10$, with smaller effects for the $J=9$ and $J=11$ levels. Consequently, 
the cross-ladder 
line at 891~GHz emerges as the strongest one, and both the 891 and 964~GHz lines exhibit very high amplifications \citep{1967ApPhL..11...62L}.
We highlight that the two vibrational states strongly coupled at $J=10$ (stronger) and $J=11$ affects the population in the (1,1$^{\rm 1e}$,0) and (0,4$^0$,0) vibrational states at neighbouring rotational levels, as discussed below.


The 964~GHz cross-ladder line and 968~GHz rotational line share the same upper level, (1,1$^{\rm 1e}$,0), $J=11$, and the former is typically stronger than the latter.
Similarly, the 891~GHz cross-ladder line, which exhibits the strongest emission, and the non-detected (1,1$^{\rm 1e}$,0), $J=10-9$ line at 879~GHz (see Fig.~\ref{fig:complete_hcn}) share the same upper level, (1,1$^{\rm 1e}$,0), $J=10$.
These suggest that for the coupled rotational levels (i.e. $J$ = 10 and 11), there is a preference for the excitation of the cross-ladder transition.
Furthermore, the non-detection of the cross-ladder line ((1,1$^{\rm 1e}$,0)--(0,4$^0$,0), $J=9-8$) at 816~GHz (see Fig.~\ref{fig:complete_hcn}) indicate that the two vibrational states are not strongly coupled at $J=9$. The absence of this cascade route favours maintaining the population inversion between the (0,4$^0$,0), $J=9$ and $J=8$ levels of the 805~GHz transition.

In the (1,1$^{\rm 1e}$,0) vibrational state, the upper level of the 891~GHz line is the lower level of the 968~GHz line. Similarly, the upper level of the 964 and 968~GHz lines acts as the lower level of the 1055~GHz line. 
Notably, the photon rate (i.e. isotropic laser luminosity) of the 891~GHz line is greater than the sum of the photon rates of the 964 and 968~GHz lines by a factor of 1.2--1.9. The photon rate of the 964~GHz line typically exceeds that of the 1055 GHz line by a factor of more than 1.5, and in certain epochs of IRC+10216, it surpasses the 1055~GHz line by more than an order of magnitude. 
These cascading effects indicate that the  $J=10$ rotational level is likely more depleted than the $J=11$ level within this vibrational state. This leads to an overpopulation of the $J=11$ level relative to the $J=10$ level, sustaining the observed 968~GHz laser. Similarly, this scenario may also account for the 1055~GHz laser. 
In addition, because of the population distribution regulated by the two cross-ladder laser emissions, the higher photon rate of the 891 GHz line compared to that of the 964 GHz line may explain why the detected 968 GHz emission is stronger than its 1055 GHz counterpart (observational finding 3).

The lower level of the 891~GHz line, (0,4$^0$,0), $J=9$,  serves as the upper level of the 805~GHz line. The bright 891~GHz laser emission causes the upper level of the 805~GHz transition to become overpopulated relative to its lower level, facilitating the coincidence of the two lasers (observational finding 4).

On the other hand, the upper and lower levels of the (0,4$^0$,0), $J=10-9$ line at 894~GHz correspond to the lower levels of the 964 and 891~GHz lines, respectively. The consistently stronger 891~GHz emission could prevent the lower level of the 894~GHz line from becoming sufficiently under-populated relative to its upper level, thereby failing to create the population inversion necessary for generating the 894~GHz laser emission. This could be the reason why we did not detect the 894~GHz line emission towards any of our targets (observational finding 5). 



\subsection{Pumping considerations}\label{sec:pumping}

As shown in Sect.~\ref{Sec:variation_description}, there appear to be two distinct patterns of variability among the five detected HCN laser transitions. 
The 891 and 964~GHz cross-ladder lines do not follow the NIR light curve, while the variations of the rotational lines at 805, 968, and 1055~GHz appear to be quasi-periodic, with a phase lag of 0.1--0.2 relative to the NIR light curve. 
Given that the cross-ladder lines are probably the main pumps of the corresponding rotational lines (see details in Sect.~\ref{Sec:exciatation}), in this section we discuss 1) the possible pumping mechanisms, radiative and chemical pumping, for the cross-ladder lines; and 2) collisional and radiative pumping may modulate the three rotational lines.

For radiative pumping, direct excitation of HCN molecules from the ground state to the excited (1,1$^{\rm 1e}$,0) state requires infrared photons at 3.6~$\mu$m \citep{1934PhRv...45..277A,2014MNRAS.437.1828B}.
To evaluate the feasibility of radiative pumping, we compare the luminosities of the brightest laser species at 891~GHz ($L_{\rm 891}$, taken from Tables~\ref{Tab:IRC+10216_laser} and \ref{Tab:6stars_laser}) with the available NIR photon luminosities that close to the 3.6~$\mu$m, for the 7 stars with HCN laser detections. 
Values for the NIR photon luminosities were calculated from flux densities listed in Table~\ref{Tab:ir-photon}, adopting a velocity range that 891~GHz HCN laser emission covered.
In addition, we assume that the infrared flux density of all stars in our sample varies by less than a factor of 4 over the stellar pulsation cycle, based on the $L-$band (at 3.5~$\mu$m) light curve of IRC+10216 measured between 1999 December 10 and 2008 November 11 \citep{2011ARep...55...31S}.
Since the NIR photon luminosities were not significantly greater than the 891~GHz HCN laser luminosities (see Table~\ref{Tab:ir-photon}), direct radiative pumping from the ground state to the (1,1$^{\rm 1e}$,0) level is probably not efficient enough to fully explain the observed laser luminosity. 
We caution that, especially for high mass-loss rate stars such as IRC+10216, the stellar emission at 3.5~$\mu$m in the inner envelope, where lasers arise, could be stronger than observed. This is because the flux is related to the emission of the dusty component of the envelope, which absorbs stellar emission and re-emits it at longer wavelengths.

Under the high temperature and gas density of the laser-emitting region, the population of HCN in vibrationally excited states such as (0,1$^{\rm 1e}$,0) becomes significant, as evidenced by various observations of vibrationally excited HCN in carbon-rich stars \citep[e.g.][]{2008ApJ...673..445F,2021A&A...651A...8F,2011A&A...529L...3C,2023Natur.617..696V}. Transitions from these states to the (1,1$^{\rm 1e}$,0) state depend on mid-infrared photons, which exhibit luminosities exceeding those of the 891 GHz HCN lasers. Thus, radiative pumping of HCN lasers through these vibrational states remains a viable possibility.


\begin{table}[htbp]
\caption{Comparison of 891~GHz HCN laser and NIR photon luminosities.}\label{Tab:ir-photon} 
\normalsize
\centering
\renewcommand\arraystretch{1.2}
\setlength{\tabcolsep}{2pt}
\begin{tabular}{lccccr}
\hline \hline 
Object  & $\lambda$ & $S_{\lambda}$ & Catalogue & $L_\lambda$ & $L_\lambda/L_{\rm 891}$\\
 & ($\mu$m) & (Jy) &  & (photons s$^{-1}$) & \\
\hline
IRC+10216\tablefootmark{*} & 3.35 & 1330 & unWISE & 2.4$\times$10$^{44}$ & 0.39 \\
CIT~6     & 3.35 & 580  & unWISE & 7.4$\times$10$^{44}$ & 0.46 \\
Y~CVn     & 3.35 & 934  & unWISE & 3.6$\times$10$^{44}$ & 2.57 \\  
          & 3.52 & 931  & DIRBE  & 3.6$\times$10$^{44}$ & 2.57 \\
S~Cep     & 3.35 & 837  & unWISE & 8.7$\times$10$^{44}$ & 1.07 \\  
          & 3.52 & 775  & DIRBE  & 8.1$\times$10$^{44}$ & 1 \\
IRC+50096 & 3.35 & 288  & unWISE & 8.7$\times$10$^{44}$ & 1.05 \\  
          & 3.52 & 313  & DIRBE  & 9.5$\times$10$^{44}$ & 1.14  \\
V~Cyg     & 3.35 & 683  & unWISE & 7.2$\times$10$^{44}$ & 0.83 \\
II~Lup    & 3.35 & 203  & unWISE & 4.7$\times$10$^{44}$ & 0.49 \\  
          & 3.52 & 295  & DIRBE  & 6.8$\times$10$^{44}$ & 0.72 \\
\hline 
\end{tabular}
\normalsize
\tablefoot{Flux densities obtained from the VizieR database \citep{2000A&AS..143...23O} via \url{http://vizier.cds.unistra.fr/vizier/sed/.} \\
References: unWISE \citep{2014AJ....147..108L}; DIRBE \citep{2004ApJS..154..673S,2010ApJS..190..203P}. \\
\tablefoottext{*}{The 891~GHz HCN laser data observed in 2010 May 13 were used here.}
}
\end{table}

Chemical pumping has been invoked to explain the laboratory HCN lasers. Various authors \citep[e.g.][]{Chantry1971,Kunstreich1975,1978OptCo..27..281R} proposed a reaction between CN and H$_2$ in a discharge (see below reaction \ref{eq:hcn}) to be the pump of the lines of the HCN Coriolis-coupled system, despite it was challenged by \citet{Skatrud1984}, who argued that the abundance of CN is much too low in such a discharge. 
As we shall further discuss, indeed sufficient HCN may be produced to make chemical pumping viable for stellar HCN lasers, as discussed by \cite{2000ApJ...528L..37S} and \cite{2003ApJ...583..446S}.
The underlying concept is that HCN molecules are preferentially formed in specific vibrational states (i.e. the Coriolis-coupled system), depending on their formation environment. 
The decay rates from the bending stack (0, $v_2$, 0) to lower vibrational states are much higher than those from the (1,1$^{\rm 1e}$,0) state \citep[see][]{Smith1981,1986ApJ...300L..19Z}, this helps sustaining any population inversion between the (1,1$^{\rm 1e}$,0) and (0,4$^0$,0) states.

Given that HCN is a parent molecule typically formed close to the stellar atmosphere \citep[e.g.][]{1964AnTok...9.....T,2006A&A...456.1001C,2014A&A...568A.111L,2020A&A...637A..59A}, a chemical pumping mechanism is possible. The excitation temperature of HCN can be as high as ${\sim}1000$~K \citep{2011A&A...529L...3C,2022A&A...666A..69J}, providing suitable conditions for the direct formation of HCN molecules in vibrationally excited states (e.g. the Coriolis-coupled system). Here, we revisit this scenario with the updated physical parameters and reaction rate coefficients. 

As noted by previous studies \citep[e.g.][]{2006A&A...456.1001C},
the formation of HCN in the innermost regions of circumstellar envelopes of AGB stars is primarily driven by the hydrogen abstraction reaction between the CN radical and H$_{2}$,  
\begin{equation}
  {\rm CN + H_2 \rightarrow HCN + H }\;,
  \label{eq:hcn}
\end{equation}
with a reaction rate coefficient of $k=4.96\times 10^{-13}(T_{\rm k}/300)^{2.6}{\rm exp}(-960/T_{\rm k})$~cm$^{3}$~s$^{-1}$, where $T_{\rm k}$ is the kinetic temperature, according to the KInetic Database for Astrochemistry (KIDA\footnote{\url{https://kida.astrochem-tools.org/}}; \citealt{2005JPCRD..34..757B}). 
Because reaction~(\ref{eq:hcn}) is the main pathway for the destruction of CN and its back reaction is impeded by a too high activation barrier of $\sim 10^4$~K, the HCN formation rate is determined by $kn({\rm H_{2}})n({\rm CN})$, where $n({\rm H_{2}})$ and $n({\rm CN})$ are the number densities of H$_{2}$ and CN, respectively.

In the innermost regions of circumstellar envelopes of C-rich stars, the fractional abundance of CN, $X({\rm CN})$, is estimated to be ${\sim}10^{-6}$ under thermal equilibrium \citep{2006A&A...456.1001C,2020A&A...637A..59A}. Assuming the H$_2$ gas density of $n({\rm H_{2}}) \sim 10^{10}$~cm$^{-3}$ and the gas temperature of $T_{\rm k} \sim 1000$~K, the rate of reaction~(\ref{eq:hcn}) is 435~cm$^{-3}$~s$^{-1}$. We then estimate whether such a condition can produce a sufficient amount of HCN molecules per unit time to account for the observed laser luminosities. 
We assume a cylindrical geometry with an aspect ratio of $a = l_{\rm los}/d=1$ for the laser-emitting zone, where $l_{\rm los}$ is the length along the line of sight and $d$ is the diameter of laser spots. 
Based on the ALMA imaging results of R~Lep \citep{2023ApJ...958...86A}, we adopt a laser spot diameter of 10~au. 
Thus, we derive an HCN production rate of $1.1 \times 10^{45}$~s$^{-1}$, which exceeds or is comparable to the observed laser photon rates as shown in Tables~\ref{Tab:IRC+10216_laser} and \ref{Tab:6stars_laser}. The HCN production rate could be enhanced if multiple spots are present in the inner regions. On the other hand, the HCN production rate may not fully correspond to the laser photon rate if not all HCN molecules contribute to laser emission. An estimate of such efficiency is beyond the scope of the current work.

However, while electronic transitions from warm CN have been detected in the optical spectra of carbon-rich stars \citep{1994ApJS...90..317B,1997A&A...323..469B}, radio CN emission has not been observed in the extended atmosphere where HCN lasers are formed \citep[e.g.][]{2018A&A...610A...4G,2024A&A...684A...4U}. This means that $X({\rm CN})$ beyond the radio photosphere must be rather low (except for the outer radii where photodissociation takes place). Non-equilibrium models predict an extremely low $X({\rm CN})$ in regions dominated by shocks \citep[see Table 4 in][]{1998A&A...330..676W}. The absence of CN close to the radio photosphere could be a result of shock-induced chemistry, although direct confirmation of shocks in emission line observations is still lacking \citep[e.g.][]{2023Natur.617..696V}.


We perform similar estimates of the HCN production rate using the parameters modelled by \citet[][their Tables 2 and 4]{1998A&A...330..676W} for a shocked region at 5$R_{\star}$, which we adopt as a lower limit estimate. The resultant CN abundance at this radius is about $X({\rm CN}) = 8.37 \times 10^{-13}$. The rates of reaction~(\ref{eq:hcn}) are 951~cm$^{-3}$~s$^{-1}$ in the shock front for $T_{\rm k} = 5079$~K and $n({\rm H_{2}}) = 1.33 \times 10^{12}$~cm$^{-3}$, and 72~cm$^{-3}$~s$^{-1}$ in the post-shock gas for $T_{\rm k} = 1544$~K and $n({\rm H_{2}}) = 2.13 \times 10^{12}$~cm$^{-3}$. 
Assuming the same cylindrical geometry as in our estimate under thermal equilibrium, we derive an HCN production rate of $2.5 \times 10^{45}$~s$^{-1}$ and $1.9 \times 10^{44}$~s$^{-1}$ 
for the shock front and post-shock region, respectively.
These rates are also comparable to the highest observed laser photon rates in our Tables~\ref{Tab:IRC+10216_laser} and \ref{Tab:6stars_laser}. 
For $R<5R_{\star}$, the modelled gas temperatures and H$_2$ number densities  in both the shock front and post-shock gas are higher than those at 5$R_{\star}$ \citep{1998A&A...330..676W}. This results in an increased rate of reaction~(\ref{eq:hcn}) and, consequently, a greater HCN production rate, suggesting that chemical pumping may, in principle, work at R$\lesssim$5$R_{\star}$.  
Hence, chemical pumping could be a viable mechanism to produce the observed HCN lasers despite a low CN abundance, provided that the gas temperature or the gas density, or both, are high enough to ensure a sufficiently high rate of reaction~(\ref{eq:hcn}).

The variations in the rotational lines at 805, 968, and 1055~GHz appear to show a quasi-periodic pattern, suggesting that additional pumping mechanisms may periodically alter the conditions of the laser gas, thereby modulating these three rotational transitions.
Cyclic stellar pulsations and convective motions are likely to generate periodic shocks, leading to the density variations in the innermost regions of circumstellar envelopes. 
Quantum dynamics calculations have shown that vibrationally elastic ($\Delta \nu=0$; purely rotational) excitation of linear molecules through collisions is almost independent of the vibrational state
\citep[e.g.][]{2013ChRv..113.8906R,2017MNRAS.469.1673B}. The collisional rate coefficients for the vibrationally inelastic process are typically orders of magnitude lower than those of the vibrationally elastic process \citep[e.g.][]{2008A&A...492..257F,2013ChRv..113.8906R,2017MNRAS.469.1673B}, making collisional excitation more likely to alter populations within the same vibrational state. If this conventional view holds true for the HCN vibrational states in the Coriolis-coupled system, collisions would influence pure rotational transitions more than cross-ladder transitions. However, the effects of collisions on the cross-ladder transitions are much less clear. While the high temperature and density in the laser-forming region may enhance the contribution of collisions to ro-vibrational transitions, it has been shown that state-to-state rate coefficients of collision-induced energy transfer may not necessarily be enhanced by intramolecular perturbations, such as Coriolis coupling \citep{Orr1987,2018MolPh.116.3666O}.
Besides collisions, infrared luminosities may contribute to the pumping of the quasi-periodic lasers because the laser variation period matches that of the NIR light curve. Therefore, we propose that periodic shocks and variations in infrared luminosity should play a role in modulating the variation of the laser luminosity. 



\subsection{Comparison with masers in O-rich AGB stars}

The pumping mechanisms responsible for SiO \citep[e.g.,][]{1994A&A...285..953B,1994A&A...285..971B,2014A&A...565A.127D}, H$_{2}$O \citep[e.g.,][]{2016MNRAS.456..374G}, OH \citep[e.g.,][]{1976ApJ...205..384E,1992ARA&A..30...75E}, and SiS masers \citep[e.g.,][]{2006ApJ...646L.127F,2017ApJ...843...54G} in evolved stars are primarily radiative and collisional pumping, with chemical pumping not being considered in their formation. On the other hand, inspired by laboratory HCN lasers and the early work of \citet{2000ApJ...528L..37S} and \citet{2003ApJ...583..446S}, chemical pumping is considered in Sect.~\ref{sec:pumping} to explain the formation of HCN lasers in C-rich stars, alongside the traditional radiative and collisional mechanisms. Considering potential differences, we performed a comparative analysis of the properties of masers and lasers in O-rich and C-rich stars.

In O-rich stars, the common molecular maser species SiO, H$_2$O, and OH are found at increasing distances from the central star \citep[e.g.][]{1981ARA&A..19..231R,1996A&ARv...7...97H}.
SiO masers occur within a few stellar radii of the stellar surface, situated between the radio photosphere and the dust formation zone \citep{1997ApJ...476..327R}.
Vibrationally excited SiO masers in the $v=1$ and $v=2$ states have been widely detected, exhibiting isotropic luminosities of a few 10$^{42-44}$ photons~s$^{-1}$ \cite[e.g.][]{2014AJ....147...22K}.
To date, SiO masers in vibrational states up to $v$ = 4 ($E_{\rm up} >$7000~K) have been detected in the red supergiant VY~CMa \citep{1993ApJ...407L..33C}, with a tentative maser detection in the $v$=6 state reported in R~Cas and $\chi$~Cyg \citep{2021ApJS..253...44R}, an O-rich and an S-type AGB star, respectively.  

In addition, vibrationally excited H$_2$O maser emission is likely to originate from the same parts of the envelope as SiO masers \citep[e.g.][]{1989ApJ...341L..91M,1995ApJ...450L..67M,2020ApJS..247...23A}, that is closer to the star than the layer harbouring 22~GHz H$_2$O masers in the vibrational ground state.
Unlike other weak vibrationally excited H$_2$O masers \citep[see review in][]{2007IAUS..242..471H}, the $J_{K_a,K_c}=1_{1,0}-1_{0,1}$ line at 658~GHz in the $v_2$ = 1 bending mode appears to be widespread and bright \citep[e.g.][]{1995ApJ...450L..67M,2007IAUS..242..481H,2018A&A...609A..25B}. 
The typical isotropic luminosity of the 658~GHz maser line is a few 10$^{43-44}$ photons~s$^{-1}$, with the exception of the extremely high luminosities of 10$^{46}$ photons~s$^{-1}$ in the red supergiants VY~CMa and VX~Sgr, for which the line is brighter than the  22~GHz transition, which is commonly the strongest H$_2$O maser line \citep{1995ApJ...450L..67M}.

We further compare the properties of HCN lasers in C-rich stars with those of masers in O-rich stars, using the brightest 891~GHz laser emission as an example.
They share three key similarities: widespread occurrence among stars, great brightness, and originating in the innermost regions of circumstellar envelopes.
The 891 GHz laser emission has been detected in seven of eight C-rich stars observed by {\it Herschel}/HIFI. Combined with the 891~GHz lasers reported by \cite{2019asrc.confE..55W}, there are a total of twelve C-rich stars known to have 891~GHz lasers. 
These observations and the high detection rates suggest that the laser emission in the 891~GHz transition is widespread in C-rich AGB stars.
The 891~GHz HCN laser has a typical isotropic luminosity of 10$^{44}$ photons~s$^{-1}$, comparable to that of the vibrationally excited SiO and H$_2$O masers in O-rich stars.
The upper energy levels of these HCN laser transitions exceed 4000~K. The lines cover velocity ranges that are much smaller than twice the terminal velocity of stellar wind. The high-angular resolution (better than 100~mas) images of the 891~GHz laser emission reveal its location within only a few au from stars \citep{2019asrc.confE..55W,2023ApJ...958...86A}. These findings support that HCN laser emission occurs in the innermost region of circumstellar envelopes of C-rich stars where dust and molecular species are forming and the molecular gas is being accelerated.
Therefore, the HCN lasers studied in this work may serve as analogues to the vibrationally excited SiO and H$_2$O masers in O-rich stars.

\section{Summary}\label{Sec:sum}

In this work, we explore the HCN lines belonging to the Coriolis-coupled system between the (1,1$^{\rm 1e}$,0) and (0,4$^0$,0) vibrational states, with frequencies above 900~GHz, for the first time in astronomical objects.
We performed SOFIA/4GREAT observations and collected all {\it Herschel}/HIFI archival data that cover all six HCN transitions in the Coriolis-coupled system for a sample of eight C-rich AGB stars. The main results are summarised as follows:

\begin{itemize}

\item[1.] The HCN lasers at 964, 968 and 1055~GHz were newly discovered towards C-rich stars.
The 805, 891 and 964~GHz HCN laser emission were detected in seven stars, the 968~GHz laser in six stars and the 1055~GHz laser in five stars. However, the 894~GHz emission was not detected towards any of the targets. \\

\item[2.]  In this Coriolis-coupled system, the cross-ladder line at 891~GHz is always the strongest, with a typical luminosity of a few 10$^{44}$ photons~s$^{-1}$. The 964~GHz laser is the second strongest laser, with a similar line profile to the 891~GHz laser. The 1055~GHz laser always has a stronger 968 GHz laser counterpart. The 805~GHz laser co-exists with the 891~GHz laser. Building on these results and the non-detection of the 894 GHz emission, we provide insights into a more complete picture of circumstellar HCN laser excitation. \\


\item[3.] 
Towards IRC+10216, all five HCN laser transitions were observed in six to eight epochs and exhibited significant variations in line profiles and intensities. The cross-ladder lines at 891 and 964~GHz exhibit similar variations, but their intensity variations do not follow a periodic light curve. In contrast, the variations of the rotational lines at 805, 968, and 1055~GHz appear to be quasi-periodic with a phase lag of 0.1 -- 0.2 relative to the NIR light curve. \\

\item[4.] We suggest chemical pumping and radiative pumping could play an important role in the production of the cross-ladder HCN lasers, while the 
quasi-periodic behaviour of the rotational HCN laser lines may be modulated by additional collisional and radiative pumping driven by periodic shocks and variations in infrared luminosity. \\


\item[5.] These HCN lasers could be an analogue of vibrationally excited SiO and H$_2$O masers in O-rich stars, sharing three key similarities: widespread occurrence, high brightness, and origin in the innermost regions of circumstellar envelopes. \\

\end{itemize}  

Given their high flux densities, existing successful self-calibration approach and band-to-band calibration \citep{2019asrc.confE..55W,2023ApJ...958...86A}, HCN lasers in the Coriolis-coupled system serve as excellent phase calibrators for high-frequency and long-baseline observations (i.e. in ALMA Band~10), which can achieve unprecedented angular resolutions. These resolutions 
allow an in-depth exploration of the physics and dynamics of  the innermost regions of circumstellar envelopes of C-rich stars.

\section*{ACKNOWLEDGMENTS}\label{sec.ack}
We sincerely thank the referee for the thorough review and highly valuable comments, which have improved the manuscript.
This work is based on observations made with the NASA/DLR Stratospheric Observatory for Infrared Astronomy (SOFIA). SOFIA is jointly operated by the Universities Space Research Association, Inc. (USRA), under NASA contract NNA17BF53C, and the Deutsches SOFIA Institut (DSI) under DLR contract 50 OK 0901 to the University of Stuttgart. We deeply appreciate the exceptional support from the SOFIA Operations Team and the GREAT Instrument Team for the observing effort.
W.Y. acknowledges the support from the National Natural Science Foundation of China (12273010, 12403027), China Postdoctoral Science Foundation (2024M751376), and Jiangsu Funding Programme for Excellent Postdoctoral Talent (2024ZB347). 
K.T.W. acknowledges support from the European Research Council (ERC) under the European Union's Horizon 2020 research and innovation programme (Grant agreement no. 883867, project EXWINGS).
Y.G. is supported by the Strategic Priority Research Program of the Chinese Academy of Sciences, Grant No. XDB0800301. 
J.C. thanks Spanish Ministry of Science and Technology for funding support under grant PID2019-107115GB-C21.
This work made use of Python libraries including SciPy\footnote{\url{https://www.scipy.org/}} \citep{jones2001scipy}, NumPy\footnote{\url{https://www.numpy.org/}} \citep{5725236}, Matplotlib\footnote{\url{https://matplotlib.org/}} \citep{Hunter:2007}, and Overleaf\footnote{\url{https://www.overleaf.com}}.

\bibliographystyle{aa}
\bibliography{references}

\begin{thebibliography}{107}
\expandafter\ifx\csname natexlab\endcsname\relax\def\natexlab#1{#1}\fi

\bibitem[{{Abia} \& {Isern}(2000)}]{2000ApJ...536..438A}
{Abia}, C. \& {Isern}, J. 2000, \apj, 536, 438

\bibitem[{{Adel} \& {Barker}(1934)}]{1934PhRv...45..277A}
{Adel}, A. \& {Barker}, E.~F. 1934, Physical Review, 45, 277

\bibitem[{{Ag{\'u}ndez} {et~al.}(2012){Ag{\'u}ndez}, {Fonfr{\'\i}a}, {Cernicharo}, {Kahane}, {Daniel}, \& {Gu{\'e}lin}}]{2012A&A...543A..48A}
{Ag{\'u}ndez}, M., {Fonfr{\'\i}a}, J.~P., {Cernicharo}, J., {et~al.} 2012, \aap, 543, A48

\bibitem[{{Ag{\'u}ndez} {et~al.}(2020){Ag{\'u}ndez}, {Mart{\'\i}nez}, {de Andres}, {Cernicharo}, \& {Mart{\'\i}n-Gago}}]{2020A&A...637A..59A}
{Ag{\'u}ndez}, M., {Mart{\'\i}nez}, J.~I., {de Andres}, P.~L., {Cernicharo}, J., \& {Mart{\'\i}n-Gago}, J.~A. 2020, \aap, 637, A59

\bibitem[{{Alcolea} {et~al.}(1999){Alcolea}, {Pardo}, {Bujarrabal}, {Bachiller}, {Barcia}, {Colomer}, {Gallego}, {G{\'o}mez-Gonz{\'a}lez}, {del Pino Cisneros}, {Planesas}, {del R{\'\i}o}, {Rodr{\'\i}guez-Franco}, {del Romero}, {Tafalla}, \& {de Vicente}}]{Alcolea1999}
{Alcolea}, J., {Pardo}, J.~R., {Bujarrabal}, V., {et~al.} 1999, \aaps, 139, 461

\bibitem[{{Asaki} {et~al.}(2020){Asaki}, {Maud}, {Fomalont}, {Phillips}, {Hirota}, {Sawada}, {Barcos-Mu{\~n}oz}, {Richards}, {Dent}, {Takahashi}, {Corder}, {Carpenter}, {Villard}, \& {Humphreys}}]{2020ApJS..247...23A}
{Asaki}, Y., {Maud}, L.~T., {Fomalont}, E.~B., {et~al.} 2020, \apjs, 247, 23

\bibitem[{{Asaki} {et~al.}(2023){Asaki}, {Maud}, {Francke}, {Nagai}, {Petry}, {Fomalont}, {Humphreys}, {Richards}, {Wong}, {Dent}, {Hirota}, {Fernandez}, {Takahashi}, \& {Hales}}]{2023ApJ...958...86A}
{Asaki}, Y., {Maud}, L.~T., {Francke}, H., {et~al.} 2023, \apj, 958, 86

\bibitem[{{Bakker} {et~al.}(1997){Bakker}, {van Dishoeck}, {Waters}, \& {Schoenmaker}}]{1997A&A...323..469B}
{Bakker}, E.~J., {van Dishoeck}, E.~F., {Waters}, L.~B.~F.~M., \& {Schoenmaker}, T. 1997, \aap, 323, 469

\bibitem[{{Balan{\c{c}}a} \& {Dayou}(2017)}]{2017MNRAS.469.1673B}
{Balan{\c{c}}a}, C. \& {Dayou}, F. 2017, \mnras, 469, 1673

\bibitem[{{Barber} {et~al.}(2014){Barber}, {Strange}, {Hill}, {Polyansky}, {Mellau}, {Yurchenko}, \& {Tennyson}}]{2014MNRAS.437.1828B}
{Barber}, R.~J., {Strange}, J.~K., {Hill}, C., {et~al.} 2014, \mnras, 437, 1828

\bibitem[{{Barnbaum}(1994)}]{1994ApJS...90..317B}
{Barnbaum}, C. 1994, \apjs, 90, 317

\bibitem[{{Baudry} {et~al.}(2018){Baudry}, {Humphreys}, {Herpin}, {Torstensson}, {Vlemmings}, {Richards}, {Gray}, {De Breuck}, \& {Olberg}}]{2018A&A...609A..25B}
{Baudry}, A., {Humphreys}, E.~M.~L., {Herpin}, F., {et~al.} 2018, \aap, 609, A25

\bibitem[{{Baulch} {et~al.}(2005){Baulch}, {Bowman}, {Cobos}, {Cox}, {Just}, {Kerr}, {Pilling}, {Stocker}, {Troe}, {Tsang}, {Walker}, \& {Warnatz}}]{2005JPCRD..34..757B}
{Baulch}, D.~L., {Bowman}, C.~T., {Cobos}, C.~J., {et~al.} 2005, Journal of Physical and Chemical Reference Data, 34, 757

\bibitem[{{Becklin} {et~al.}(1969){Becklin}, {Frogel}, {Hyland}, {Kristian}, \& {Neugebauer}}]{1969ApJ...158L.133B}
{Becklin}, E.~E., {Frogel}, J.~A., {Hyland}, A.~R., {Kristian}, J., \& {Neugebauer}, G. 1969, \apjl, 158, L133

\bibitem[{{Brown} {et~al.}(1975){Brown}, {Hougen}, {Huber}, {Johns}, {Kopp}, {Lefebvre-Brion}, {Merer}, {Ramsay}, {Rostas}, \& {Zare}}]{1975JMoSp..55..500B}
{Brown}, J.~M., {Hougen}, J.~T., {Huber}, K.~P., {et~al.} 1975, Journal of Molecular Spectroscopy, 55, 500

\bibitem[{{Bujarrabal}(1994{\natexlab{a}})}]{1994A&A...285..953B}
{Bujarrabal}, V. 1994{\natexlab{a}}, \aap, 285, 953

\bibitem[{{Bujarrabal}(1994{\natexlab{b}})}]{1994A&A...285..971B}
{Bujarrabal}, V. 1994{\natexlab{b}}, \aap, 285, 971

\bibitem[{{Cernicharo} {et~al.}(2011){Cernicharo}, {Ag{\'u}ndez}, {Kahane}, {Gu{\'e}lin}, {Goicoechea}, {Marcelino}, {De Beck}, \& {Decin}}]{2011A&A...529L...3C}
{Cernicharo}, J., {Ag{\'u}ndez}, M., {Kahane}, C., {et~al.} 2011, \aap, 529, L3

\bibitem[{{Cernicharo} {et~al.}(1996){Cernicharo}, {Barlow}, {Gonzalez-Alfonso}, {Cox}, {Clegg}, {Nguyen-Q-Rieu}, {Omont}, {Guelin}, {Liu}, {Sylvester}, {Lim}, {Griffin}, {Swinyard}, {Unger}, {Ade}, {Baluteau}, {Caux}, {Cohen}, {Emery}, {Fischer}, {Furniss}, {Glencross}, {Greenhouse}, {Gry}, {Joubert}, {Lorenzetti}, {Nisini}, {Orfei}, {Pequignot}, {Saraceno}, {Serra}, {Skinner}, {Smith}, {Towlson}, {Walker}, {Armand}, {Burgdorf}, {Ewart}, {di Giorgio}, {Molinari}, {Price}, {Sidher}, {Texier}, \& {Trams}}]{1996A&A...315L.201C}
{Cernicharo}, J., {Barlow}, M.~J., {Gonzalez-Alfonso}, E., {et~al.} 1996, \aap, 315, L201

\bibitem[{{Cernicharo} {et~al.}(1993){Cernicharo}, {Bujarrabal}, \& {Santaren}}]{1993ApJ...407L..33C}
{Cernicharo}, J., {Bujarrabal}, V., \& {Santaren}, J.~L. 1993, \apjl, 407, L33

\bibitem[{{Cernicharo} {et~al.}(2013){Cernicharo}, {Daniel}, {Castro-Carrizo}, {Agundez}, {Marcelino}, {Joblin}, {Goicoechea}, \& {Gu{\'e}lin}}]{2013ApJ...778L..25C}
{Cernicharo}, J., {Daniel}, F., {Castro-Carrizo}, A., {et~al.} 2013, \apjl, 778, L25

\bibitem[{{Cernicharo} {et~al.}(2000){Cernicharo}, {Gu{\'e}lin}, \& {Kahane}}]{2000A&AS..142..181C}
{Cernicharo}, J., {Gu{\'e}lin}, M., \& {Kahane}, C. 2000, \aaps, 142, 181

\bibitem[{{Cernicharo} {et~al.}(2014){Cernicharo}, {Teyssier}, {Quintana-Lacaci}, {Daniel}, {Ag{\'u}ndez}, {Velilla-Prieto}, {Decin}, {Gu{\'e}lin}, {Encrenaz}, {Garc{\'\i}a-Lario}, {de Beck}, {Barlow}, {Groenewegen}, {Neufeld}, \& {Pearson}}]{2014ApJ...796L..21C}
{Cernicharo}, J., {Teyssier}, D., {Quintana-Lacaci}, G., {et~al.} 2014, \apjl, 796, L21

\bibitem[{{Cernicharo} {et~al.}(2010){Cernicharo}, {Waters}, {Decin}, {Encrenaz}, {Tielens}, {Ag{\'u}ndez}, {De Beck}, {M{\"u}ller}, {Goicoechea}, {Barlow}, {Benz}, {Crimier}, {Daniel}, {di Giorgio}, {Fich}, {Gaier}, {Garc{\'\i}a-Lario}, {de Koter}, {Khouri}, {Liseau}, {Lombaert}, {Erickson}, {Pardo}, {Pearson}, {Shipman}, {S{\'a}nchez Contreras}, \& {Teyssier}}]{2010A&A...521L...8C}
{Cernicharo}, J., {Waters}, L.~B.~F.~M., {Decin}, L., {et~al.} 2010, \aap, 521, L8

\bibitem[{{Chantry}(1971)}]{Chantry1971}
{Chantry}, G.~W. 1971, {Submillimetre spectroscopy} (London: Academic), 241

\bibitem[{{Cherchneff}(2006)}]{2006A&A...456.1001C}
{Cherchneff}, I. 2006, \aap, 456, 1001

\bibitem[{{de Graauw} {et~al.}(2010){de Graauw}, {Helmich}, {Phillips}, {Stutzki}, {Caux}, {Whyborn}, {Dieleman}, {Roelfsema}, {Aarts}, {Assendorp}, {Bachiller}, {Baechtold}, {Barcia}, {Beintema}, {Belitsky}, {Benz}, {Bieber}, {Boogert}, {Borys}, {Bumble}, {Ca{\"\i}s}, {Caris}, {Cerulli-Irelli}, {Chattopadhyay}, {Cherednichenko}, {Ciechanowicz}, {Coeur-Joly}, {Comito}, {Cros}, {de Jonge}, {de Lange}, {Delforges}, {Delorme}, {den Boggende}, {Desbat}, {Diez-Gonz{\'a}lez}, {di Giorgio}, {Dubbeldam}, {Edwards}, {Eggens}, {Erickson}, {Evers}, {Fich}, {Finn}, {Franke}, {Gaier}, {Gal}, {Gao}, {Gallego}, {Gauffre}, {Gill}, {Glenz}, {Golstein}, {Goulooze}, {Gunsing}, {G{\"u}sten}, {Hartogh}, {Hatch}, {Higgins}, {Honingh}, {Huisman}, {Jackson}, {Jacobs}, {Jacobs}, {Jarchow}, {Javadi}, {Jellema}, {Justen}, {Karpov}, {Kasemann}, {Kawamura}, {Keizer}, {Kester}, {Klapwijk}, {Klein}, {Kollberg}, {Kooi}, {Kooiman}, {Kopf}, {Krause}, {Krieg}, {Kramer}, {Kruizenga}, {Kuhn}, {Laauwen}, {Lai}, {Larsson}, {Leduc}, {Leinz}, {Lin},
  {Liseau}, {Liu}, {Loose}, {L{\'o}pez-Fernandez}, {Lord}, {Luinge}, {Marston}, {Mart{\'\i}n-Pintado}, {Maestrini}, {Maiwald}, {McCoey}, {Mehdi}, {Megej}, {Melchior}, {Meinsma}, {Merkel}, {Michalska}, {Monstein}, {Moratschke}, {Morris}, {Muller}, {Murphy}, {Naber}, {Natale}, {Nowosielski}, {Nuzzolo}, {Olberg}, {Olbrich}, {Orfei}, {Orleanski}, {Ossenkopf}, {Peacock}, {Pearson}, {Peron}, {Phillip-May}, {Piazzo}, {Planesas}, {Rataj}, {Ravera}, {Risacher}, {Salez}, {Samoska}, {Saraceno}, {Schieder}, {Schlecht}, {Schl{\"o}der}, {Schm{\"u}lling}, {Schultz}, {Schuster}, {Siebertz}, {Smit}, {Szczerba}, {Shipman}, {Steinmetz}, {Stern}, {Stokroos}, {Teipen}, {Teyssier}, {Tils}, {Trappe}, {van Baaren}, {van Leeuwen}, {van de Stadt}, {Visser}, {Wildeman}, {Wafelbakker}, {Ward}, {Wesselius}, {Wild}, {Wulff}, {Wunsch}, {Tielens}, {Zaal}, {Zirath}, {Zmuidzinas}, \& {Zwart}}]{2010A&A...518L...6D}
{de Graauw}, T., {Helmich}, F.~P., {Phillips}, T.~G., {et~al.} 2010, \aap, 518, L6

\bibitem[{{Desmurs} {et~al.}(2014){Desmurs}, {Bujarrabal}, {Lindqvist}, {Alcolea}, {Soria-Ruiz}, \& {Bergman}}]{2014A&A...565A.127D}
{Desmurs}, J.~F., {Bujarrabal}, V., {Lindqvist}, M., {et~al.} 2014, \aap, 565, A127

\bibitem[{{Dur{\'a}n} {et~al.}(2021){Dur{\'a}n}, {Gusten}, {Risacher}, {Gorlitz}, {Klein}, {Reyes}, {Ricken}, {Wunsch}, {Graf}, {Jacobs}, {Honingh}, {Stutzki}, {de Lange}, {Delorme}, {Krieg}, \& {Lis}}]{2021ITTST..11..194D}
{Dur{\'a}n}, C.~A., {Gusten}, R., {Risacher}, C., {et~al.} 2021, IEEE Transactions on Terahertz Science and Technology, 11, 194

\bibitem[{{Elitzur}(1992)}]{1992ARA&A..30...75E}
{Elitzur}, M. 1992, \araa, 30, 75

\bibitem[{{Elitzur} {et~al.}(1976){Elitzur}, {Goldreich}, \& {Scoville}}]{1976ApJ...205..384E}
{Elitzur}, M., {Goldreich}, P., \& {Scoville}, N. 1976, \apj, 205, 384

\bibitem[{{Faure} \& {Josselin}(2008)}]{2008A&A...492..257F}
{Faure}, A. \& {Josselin}, E. 2008, \aap, 492, 257

\bibitem[{{Feast} {et~al.}(2003){Feast}, {Whitelock}, \& {Marang}}]{2003MNRAS.346..878F}
{Feast}, M.~W., {Whitelock}, P.~A., \& {Marang}, F. 2003, \mnras, 346, 878

\bibitem[{{Fonfr{\'\i}a} {et~al.}(2008){Fonfr{\'\i}a}, {Cernicharo}, {Richter}, \& {Lacy}}]{2008ApJ...673..445F}
{Fonfr{\'\i}a}, J.~P., {Cernicharo}, J., {Richter}, M.~J., \& {Lacy}, J.~H. 2008, \apj, 673, 445

\bibitem[{{Fonfr{\'\i}a} {et~al.}(2018){Fonfr{\'\i}a}, {Fern{\'a}ndez-L{\'o}pez}, {Pardo}, {Ag{\'u}ndez}, {S{\'a}nchez Contreras}, {Velilla Prieto}, {Cernicharo}, {Santander-Garc{\'\i}a}, {Quintana-Lacaci}, {Castro-Carrizo}, \& {Curiel}}]{2018ApJ...860..162F}
{Fonfr{\'\i}a}, J.~P., {Fern{\'a}ndez-L{\'o}pez}, M., {Pardo}, J.~R., {et~al.} 2018, \apj, 860, 162

\bibitem[{{Fonfr{\'\i}a} {et~al.}(2021){Fonfr{\'\i}a}, {Montiel}, {Cernicharo}, {DeWitt}, {Richter}, {Lacy}, {Greathouse}, {Santander-Garc{\'\i}a}, {Ag{\'u}ndez}, \& {Massalkhi}}]{2021A&A...651A...8F}
{Fonfr{\'\i}a}, J.~P., {Montiel}, E.~J., {Cernicharo}, J., {et~al.} 2021, \aap, 651, A8

\bibitem[{{Fonfr{\'\i}a Exp{\'o}sito} {et~al.}(2006){Fonfr{\'\i}a Exp{\'o}sito}, {Ag{\'u}ndez}, {Tercero}, {Pardo}, \& {Cernicharo}}]{2006ApJ...646L.127F}
{Fonfr{\'\i}a Exp{\'o}sito}, J.~P., {Ag{\'u}ndez}, M., {Tercero}, B., {Pardo}, J.~R., \& {Cernicharo}, J. 2006, \apjl, 646, L127

\bibitem[{{Gebbie} {et~al.}(1964){Gebbie}, {Stone}, \& {Findlay}}]{1964Natur.202..685G}
{Gebbie}, H.~A., {Stone}, N.~W.~B., \& {Findlay}, F.~D. 1964, \nat, 202, 685

\bibitem[{{Gong} {et~al.}(2017){Gong}, {Henkel}, {Ott}, {Menten}, {Morris}, {Keller}, {Claussen}, {Grasshoff}, \& {Mao}}]{2017ApJ...843...54G}
{Gong}, Y., {Henkel}, C., {Ott}, J., {et~al.} 2017, \apj, 843, 54

\bibitem[{{Gong} {et~al.}(2015){Gong}, {Henkel}, {Spezzano}, {Thorwirth}, {Menten}, {Wyrowski}, {Mao}, \& {Klein}}]{2015A&A...574A..56G}
{Gong}, Y., {Henkel}, C., {Spezzano}, S., {et~al.} 2015, \aap, 574, A56

\bibitem[{{Gray} {et~al.}(2016){Gray}, {Baudry}, {Richards}, {Humphreys}, {Sobolev}, \& {Yates}}]{2016MNRAS.456..374G}
{Gray}, M.~D., {Baudry}, A., {Richards}, A.~M.~S., {et~al.} 2016, \mnras, 456, 374

\bibitem[{{Guan} {et~al.}(2012){Guan}, {Stutzki}, {Graf}, {G{\"u}sten}, {Okada}, {Requena-Torres}, {Simon}, \& {Wiesemeyer}}]{2012A&A...542L...4G}
{Guan}, X., {Stutzki}, J., {Graf}, U.~U., {et~al.} 2012, \aap, 542, L4

\bibitem[{{Gu{\'e}lin} {et~al.}(2018){Gu{\'e}lin}, {Patel}, {Bremer}, {Cernicharo}, {Castro-Carrizo}, {Pety}, {Fonfr{\'\i}a}, {Ag{\'u}ndez}, {Santander-Garc{\'\i}a}, {Quintana-Lacaci}, {Velilla Prieto}, {Blundell}, \& {Thaddeus}}]{2018A&A...610A...4G}
{Gu{\'e}lin}, M., {Patel}, N.~A., {Bremer}, M., {et~al.} 2018, \aap, 610, A4

\bibitem[{{Guilloteau} {et~al.}(1987){Guilloteau}, {Omont}, \& {Lucas}}]{1987A&A...176L..24G}
{Guilloteau}, S., {Omont}, A., \& {Lucas}, R. 1987, \aap, 176, L24

\bibitem[{{Habing}(1996)}]{1996A&ARv...7...97H}
{Habing}, H.~J. 1996, \aapr, 7, 97

\bibitem[{{Harris} {et~al.}(2006){Harris}, {Tennyson}, {Kaminsky}, {Pavlenko}, \& {Jones}}]{2006MNRAS.367..400H}
{Harris}, G.~J., {Tennyson}, J., {Kaminsky}, B.~M., {Pavlenko}, Y.~V., \& {Jones}, H.~R.~A. 2006, \mnras, 367, 400

\bibitem[{{He} {et~al.}(2019){He}, {Kami{\'n}ski}, {Mennickent}, {Shenavrin}, {Mardones}, {Wang}, {Tang}, {Schmidt}, {Szczerba}, \& {Ge}}]{2019ApJ...883..165H}
{He}, J.~H., {Kami{\'n}ski}, T., {Mennickent}, R.~E., {et~al.} 2019, \apj, 883, 165

\bibitem[{{Henkel} {et~al.}(1983){Henkel}, {Matthews}, \& {Morris}}]{1983ApJ...267..184H}
{Henkel}, C., {Matthews}, H.~E., \& {Morris}, M. 1983, \apj, 267, 184

\bibitem[{{Hocker} \& {Javan}(1967)}]{1967PhLA...25..489H}
{Hocker}, L.~O. \& {Javan}, A. 1967, Physics Letters A, 25, 489

\bibitem[{{Humphreys}(2007)}]{2007IAUS..242..471H}
{Humphreys}, E.~M.~L. 2007, in Astrophysical Masers and their Environments, ed. J.~M. {Chapman} \& W.~A. {Baan}, Vol. 242, 471--480

\bibitem[{Hunter(2007)}]{Hunter:2007}
Hunter, J.~D. 2007, Computing in Science \& Engineering, 9, 90

\bibitem[{{Hunter} {et~al.}(2007){Hunter}, {Young}, {Christensen}, \& {Gurwell}}]{2007IAUS..242..481H}
{Hunter}, T.~R., {Young}, K.~H., {Christensen}, R.~D., \& {Gurwell}, M.~A. 2007, in Astrophysical Masers and their Environments, ed. J.~M. {Chapman} \& W.~A. {Baan}, Vol. 242, 481--488

\bibitem[{{Jeste} {et~al.}(2022){Jeste}, {Gong}, {Wong}, {Menten}, {Kami{\'n}ski}, \& {Wyrowski}}]{2022A&A...666A..69J}
{Jeste}, M., {Gong}, Y., {Wong}, K.~T., {et~al.} 2022, \aap, 666, A69

\bibitem[{Jones {et~al.}(2001)Jones, Oliphant, Peterson, {et~al.}}]{jones2001scipy}
Jones, E., Oliphant, T., Peterson, P., {et~al.} 2001, {SciPy}: Open source scientific tools for {Python}

\bibitem[{{Kim} {et~al.}(2014){Kim}, {Cho}, \& {Kim}}]{2014AJ....147...22K}
{Kim}, J., {Cho}, S.-H., \& {Kim}, S.~J. 2014, \aj, 147, 22

\bibitem[{{Klein} {et~al.}(2012){Klein}, {Hochg{\"u}rtel}, {Kr{\"a}mer}, {Bell}, {Meyer}, \& {G{\"u}sten}}]{2012A&A...542L...3K}
{Klein}, B., {Hochg{\"u}rtel}, S., {Kr{\"a}mer}, I., {et~al.} 2012, \aap, 542, L3

\bibitem[{{Kunstreich} \& {Lesieur}(1975)}]{Kunstreich1975}
{Kunstreich}, S. \& {Lesieur}, J.~P. 1975, Optics Communications, 13, 17

\bibitem[{{Lang}(2014)}]{2014AJ....147..108L}
{Lang}, D. 2014, \aj, 147, 108

\bibitem[{{Li} {et~al.}(2014){Li}, {Millar}, {Walsh}, {Heays}, \& {van Dishoeck}}]{2014A&A...568A.111L}
{Li}, X., {Millar}, T.~J., {Walsh}, C., {Heays}, A.~N., \& {van Dishoeck}, E.~F. 2014, \aap, 568, A111

\bibitem[{{Lide} \& {Maki}(1967)}]{1967ApPhL..11...62L}
{Lide}, David~R., J. \& {Maki}, A.~G. 1967, Applied Physics Letters, 11, 62

\bibitem[{{Lucas} \& {Cernicharo}(1989)}]{1989A&A...218L..20L}
{Lucas}, R. \& {Cernicharo}, J. 1989, \aap, 218, L20

\bibitem[{{Lucas} {et~al.}(1988){Lucas}, {Guilloteau}, \& {Omont}}]{1988A&A...194..230L}
{Lucas}, R., {Guilloteau}, S., \& {Omont}, A. 1988, \aap, 194, 230

\bibitem[{{Maki} {et~al.}(1996){Maki}, {Quapp}, {Klee}, {Mellau}, \& {Albert}}]{1996JMoSp.180..323M}
{Maki}, A., {Quapp}, W., {Klee}, S., {Mellau}, G.~C., \& {Albert}, S. 1996, Journal of Molecular Spectroscopy, 180, 323

\bibitem[{{Maki} \& {Blaine}(1964)}]{1964JMoSp..12...45M}
{Maki}, A.~G. \& {Blaine}, L.~R. 1964, Journal of Molecular Spectroscopy, 12, 45

\bibitem[{{Massalkhi} {et~al.}(2018){Massalkhi}, {Ag{\'u}ndez}, {Cernicharo}, {Velilla Prieto}, {Goicoechea}, {Quintana-Lacaci}, {Fonfr{\'\i}a}, {Alcolea}, \& {Bujarrabal}}]{2018A&A...611A..29M}
{Massalkhi}, S., {Ag{\'u}ndez}, M., {Cernicharo}, J., {et~al.} 2018, \aap, 611, A29

\bibitem[{{Mathias} {et~al.}(1965){Mathias}, {Crocker}, \& {Wills}}]{1965ElL.....1...45M}
{Mathias}, L.~E.~S., {Crocker}, A., \& {Wills}, M.~S. 1965, Electronics Letters, 1, 45

\bibitem[{{Meadows} {et~al.}(1987){Meadows}, {Good}, \& {Wolstencroft}}]{1987MNRAS.225P..43M}
{Meadows}, P.~J., {Good}, A.~R., \& {Wolstencroft}, R.~D. 1987, \mnras, 225, 43P

\bibitem[{{Menten} \& {Melnick}(1989)}]{1989ApJ...341L..91M}
{Menten}, K.~M. \& {Melnick}, G.~J. 1989, \apjl, 341, L91

\bibitem[{{Menten} {et~al.}(2012){Menten}, {Reid}, {Kami{\'n}ski}, \& {Claussen}}]{2012A&A...543A..73M}
{Menten}, K.~M., {Reid}, M.~J., {Kami{\'n}ski}, T., \& {Claussen}, M.~J. 2012, \aap, 543, A73

\bibitem[{{Menten} {et~al.}(2018){Menten}, {Wyrowski}, {Keller}, \& {Kami{\'n}ski}}]{2018A&A...613A..49M}
{Menten}, K.~M., {Wyrowski}, F., {Keller}, D., \& {Kami{\'n}ski}, T. 2018, \aap, 613, A49

\bibitem[{{Menten} \& {Young}(1995)}]{1995ApJ...450L..67M}
{Menten}, K.~M. \& {Young}, K. 1995, \apjl, 450, L67

\bibitem[{{Ochsenbein} {et~al.}(2000){Ochsenbein}, {Bauer}, \& {Marcout}}]{2000A&AS..143...23O}
{Ochsenbein}, F., {Bauer}, P., \& {Marcout}, J. 2000, \aaps, 143, 23

\bibitem[{{Orr}(2018)}]{2018MolPh.116.3666O}
{Orr}, B.~J. 2018, Molecular Physics, 116, 3666

\bibitem[{Orr \& Smith(1987)}]{Orr1987}
Orr, B.~J. \& Smith, I. W.~M. 1987, The Journal of Physical Chemistry, 91, 6106

\bibitem[{{Paine}(2022)}]{2022zndo...6774378P}
{Paine}, S. 2022, {The am atmospheric model}

\bibitem[{{Pardo} {et~al.}(2004){Pardo}, {Alcolea}, {Bujarrabal}, {Colomer}, {del Romero}, \& {de Vicente}}]{2004A&A...424..145P}
{Pardo}, J.~R., {Alcolea}, J., {Bujarrabal}, V., {et~al.} 2004, \aap, 424, 145

\bibitem[{{Pardo} {et~al.}(2018){Pardo}, {Cernicharo}, {Velilla Prieto}, {Fonfr{\'\i}a}, {Ag{\'u}ndez}, {Quintana-Lacaci}, {Massalkhi}, {Tercero}, {G{\'o}mez-Garrido}, {de Vicente}, {Gu{\'e}lin}, {Kramer}, {Marka}, {Teyssier}, \& {Neufeld}}]{2018A&A...615L...4P}
{Pardo}, J.~R., {Cernicharo}, J., {Velilla Prieto}, L., {et~al.} 2018, \aap, 615, L4

\bibitem[{{Pety}(2005)}]{2005sf2a.conf..721P}
{Pety}, J. 2005, in SF2A-2005: Semaine de l'Astrophysique Francaise, ed. F.~{Casoli}, T.~{Contini}, J.~M. {Hameury}, \& L.~{Pagani}, 721

\bibitem[{{Pilbratt} {et~al.}(2010){Pilbratt}, {Riedinger}, {Passvogel}, {Crone}, {Doyle}, {Gageur}, {Heras}, {Jewell}, {Metcalfe}, {Ott}, \& {Schmidt}}]{2010A&A...518L...1P}
{Pilbratt}, G.~L., {Riedinger}, J.~R., {Passvogel}, T., {et~al.} 2010, \aap, 518, L1

\bibitem[{{Price} {et~al.}(2010){Price}, {Smith}, {Kuchar}, {Mizuno}, \& {Kraemer}}]{2010ApJS..190..203P}
{Price}, S.~D., {Smith}, B.~J., {Kuchar}, T.~A., {Mizuno}, D.~R., \& {Kraemer}, K.~E. 2010, \apjs, 190, 203

\bibitem[{{Ramstedt} \& {Olofsson}(2014)}]{2014A&A...566A.145R}
{Ramstedt}, S. \& {Olofsson}, H. 2014, \aap, 566, A145

\bibitem[{{Reid} \& {Menten}(1997)}]{1997ApJ...476..327R}
{Reid}, M.~J. \& {Menten}, K.~M. 1997, \apj, 476, 327

\bibitem[{{Reid} \& {Moran}(1981)}]{1981ARA&A..19..231R}
{Reid}, M.~J. \& {Moran}, J.~M. 1981, \araa, 19, 231

\bibitem[{{Rizzo} {et~al.}(2021){Rizzo}, {Cernicharo}, \& {Garc{\'\i}a-Mir{\'o}}}]{2021ApJS..253...44R}
{Rizzo}, J.~R., {Cernicharo}, J., \& {Garc{\'\i}a-Mir{\'o}}, C. 2021, \apjs, 253, 44

\bibitem[{{Robinson}(1978)}]{1978OptCo..27..281R}
{Robinson}, D.~W. 1978, Optics Communications, 27, 281

\bibitem[{{Roueff} \& {Lique}(2013)}]{2013ChRv..113.8906R}
{Roueff}, E. \& {Lique}, F. 2013, Chemical Reviews, 113, 8906

\bibitem[{{Schilke} {et~al.}(2000){Schilke}, {Mehringer}, \& {Menten}}]{2000ApJ...528L..37S}
{Schilke}, P., {Mehringer}, D.~M., \& {Menten}, K.~M. 2000, \apjl, 528, L37

\bibitem[{{Schilke} \& {Menten}(2003)}]{2003ApJ...583..446S}
{Schilke}, P. \& {Menten}, K.~M. 2003, \apj, 583, 446

\bibitem[{{Schmidt} {et~al.}(2002){Schmidt}, {Hines}, \& {Swift}}]{2002ApJ...576..429S}
{Schmidt}, G.~D., {Hines}, D.~C., \& {Swift}, S. 2002, \apj, 576, 429

\bibitem[{{Sch{\"o}ier} {et~al.}(2013){Sch{\"o}ier}, {Ramstedt}, {Olofsson}, {Lindqvist}, {Bieging}, \& {Marvel}}]{2013A&A...550A..78S}
{Sch{\"o}ier}, F.~L., {Ramstedt}, S., {Olofsson}, H., {et~al.} 2013, \aap, 550, A78

\bibitem[{{Shenavrin} {et~al.}(2011){Shenavrin}, {Taranova}, \& {Nadzhip}}]{2011ARep...55...31S}
{Shenavrin}, V.~I., {Taranova}, O.~G., \& {Nadzhip}, A.~E. 2011, Astronomy Reports, 55, 31

\bibitem[{{Shipman} {et~al.}(2017){Shipman}, {Beaulieu}, {Teyssier}, {Morris}, {Rengel}, {McCoey}, {Edwards}, {Kester}, {Lorenzani}, {Coeur-Joly}, {Melchior}, {Xie}, {Sanchez}, {Zaal}, {Avruch}, {Borys}, {Braine}, {Comito}, {Delforge}, {Herpin}, {Hoac}, {Kwon}, {Lord}, {Marston}, {Mueller}, {Olberg}, {Ossenkopf}, {Puga}, \& {Akyilmaz-Yabaci}}]{2017A&A...608A..49S}
{Shipman}, R.~F., {Beaulieu}, S.~F., {Teyssier}, D., {et~al.} 2017, \aap, 608, A49

\bibitem[{{Skatrud} \& {De Lucia}(1984)}]{Skatrud1984}
{Skatrud}, D.~D. \& {De Lucia}, F.~C. 1984, Applied Physics B: Lasers and Optics, 35, 179

\bibitem[{{Smith} {et~al.}(2004){Smith}, {Price}, \& {Baker}}]{2004ApJS..154..673S}
{Smith}, B.~J., {Price}, S.~D., \& {Baker}, R.~I. 2004, \apjs, 154, 673

\bibitem[{Smith(1981)}]{Smith1981}
Smith, I. W.~M. 1981, J. Chem. Soc.{,} Faraday Trans. 2, 77, 2357

\bibitem[{{Tennyson} {et~al.}(2024){Tennyson}, {Yurchenko}, {Zhang}, {Bowesman}, {Brady}, {Buldyreva}, {Chubb}, {Gamache}, {Gorman}, {Guest}, {Hill}, {Kefala}, {Lynas-Gray}, {Mellor}, {McKemmish}, {Mitev}, {Mizus}, {Owens}, {Peng}, {Perri}, {Pezzella}, {Polyansky}, {Qu}, {Semenov}, {Smola}, {Solokov}, {Somogyi}, {Upadhyay}, {Wright}, \& {Zobov}}]{2024JQSRT.32609083T}
{Tennyson}, J., {Yurchenko}, S.~N., {Zhang}, J., {et~al.} 2024, \jqsrt, 326, 109083

\bibitem[{{Trammell} {et~al.}(1994){Trammell}, {Dinerstein}, \& {Goodrich}}]{1994AJ....108..984T}
{Trammell}, S.~R., {Dinerstein}, H.~L., \& {Goodrich}, R.~W. 1994, \aj, 108, 984

\bibitem[{{Tsuji}(1964)}]{1964AnTok...9.....T}
{Tsuji}, T. 1964, Annals of the Tokyo Astronomical Observatory, 9, 1

\bibitem[{{Unnikrishnan} {et~al.}(2024){Unnikrishnan}, {De Beck}, {Nyman}, {Olofsson}, {Vlemmings}, {Tafoya}, {Maercker}, {Charnley}, {Cordiner}, {de Gregorio}, {Humphreys}, {Millar}, \& {Rawlings}}]{2024A&A...684A...4U}
{Unnikrishnan}, R., {De Beck}, E., {Nyman}, L.~{\r{A}}., {et~al.} 2024, \aap, 684, A4

\bibitem[{{van der Walt} {et~al.}(2011){van der Walt}, {Colbert}, \& {Varoquaux}}]{5725236}
{van der Walt}, S., {Colbert}, S.~C., \& {Varoquaux}, G. 2011, Computing in Science Engineering, 13, 22

\bibitem[{{Velilla-Prieto} {et~al.}(2023){Velilla-Prieto}, {Fonfr{\'\i}a}, {Ag{\'u}ndez}, {Castro-Carrizo}, {Gu{\'e}lin}, {Quintana-Lacaci}, {Cherchneff}, {Joblin}, {McCarthy}, {Mart{\'\i}n-Gago}, \& {Cernicharo}}]{2023Natur.617..696V}
{Velilla-Prieto}, L., {Fonfr{\'\i}a}, J.~P., {Ag{\'u}ndez}, M., {et~al.} 2023, \nat, 617, 696

\bibitem[{{Willacy} \& {Cherchneff}(1998)}]{1998A&A...330..676W}
{Willacy}, K. \& {Cherchneff}, I. 1998, \aap, 330, 676

\bibitem[{{Wong}(2019)}]{2019asrc.confE..55W}
{Wong}, K.~T. 2019, in ALMA2019: Science Results and Cross-Facility Synergies, 55

\bibitem[{{Yang} {et~al.}(2024){Yang}, {Wu}, {Gong}, {Mauron}, {Zhang}, {Menten}, {Mai}, {Liu}, {Li}, \& {Li}}]{2024ApJ...961..190Y}
{Yang}, W., {Wu}, Y., {Gong}, Y., {et~al.} 2024, \apj, 961, 190

\bibitem[{{Young} {et~al.}(2012){Young}, {Becklin}, {Marcum}, {Roellig}, {De Buizer}, {Herter}, {G{\"u}sten}, {Dunham}, {Temi}, {Andersson}, {Backman}, {Burgdorf}, {Caroff}, {Casey}, {Davidson}, {Erickson}, {Gehrz}, {Harper}, {Harvey}, {Helton}, {Horner}, {Howard}, {Klein}, {Krabbe}, {McLean}, {Meyer}, {Miles}, {Morris}, {Reach}, {Rho}, {Richter}, {Roeser}, {Sandell}, {Sankrit}, {Savage}, {Smith}, {Shuping}, {Vacca}, {Vaillancourt}, {Wolf}, \& {Zinnecker}}]{2012ApJ...749L..17Y}
{Young}, E.~T., {Becklin}, E.~E., {Marcum}, P.~M., {et~al.} 2012, \apjl, 749, L17

\bibitem[{{Zhang} {et~al.}(2009){Zhang}, {Kwok}, \& {Nakashima}}]{2009ApJ...700.1262Z}
{Zhang}, Y., {Kwok}, S., \& {Nakashima}, J.-i. 2009, \apj, 700, 1262

\bibitem[{{Ziurys} \& {Turner}(1986)}]{1986ApJ...300L..19Z}
{Ziurys}, L.~M. \& {Turner}, B.~E. 1986, \apjl, 300, L19

\end{thebibliography}

\begin{appendix}

\section{{\it Herschel}/HIFI observations}\label{sec:appendix-a}
Table~\ref{Tab:hifi} lists details of all the {\it Herschel}/HIFI observations of HCN transitions used in this study.

\begin{table}[htbp]
\caption{Summary of {\it Herschel}/HIFI observations.}\label{Tab:hifi} 
\tiny
\centering
\setlength{\tabcolsep}{4pt}
\begin{tabular}{llll}
\hline \hline 
Obs. date  & {\it Herschel} & Line & Obs. mode \\
(yyyy-mm-dd) & OBSID & (GHz) & \\
\hline
\multicolumn{4}{c}{IRC+10216} \\
\hline
2010-05-12 & 1342196473 & 805 & Spectral scan\\
2011-06-09 & 1342222323 & 805$^*$ & Single point\\
2012-05-03 & 1342245305 & 805$^*$ & Single point\\
2012-10-24 & 1342254408 & 805$^*$ & Single point\\
2012-11-30 & 1342256274 & 805$^*$ & Single point\\
2013-04-29 & 1342271240 & 805$^*$ & Single point\\
2010-05-13 & 1342196518 & 891,894 & Spectral scan\\
2011-06-09 & 1342222292 & 891$^\dagger$ & Single point\\
2012-05-04 & 1342245342 & 891$^\dagger$ & Single point\\
2012-10-26 & 1342253951 & 891$^\dagger$ & Single point\\
2012-11-30 & 1342256278 & 891$^\dagger$ & Single point\\
2013-04-28 & 1342271201 & 891$^\dagger$ & Single point\\
2011-06-09 & 1342222297 & 894$^*$ & Single point\\
2012-05-04 & 1342245347 & 894$^*$ & Single point\\
2012-10-26 & 1342253956 & 894$^*$ & Single point\\
2012-11-30 & 1342256283 & 894$^*$ & Single point\\
2013-04-28 & 1342271206 & 894$^*$ & Single point\\
2010-05-16 & 1342196590 & 964,968,1055 & Spectral scan\\
2011-06-09 & 1342222351 & 964 & Single point \\
2012-05-04 & 1342245359 & 964 & Single point \\
2012-10-24 & 1342254419 & 964 & Single point \\
2012-11-30 & 1342256285 & 964 & Single point \\
2013-04-29 & 1342271230 & 964 & Single point \\
2011-06-09 & 1342222352 & 968 & Single point \\
2012-05-04 & 1342245360 & 968 & Single point \\
2012-10-24 & 1342254420 & 968 & Single point \\
2012-11-30 & 1342256286 & 968 & Single point \\
2013-04-29 & 1342271231 & 968 & Single point \\
2010-05-11 & 1342196423 & 1055 & Spectral scan\\
2011-06-09 & 1342222353 & 1055 & Single point\\
2012-05-04 & 1342245361 & 1055 & Single point\\
2012-10-24 & 1342254421 & 1055 & Single point\\
2012-11-30 & 1342256287 & 1055 & Single point\\
2013-04-29 & 1342271232 & 1055 & Single point\\
\hline
\multicolumn{4}{c}{CIT 6} \\
\hline
2012-06-16 & 1342247079 & 805     & Spectral scan \\
2012-05-04 & 1342245341 & 891,894 & Spectral scan \\
2012-05-04 & 1342245367 & 964,968 & Spectral scan \\
2012-05-04 & 1342245374 & 1055    & Spectral scan \\
\hline
\multicolumn{4}{c}{Y CVn} \\
\hline
2012-06-16 & 1342247080 & 805         & Spectral scan \\
2012-06-14 & 1342247025 & 891,894     & Spectral scan \\
2012-06-15 & 1342247044 & 964,968     & Spectral scan \\
2012-05-18 & 1342246497 & 1055        & Spectral scan \\
\hline
\multicolumn{4}{c}{S Cep} \\
\hline
2012-05-23 & 1342246032 & 805         & Spectral scan \\
2012-05-24 & 1342246045 & 891,894     & Spectral scan \\
2012-05-24 & 1342246330 & 964,968     & Spectral scan \\
2012-05-24 & 1342246338 & 1055        & Spectral scan \\
\hline
\multicolumn{4}{c}{IRC+50096} \\
\hline
2012-08-25 & 1342250216 & 805         & Spectral scan \\
2012-08-11 & 1342249438 & 891,894     & Spectral scan \\
2012-08-16 & 1342249614 & 964,968     & Spectral scan \\
2012-08-25 & 1342250210 & 1055        & Spectral scan \\
\hline
\multicolumn{4}{c}{V Cyg} \\
\hline
2012-05-23 & 1342246035 & 805         & Spectral scan \\
2012-05-24 & 1342246049 & 891,894     & Spectral scan \\
2012-05-03 & 1342245331 & 964,968     & Spectral scan \\
2012-06-21 & 1342247175 & 1055        & Spectral scan \\
\hline
\multicolumn{4}{c}{CRL~3068} \\
\hline
2012-06-21 & 1342247178 & 805         & Spectral scan \\
2012-05-24 & 1342246047 & 891,894     & Spectral scan \\
2012-05-31 & 1342246518 & 964,968     & Spectral scan \\
2012-06-21 & 1342247176 & 1055        & Spectral scan \\
\hline
\multicolumn{4}{c}{II Lup} \\
\hline
2012-09-12 & 1342251014 & 805         & Spectral scan \\
2012-09-18 & 1342251111 & 891,894     & Spectral scan \\
2012-09-07 & 1342250707 & 964,968     & Spectral scan \\
2012-09-21 & 1342251491 & 1055        & Spectral scan \\
\hline
\hline
\end{tabular}
\normalsize
\tablefoot{The signal recorded in the other sideband is marked by a star. 
The 891~GHz laser spectra of IRC+10216  indicated by a dagger exhibited contamination from $^{13}$CO ($J=8-7$) emission in the other sideband.
For IRC+10216, the data processed by pipeline were used, while for the other six stars, the Highly Processed Data Product (HPDP) with improved baselines were utilised.
}
\end{table}

\section{Contamination Removal for {\it Herschel}/HIFI 891~GHz Spectra}\label{sec:appendix-b}
Through examining the spectra observed at different epochs and scrutinising the possible contamination that arise from signals in the other sideband, we found that the 891~GHz laser line spectra from the upper sideband (USB) observed towards IRC+10216 between 2011 and 2013 were indeed contaminated.
The contamination is attributed to the $^{13}$CO~($8-7$) line (centred at 881.272808~GHz) originating from the lower sideband (LSB).

\begin{figure}[htbp]
\center
\includegraphics[width=0.24\textwidth]{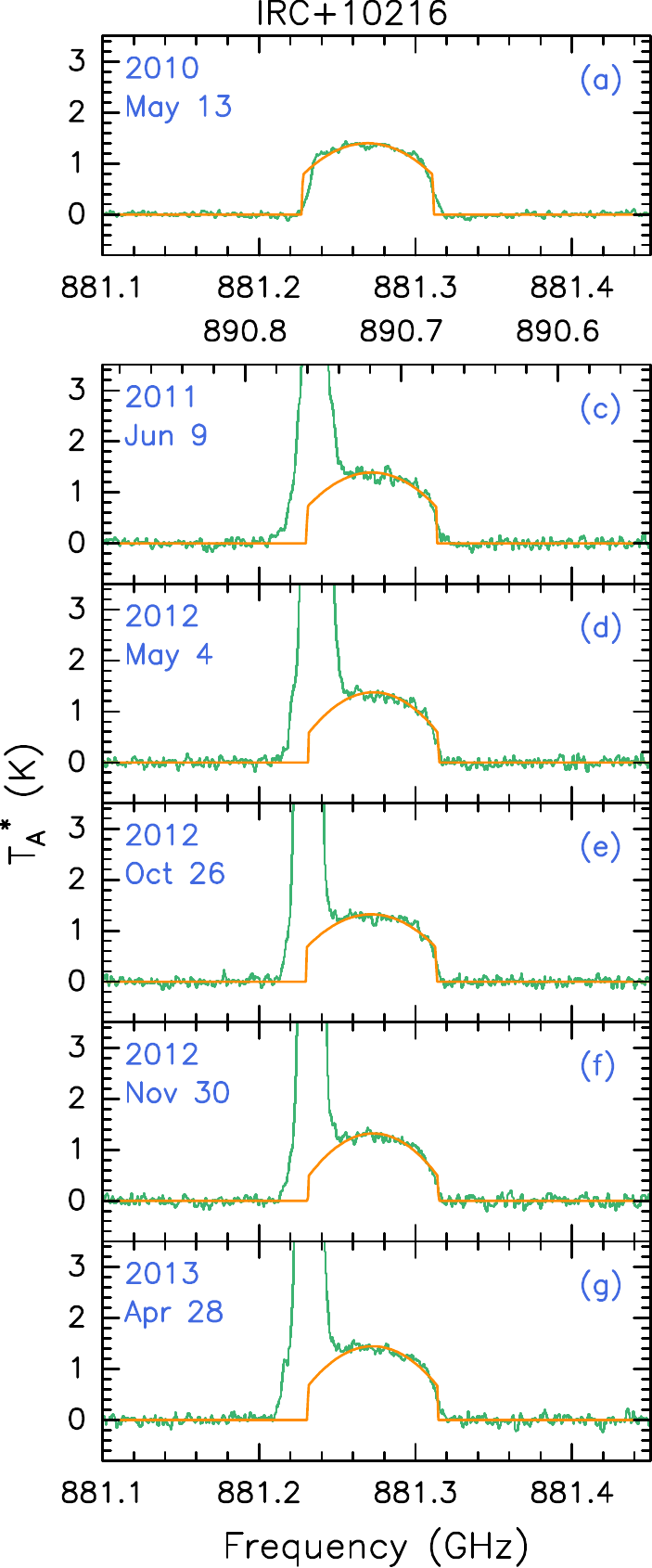}
\includegraphics[width=0.24\textwidth]{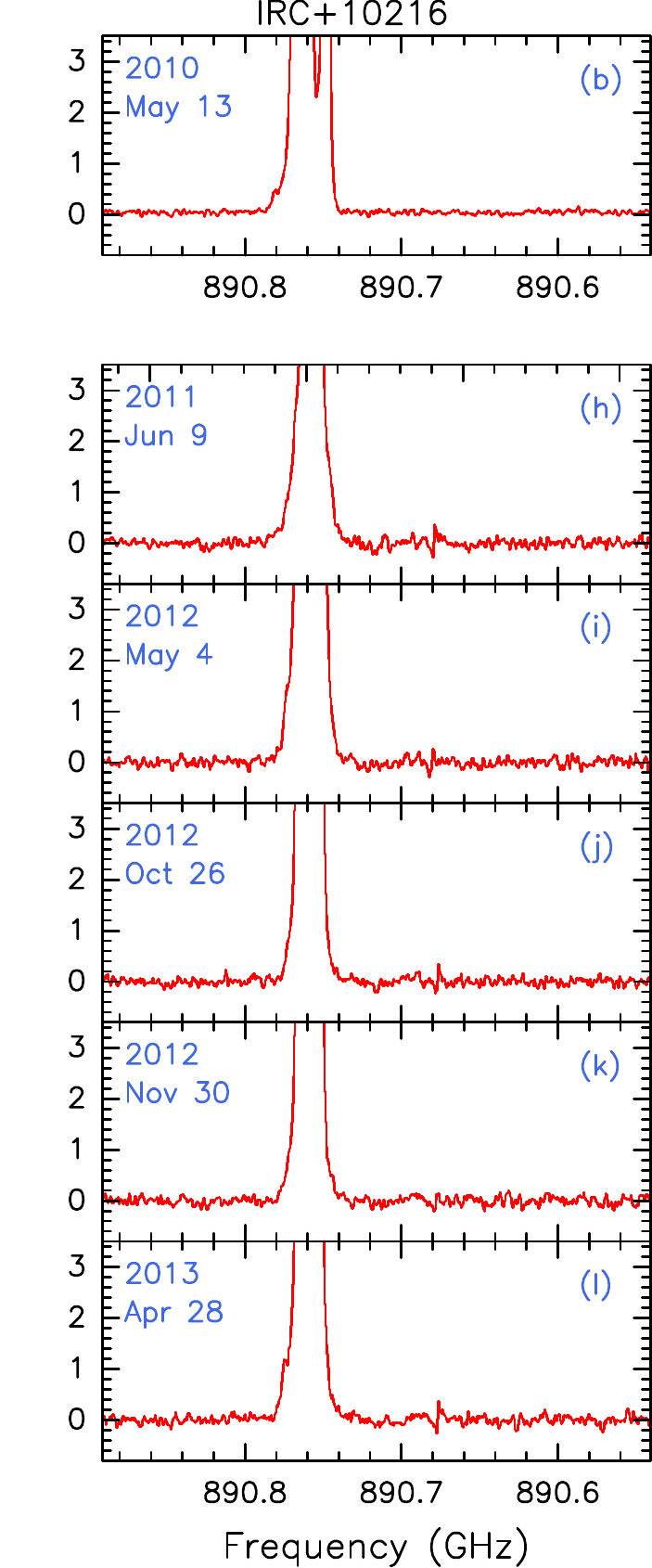}
\caption{Panels (a) and (b) show the spectra of $^{13}$CO~($8-7$) and HCN laser emission at 891~GHz obtained in 2010 May 15. 
Panels (c--g) show the original 891~GHz laser spectra in the USB that were contaminated by $^{13}$CO~($8-7$) from the LSB during 2011 to 2013 (in green). The upper and lower axes indicate the frequency ranges of the blended emission in the USB and LSB, respectively. The orange curves present the fitting results for the $^{13}$CO contamination. 
Panels (h--l) show the 891~GHz laser spectra (in red) that have been removed the $^{13}$CO contamination.
\label{fig:irc10216_891_contamination}}
\end{figure}

The 891~GHz spectra observed in 2010 May 15 were acquired using a different observing mode and frequency settings compared to those spectra obtained between 2011 and 2013.
There is no evidence of contamination for both $^{13}$CO~($8-7$) and the HCN laser emission at 891~GHz (see Figs~\ref{fig:irc10216_891_contamination}a,b). We adopted the SHELL method in the GILDAS/CLASS software to fit the $^{13}$CO~($8-7$) emission.
To highlight the contamination of the laser emission by $^{13}$CO~($8-7$) (see panels c--f), we folded the blended emission into the LSB and used the same frequency range as shown in panel (a). The corresponding frequency range of the USB is also shown on the upper x-axis to label the strong emission component's frequency at 890.76~GHz.

Unfortunately, the blended HCN laser emission components occur at the edge of the $^{13}$CO~($8-7$) line's profile, making it challenging to determine the frequency coverage of $^{13}$CO emission directly from spectra.
Since the velocity ranges of thermal emissions typically remain consistent over time \citep[see, e.g.][]{2014ApJ...796L..21C}, we fixed the frequency coverage of the $^{13}$CO~($8-7$) emission (corresponding to a constant expansion velocity $V_{\rm exp}=$ 14.13~\kms) obtained from observations in 2010 (Fig.~\ref{fig:irc10216_891_contamination}a) to fit the contributions of $^{13}$CO ($8-7$) in the contaminated spectra (see Figs~\ref{fig:irc10216_891_contamination}c--g). The fitted $^{13}$CO ($8-7$) spectra were then subtracted, resulting in the 891~GHz HCN laser spectra for our analysis (see Figs~\ref{fig:irc10216_891_contamination}h--l).

\section{Parameters of the detected HCN lasers and comments on individual stars} \label{Sec:individual}

Table~\ref{Tab:IRC+10216_laser} summarises the observed parameters of HCN lasers in multiple epochs towards IRC+10216. Table~\ref{Tab:6stars_laser} presents the observed parameters of HCN lasers towards CIT~6, Y~CVn, S~Cep, IRC+50096, V~Cyg, II~Lup, and CRL~3068.
In the following, we focus on the specificities of HCN lasers and masers that have been detected towards each star.

\begin{table*}[!hbt]
\caption{Observed parameters of HCN laser lines towards IRC+10216.}\label{Tab:IRC+10216_laser}
\normalsize
\centering
\setlength{\tabcolsep}{5pt}
\begin{tabular}{lllccrrrr}
\hline \hline 
Line & Obs. date & $\phi_{\rm IR}$ & $V_{\rm pk}$ & $V_{\rm range}$ & $\int S{\rm d}v$ & $S_{\rm pk}$ & $L_{\rm HCN}$ & 1$\sigma$ @ $V_{\rm ch}$\\
(GHz) & (yyyy-mm-dd) & & (\kms) & (\kms) & (Jy~\kms) & (Jy) & (photons~s$^{-1}$) & (Jy @ \kms)\\
\hline
805  & 2010-05-12           & 0.23 & $-$26.7 & [$-$34.2, $-$21.1] & 9273.1  & 2209.2  & 1.1$\times$10$^{44}$ & 11.2 @ 0.19\\
     & 2011-06-09           & 0.85 & $-$26.9 & [$-$32.6, $-$20.2] & 3963.6  & 1455.4  & 4.7$\times$10$^{43}$ & 11.1 @ 0.19\\
     & 2012-05-03           & 0.38 & $-$26.3 & [$-$32.1, $-$20.6] & 9625.8  & 2673.0  & 1.1$\times$10$^{44}$ & 11.6 @ 0.19\\
     & 2012-10-24           & 0.65 & $-$26.1 & [$-$32.8, $-$20.6] & 3478.8  & 895.0   & 4.1$\times$10$^{43}$ & 10.8 @ 0.19\\
     & 2012-11-30           & 0.71 & $-$26.4 & [$-$29.8, $-$21.6] & 2776.8  & 728.3   & 3.3$\times$10$^{43}$ & 11.4 @ 0.19\\
     & 2013-04-29           & 0.95 & $-$25.0 & [$-$31.8, $-$20.8] & 8106.7  & 2213.9  & 9.6$\times$10$^{43}$ & 10.4 @ 0.19\\
891  & 2010-05-13           & 0.23 & $-$26.7 & [$-$34.8, $-$19.6] & 51616.9 & 13702.8 & 6.1$\times$10$^{44}$ & 18.0 @ 0.17\\
     & 2011-06-09$^\dagger$ & 0.85 & $-$26.5 & [$-$34.8, $-$19.6] & 27596.9 & 8814.5  & 3.3$\times$10$^{44}$ & 34.1 @ 0.17\\
     & 2012-05-04$^\dagger$ & 0.38 & $-$26.0 & [$-$31.8, $-$20.0] & 62405.5 & 16751.3 & 7.4$\times$10$^{44}$ & 34.7 @ 0.17\\
     & 2012-10-26$^\dagger$ & 0.66 & $-$26.2 & [$-$32.0, $-$20.6] & 60922.2 & 21214.6 & 7.2$\times$10$^{44}$ & 31.0 @ 0.17\\
     & 2012-11-30$^\dagger$ & 0.71 & $-$26.2 & [$-$32.6, $-$19.8] & 61743.9 & 21955.3 & 7.3$\times$10$^{44}$ & 37.5 @ 0.17\\
     & 2013-04-28$^\dagger$ & 0.95 & $-$26.0 & [$-$33.2, $-$20.4] & 80099.9 & 26370.0 & 9.5$\times$10$^{44}$ & 33.7 @ 0.17\\
     & 2018-12-17           & 0.22 & $-$23.9 & [$-$30.2, $-$21.8] & 52621.7 & 14023.6 & 6.2$\times$10$^{44}$ & 666.4 @ 0.16\\
964  & 2010-05-16           & 0.24 & $-$26.5 & [$-$34.0, $-$20.2] & 19870.9 & 7057.0  & 2.3$\times$10$^{44}$ & 21.8 @ 0.16\\
     & 2011-06-09           & 0.85 & $-$26.6 & [$-$32.0, $-$20.5] & 11239.3 & 4642.1  & 1.3$\times$10$^{44}$ & 36.5 @ 0.16\\
     & 2012-05-04           & 0.38 & $-$26.0 & [$-$32.1, $-$20.6] & 35326.3 & 10469.9 & 4.2$\times$10$^{44}$ & 34.7 @ 0.16\\
     & 2012-10-24           & 0.65 & $-$25.8 & [$-$32.0, $-$21.0] & 36280.0 & 13332.5 & 4.3$\times$10$^{44}$ & 35.2 @ 0.16\\
     & 2012-11-30           & 0.71 & $-$25.8 & [$-$31.8, $-$20.6] & 36638.3 & 13940.8 & 4.3$\times$10$^{44}$ & 34.3 @ 0.16\\
     & 2013-04-29           & 0.95 & $-$25.8 & [$-$32.8, $-$20.6] & 48059.8 & 18223.7 & 5.7$\times$10$^{44}$ & 33.9 @ 0.16\\
     & 2018-12-17           & 0.22 & $-$23.8 & [$-$29.8, $-$21.0] & 33876.5 & 10337.9 & 4.0$\times$10$^{44}$ & 260.8 @ 0.15\\
968  & 2010-05-16           & 0.24 & $-$26.4 & [$-$34.0, $-$19.8] & 15959.7 & 3310.8  & 1.9$\times$10$^{44}$ & 22.8 @ 0.15\\
     & 2011-06-09           & 0.85 & $-$26.9 & [$-$34.0, $-$19.0] & 9667.0  & 2473.0  & 1.1$\times$10$^{44}$ & 37.7 @ 0.15\\
     & 2012-05-04           & 0.38 & $-$26.9 & [$-$32.1, $-$20.4] & 9374.2  & 2124.6  & 1.1$\times$10$^{44}$ & 32.4 @ 0.15\\
     & 2012-10-24           & 0.65 & $-$27.0 & [$-$34.0, $-$20.0] & 4126.0  & 957.8   & 4.9$\times$10$^{43}$ & 34.8 @ 0.15\\
     & 2012-11-30           & 0.71 & $-$27.0 & [$-$34.5, $-$21.0] & 4073.3  & 903.8   & 4.8$\times$10$^{43}$ & 35.2 @ 0.15\\
     & 2013-04-29           & 0.95 & $-$26.5 & [$-$33.8, $-$19.0] & 11193.1 & 2324.4  & 1.3$\times$10$^{44}$ & 35.3 @ 0.15\\
     & 2018-12-17           & 0.22 & $-$26.1 & [$-$29.0, $-$20.0] & 9533.8  & 3051.2  & 1.1$\times$10$^{44}$ & 285.0 @ 0.15\\
1055 & 2010-05-11           & 0.23 & $-$27.4 & [$-$34.0, $-$19.8] & 9108.4  & 1695.5  & 1.1$\times$10$^{44}$ & 33.4 @ 0.14\\
     & 2010-05-16           & 0.24 & $-$25.6 & [$-$32.6, $-$20.4] & 8718.2  & 1688.8  & 1.0$\times$10$^{44}$ & 36.1 @ 0.14\\
     & 2011-06-09           & 0.85 & $-$26.3 & [$-$32.1, $-$20.7] & 4713.9  & 1143.8  & 5.6$\times$10$^{43}$ & 33.1 @ 0.14\\
     & 2012-05-04           & 0.38 & $-$25.7 & [$-$32.0, $-$21.0] & 6285.6  & 1951.0  & 7.4$\times$10$^{43}$ & 33.4 @ 0.14\\
     & 2012-10-24           & 0.65 & $-$25.3 & [$-$32.0, $-$21.0] & 2953.9  & 895.8   & 3.5$\times$10$^{43}$ & 33.7 @ 0.14\\
     & 2012-11-30           & 0.71 & $-$25.1 & [$-$29.5, $-$21.0] & 2274.2  & 707.2   & 2.7$\times$10$^{43}$ & 30.1 @ 0.14\\
     & 2013-04-29           & 0.95 & $-$25.0 & [$-$32.1, $-$20.2] & 6416.7  & 1656.4  & 7.6$\times$10$^{43}$ & 33.0 @ 0.14\\
     & 2018-12-17           & 0.22 & $-$25.6 & [$-$28.1, $-$20.9] & 5136.4  & 1928.6  & 6.1$\times$10$^{43}$ & 177.1 @ 0.14\\
\hline
\end{tabular}
\normalsize
\tablefoot{Columns 2--9 list the observing date, phase ($\phi_{\rm IR}$), peak velocity ($V_{\rm pk}$), velocity range ($V_{\rm range}$), integrated flux density ($\int S{\rm d}v$), peak intensity ($S_{\rm pk}$), HCN laser isotropic luminosity ($L_{\rm HCN}$), and 1$\sigma$ noise level at the channel spacing ($V_{\rm ch}$) for the corresponding HCN laser transition.
The phases, where $\phi_{\rm IR}$= 0 corresponds to the maximum NIR brightness, were estimated using a period of 630 days and the epoch of the maximum, 2454554, in Julian Day format \citep{2012A&A...543A..73M}. These values were derived based on $H$, $K$, $L$ and $M$ infrared light curves of IRC+10216 \citep{2011ARep...55...31S}. 
The 891~GHz laser spectra exhibit contamination from $^{13}$CO ($J=8-7$) emission in the other sideband, indicated by a dagger. The contamination removal spectra 
(see Figs.~\ref{fig:irc10216_891_contamination}h--l) were used to derive the parameters.}
\end{table*}

\begin{table*}[!hbt]
\caption{Observed parameters of HCN laser lines towards CIT~6, Y~CVn, S~Cep, IRC+50096, V~Cyg, II~Lup, and CRL~3068.}\label{Tab:6stars_laser}
\normalsize
\centering
\setlength{\tabcolsep}{5pt}
\begin{tabular}{lllcrrrrr}
\hline \hline 
Line & Obs. date & $\phi$ & $V_{\rm pk}$ & $V_{\rm range}$ & $\int S{\rm d}v$ & $S_{\rm pk}$ & $L_{\rm HCN}$ & 1$\sigma$ @ $V_{\rm ch}$\\
(GHz) & (yyyy-mm-ss) & & (\kms) & (\kms) & (Jy~\kms) & (Jy) & (photons s$^{-1}$) & (Jy @ \kms)\\
\hline
\multicolumn{9}{c}{CIT~6} \\
\hline
805  & 2012-06-16 & 0.21 & $-$6.7 & [$-$10.0, $-$0.6] & 2501.4  & 463.7  & 3.2$\times$10$^{44}$ & 27.3 @ 0.19\\
891  & 2012-05-04 & 0.14 & $-$4.8 & [$-$10.0, 0.0]    & 12185.4 & 2909.8 & 1.6$\times$10$^{45}$ & 28.1 @ 0.17\\
964  & 2012-05-04 & 0.14 & $-$4.7 & [$-$10.0, 0.0]    & 6608.6 & 1633.4 & 8.4$\times$10$^{44}$  & 49.3 @ 0.16\\
     & 2018-12-17 & 0.92 & $-$5.6 & [$-$7.6, $-$1.2]  & 6689.1 & 2308.4 & 8.5$\times$10$^{44}$  & 172.1 @ 0.30\\
968  & 2012-05-04 & 0.14 & $-$4.6 & [$-$10.0, $-$0.7] & 4167.7 & 760.2 & 5.3$\times$10$^{44}$   & 45.5 @ 0.15\\
     & 2018-12-17 & 0.92 & $-$4.9 & [$-$7.8, $-$1.1]  & 3796.0 & 890.4 & 4.8$\times$10$^{44}$   & 171.8 @ 0.30\\
1055 & 2012-05-04 & 0.14 & $-$4.1 & [$-$7.6, $-$1.4]  & 2098.0 & 786.5 & 2.7$\times$10$^{44}$   & 98.7 @ 0.14\\
\hline
\multicolumn{9}{c}{Y~CVn} \\
\hline
805  & 2012-06-16 & & +19.9 & [+14.2, +25.4] & 1446.5 & 245.4 & 4.2$\times$10$^{43}$ & 28.4 @ 0.19\\
891  & 2012-06-14 & & +21.1 & [+14.8, +28.0] & 4877.9 & 825.8 & 1.4$\times$10$^{44}$ & 27.6 @ 0.17\\
964  & 2012-06-15 & & \nodata & \nodata & \nodata & $<$152.1 & \nodata & 50.7 @ 0.16\\
     & 2018-12-17 & & +20.8 & [+18.4, +21.6] & 1342.6 & 534.0 & 3.9$\times$10$^{43}$ & 103.8 @ 0.30\\
968  & 2012-06-15 & & +21.0 & [+15.4, +25.6] & 2238.6 & 387.6 & 6.5$\times$10$^{43}$ & 66.1 @ 0.15\\ 
     & 2018-12-17 & & +18.4 & [+15.1, +23.0] & 3238.2 & 989.1 & 9.4$\times$10$^{43}$ & 149.7 @ 0.30\\
1055 & 2012-05-18 & & \nodata & \nodata & \nodata & $<$310.9 & \nodata & 103.6 @ 0.14\\
\hline
\multicolumn{9}{c}{S~Cep} \\
\hline
805  & 2012-05-23 & 0.40 & $-$16.8 & [$-$22.2, $-$9.4]  & 2729.3 & 547.1 & 2.4$\times$10$^{44}$  & 29.5 @ 0.19\\
891  & 2012-05-24 & 0.40 & $-$15.2 & [$-$22.0, $-$10.0] & 9272.6 & 1983.0 & 8.1$\times$10$^{44}$ & 26.6 @ 0.17\\
964  & 2012-05-24 & 0.40 & $-$16.7 & [$-$20.4, $-$11.6] & 4417.5 & 1077.8 & 3.8$\times$10$^{44}$ & 50.7 @ 0.16\\
     & 2018-12-17 & 0.35 & $-$17.5 & [$-$19.6, $-$14.2] & 4132.0 & 1401.3 & 3.6$\times$10$^{44}$ & 135.5 @ 0.30\\
968  & 2012-05-24 & 0.40 & $-$13.8 & [$-$20.4, $-$11.2] & 2071.2 & 407.4 & 1.8$\times$10$^{44}$  & 61.0 @ 0.15\\
     & 2018-12-17 & 0.35 & $-$15.9 & [$-$18.3, $-$12.4] & 2132.6 & 640.3 & 1.9$\times$10$^{44}$  & 152.6 @ 0.60\\
1055 & 2012-05-24 & 0.40 & $-$16.5 & [$-$17.8, $-$12.4] & 2121.8 & 581.9 & 1.8$\times$10$^{44}$  & 90.4 @ 0.14\\
\hline
\multicolumn{9}{c}{IRC+50096} \\
\hline
805  & 2012-08-25 & 0.84 & $-$12.1 & [$-$21.0, $-$10.0] & 916.3 & 156.7 & 1.7$\times$10$^{44}$  & 30.0 @ 0.19\\
891  & 2012-08-11 & 0.81 & $-$18.6 & [$-$23.0, $-$7.0] & 4391.2 & 790.4 & 8.3$\times$10$^{44}$  & 29.4 @ 0.17\\
964  & 2012-08-16 & 0.82 & $-$18.4 & [$-$20.2, $-$8.0] & 1874.0 & 346.8 & 3.5$\times$10$^{44}$  & 52.2 @ 0.16\\
968  & 2012-08-16 & 0.82 & \nodata & \nodata & \nodata & $<$153.9 & \nodata  & 51.3 @ 0.15\\
1055 & 2012-08-25 & 0.82 & \nodata & \nodata & \nodata & $<$291.3 & \nodata  & 97.1 @ 0.14\\
\hline
\multicolumn{9}{c}{V Cyg} \\
\hline
805  & 2012-05-23 & 0.56 & +14.3 & [+6.8, +21.4] & 2292.8 & 471.8 & 1.9$\times$10$^{44}$    & 29.3 @ 0.19\\
891  & 2012-05-24 & 0.57 & +11.0 & [+8.0, +21.0] & 10737.9 & 2215.4 & 8.7$\times$10$^{44}$  & 28.8 @ 0.17\\
964  & 2012-05-03 & 0.52 & +11.2 & [+9.0, +21.4] & 5975.4 & 1321.8 & 4.8$\times$10$^{44}$   & 56.1 @ 0.16\\
968  & 2012-05-03 & 0.52 & +14.5 & [+7.5, +19.6] & 2182.9 & 464.8 & 1.8$\times$10$^{44}$    & 52.1 @ 0.15\\
1055 & 2012-06-21 & 0.63 & +15.1 & [+8.3, +18.2] & 2093.1 & 527.3 & 1.7$\times$10$^{44}$    & 97.6 @ 0.14\\
\hline
\multicolumn{9}{c}{II Lup} \\
\hline
805  & 2012-09-12 & 0.43 & $-$15.9 & [$-$19.0, $-$11.0] & 447.1 & 148.6 & 1.1$\times$10$^{44}$   & 7.2 @ 0.19\\
891  & 2012-09-18 & 0.45 & $-$16.1 & [$-$20.4, $-$11.0] & 3830.4 & 1042.5 & 9.5$\times$10$^{44}$ & 11.8 @ 0.17\\
964  & 2012-09-07 & 0.42 & $-$14.6 & [$-$18.6, $-$11.0] & 2275.7 & 739.4 & 5.6$\times$10$^{44}$  & 12.0 @ 0.16\\
968  & 2012-09-07 & 0.42 & $-$15.9 & [$-$18.0, $-$12.0] & 654.7 & 190.2 & 1.6$\times$10$^{44}$   & 13.0 @ 0.15\\
1055 & 2012-09-21 & 0.46 & $-$14.6 & [$-$16.4, $-$12.4] & 324.1 & 120.9 & 8.0$\times$10$^{43}$   & 20.3 @ 0.14\\
\hline
\multicolumn{9}{c}{CRL~3068} \\
\hline
805  & 2012-06-21 &  & \nodata & \nodata & \nodata & $<$69.3 & \nodata  & 23.1 @ 0.19\\
891  & 2012-05-24 &  & \nodata & \nodata & \nodata & $<$76.2 & \nodata  & 25.4 @ 0.17\\
964  & 2012-05-31 &  & \nodata & \nodata & \nodata & $<$179.7 & \nodata  & 59.9 @ 0.16\\
968  & 2012-05-31 &  & \nodata & \nodata & \nodata & $<$139.8 & \nodata  & 46.6 @ 0.15\\
1055 & 2012-06-21 &  & \nodata & \nodata & \nodata & $<$277.8 & \nodata  & 92.6 @ 0.14\\
\hline
\end{tabular}
\normalsize
\tablefoot{The descriptions of the table are the same as in Table~\ref{Tab:IRC+10216_laser}.
For the non-detection, the upper limits of the peak intensity ($<3 \sigma$) are given. 
No ephemeris information for Y~CVn , II~Lup and CRL~3068 are provided by AAVSO. We additionally adopted the data from \cite{2003MNRAS.346..878F} for II~Lup to estimate their stellar phases for the observing dates.}
\end{table*}

{\it \textsf{IRC+10216 (CW~Leo).}} 
As an archetypal C-rich long period variable AGB star, IRC+10216 has been extensively studied due to its proximity, high mass-loss rate, extreme infrared brightness, and its role as a rich repository of diverse molecular species \citep[e.g.][]{1969ApJ...158L.133B,2000A&AS..142..181C,2015A&A...574A..56G}.
IRC+10216 is believed to undergo a transition phase between the AGB and the formation of a planetary nebula \citep{1994AJ....108..984T}, and is the source hosting the by far largest number of known HCN maser lines to date \citep[see Table A.1 in][]{2022A&A...666A..69J}.
The detected 805, 891, 964, 968, and 1055~GHz HCN laser emission lines cover a similar velocity range between $-$35 and $-$20~\kms\, in the LSR frame, which is nearly half of the velocity range of most thermal lines such as HCN transitions in the vibrational ground state \citep{2018A&A...613A..49M,2022A&A...666A..69J}.
The strongest components of these detected laser lines are observed near the systemic velocity (see Fig.~\ref{fig:IRC10216_5laser}), and show very different line profiles compared with the (0,1$^{\rm 1f}$,0), $J=4-3$ HCN maser \citep{2022A&A...666A..69J} and SiS masers \citep{1983ApJ...267..184H,2006ApJ...646L.127F,2017ApJ...843...54G,2018ApJ...860..162F}, which could be due to infrared line overlaps or foreground masers amplifying background emission. 
The maser emission in the (0,1$^{\rm 1e}$,0), $J=2-1$ line also show drastic variations in line profile (see Fig. 7 in \citealt{2022A&A...666A..69J}). 
The line observed on 2018 December show multiple maser features with the strongest one aligning with the systemic velocity, which is different from the line profiles of the laser emissions detected by SOFIA on a nearby date.


{\it \textsf{CIT~6 (RW LMi).}} 
CIT~6 has an exceptionally rich circumstellar envelope \citep[e.g.][]{2002ApJ...576..429S}, displaying many characteristics similar to those of IRC+10216. It is, however, at a much larger distance (see Table~\ref{Tab:obs_source}).
Although CIT~6 has a period and a terminal velocity comparable to IRC+10216, its mass-loss rate is several times lower.
After scaling the two sources to the same distance, the fluxes of the HCN laser transitions detected towards CIT~6 are slightly higher than those for IRC+10216.
In addition, the known HCN maser emissions in the (0,2$^{\rm 0}$,0), $J=1-0$ and (0,1$^{\rm 1e}$,0), $J=2-1$ lines were also stronger in CIT~6 than in IRC+10216 \citep{1987A&A...176L..24G,2018A&A...613A..49M}.
It is worth noting that all detected HCN laser and maser transitions, including the 805 and 891~GHz lines discovered by \cite{2003ApJ...583..446S} and the two maser lines mentioned above, exhibited peak velocities clearly blueshifted ($>$2~\kms) with respect to the systemic velocity (see the left panels of Fig.~\ref{fig:3stars_laser}).
Furthermore, the velocity centroid of emission is also significantly blueshifted, highlighting a notable difference between CIT~6 and other sources.

{\it \textsf{Y~CVn.}} 
Y~CVn is a J-type C-rich semiregular star with a $^{12}$C/$^{13}$C ratio of $<$15 \cite[e.g.][]{2000ApJ...536..438A}, and this star has the smallest mass-loss rate, the shortest period, and the weakest detected laser emission in our sample.
HCN emission in the (1,1$^{\rm 1e}$,0)--(0,4$^0$,0), $J=11-10$ line at 964~GHz was not detected on 2012 June 15 (see the middle panels of Fig.~\ref{fig:3stars_laser}), while weak emission was observed on 2018 December 17.
All the detected HCN laser emission is blueshifted with respect to the systemic velocity, but not as prominent as in CIT~6.

{\it \textsf{S~Cep.}} 
The 805, 891, and 964~GHz HCN laser emissions detected in 2012 May show three peaks at $\sim$ $-$15.1, $-$16.8, and $-$19.1~\kms, which are aligned in velocity (see the right panels of Fig.~\ref{fig:3stars_laser}).
The peak velocity of the strongest feature varies: the laser component at approximately $-$16.8~\kms, is the strongest for the 805 and 964~GHz lines, whereas the component at $-$15.1~\kms, is the strongest at 891~GHz.
Before this work, the only HCN maser detected towards this star was in the (0,2$^0$,0), $J=1-0$ transition \citep{1988A&A...194..230L}, which showed a single narrow maser feature at $-$15.8~\kms.

{\it \textsf{IRC+50096 (V~384~Per).}} 
The observations were conducted at stellar pulsation phases ranging from 0.81 to 0.84, and HCN laser emissions were detected at 805, 891, and 964~GHz (see the leftmost panels of Fig.~\ref{fig:3stars_laser_2}).
The HCN laser emission at 891~GHz shows two peaks: the stronger peak at $-$18.6~\kms\, is blueshifted with respect to the systemic velocity of 1.5~\kms, while the secondary peak at $-$10.4~\kms\, is significantly redshifted from the systemic velocity of $\sim$6~\kms.
The line profile of the 964~GHz emission is similar to that of the 891~GHz emission, with the strongest emission aligned in velocity.
In contrast, the line profile of the 805~GHz emission appears to be flipped from the 964~GHz line profile, where the strongest emission is redshifted with respect to the systemic velocity of $\sim$5~\kms.
Prior to this study, the only known HCN maser towards this star was in the (0,2$^0$,0), $J=1-0$ transition of HCN \citep{1988A&A...194..230L}, which exhibited a single narrow maser feature at $-$15.4~\kms.

{\it \textsf{V~Cyg.} }
The observations were made at stellar phases ranging from 0.52 to 0.63, during which HCN laser emissions were detected in all five transitions (see the second panels from left of Fig.~\ref{fig:3stars_laser_2}).
HCN laser emissions in the 805, 891 and 964~GHz lines exhibit a double-peak profile, with two peaks separated on either side of the systemic velocity and a low intensity at the systemic velocity.
Stronger peaks were detected on the blueshifted side in the 891 and 964~GHz lines, while the redshifted side is stronger for the 805~GHz line.
In contrast, HCN laser emissions at 968 and 1055~GHz are dominated by a component at $\sim$15~\kms, which is relatively close to the systemic velocity.
Before this work, the only known HCN maser transitions detected in this star were the (0,1$^{\rm 1e}$,0), $J=3-2$, and (0,1$^{\rm 1e}$,0), $J=4-3$ lines \citep{1988A&A...194..230L}, both of which showed a weak maser feature.

{\it \textsf{II~Lup (IRAS~15194$-$5115).}} 
II~Lup is a high mass-loss rate Mira variable, and the third brightest C-rich star in the sky at 12~$\mu$m \citep{1987MNRAS.225P..43M}.
HCN laser emission was detected in all five transitions, with each emission dominated by a single, narrow component close to the systemic velocity (see the third panels from left of Fig.~\ref{fig:3stars_laser_2}).
These line profiles differ from the previously known HCN maser emissions in the $J=2-1$, (0,1$^{\rm 1e}$,0) and $J=4-3$, (0,1$^{\rm 1f}$,0) lines \citep{2018A&A...613A..49M,2022A&A...666A..69J}, which exhibit broader velocity coverage and multiple ($\geq$2) emission features.

{\it \textsf{CRL~3068 (LL~Peg).}}
The extreme C-rich star CRL~3068 has comparable high mass-loss rate, long period and thick dusty shells as IRC+10216. A comparison of isotopic ratios indicates that this star is a more evolved object than IRC+10216 \citep{2009ApJ...700.1262Z}. Two HCN transitions, that are the (0,1$^{\rm 1f}$,0), $J=4-3$ and the (0,1$^{\rm 1e}$,0), $J=2-1$ lines, are known to show maser actions \citep{2022A&A...666A..69J}. 
In contrast, there is no emission from any of the HCN laser transitions detected in this star. 
Except for the 891 GHz line, the non-detection of most HCN lasers may be attributed to the star's distance, which is the largest among our sample (Table~\ref{Tab:obs_source}), as well as its relatively low HCN abundance as modelled by \citet{2013A&A...550A..78S}. The circumstellar chemistry of CRL~3068 may be different from that of a typical C-rich AGB star \citep{2009ApJ...700.1262Z}. Spatially-resolved observations will allow a better understanding of the spatial distribution and physical conditions of HCN at inner radii. 

\section{A more comprehensive view of lines between (1,1$^{\rm 1e}$,0) and (0,4$^0$,0) vibrational states} \label{Sec:appendix-d}

\begin{figure*}[htbp]
\center
\includegraphics[width=0.6\textwidth]{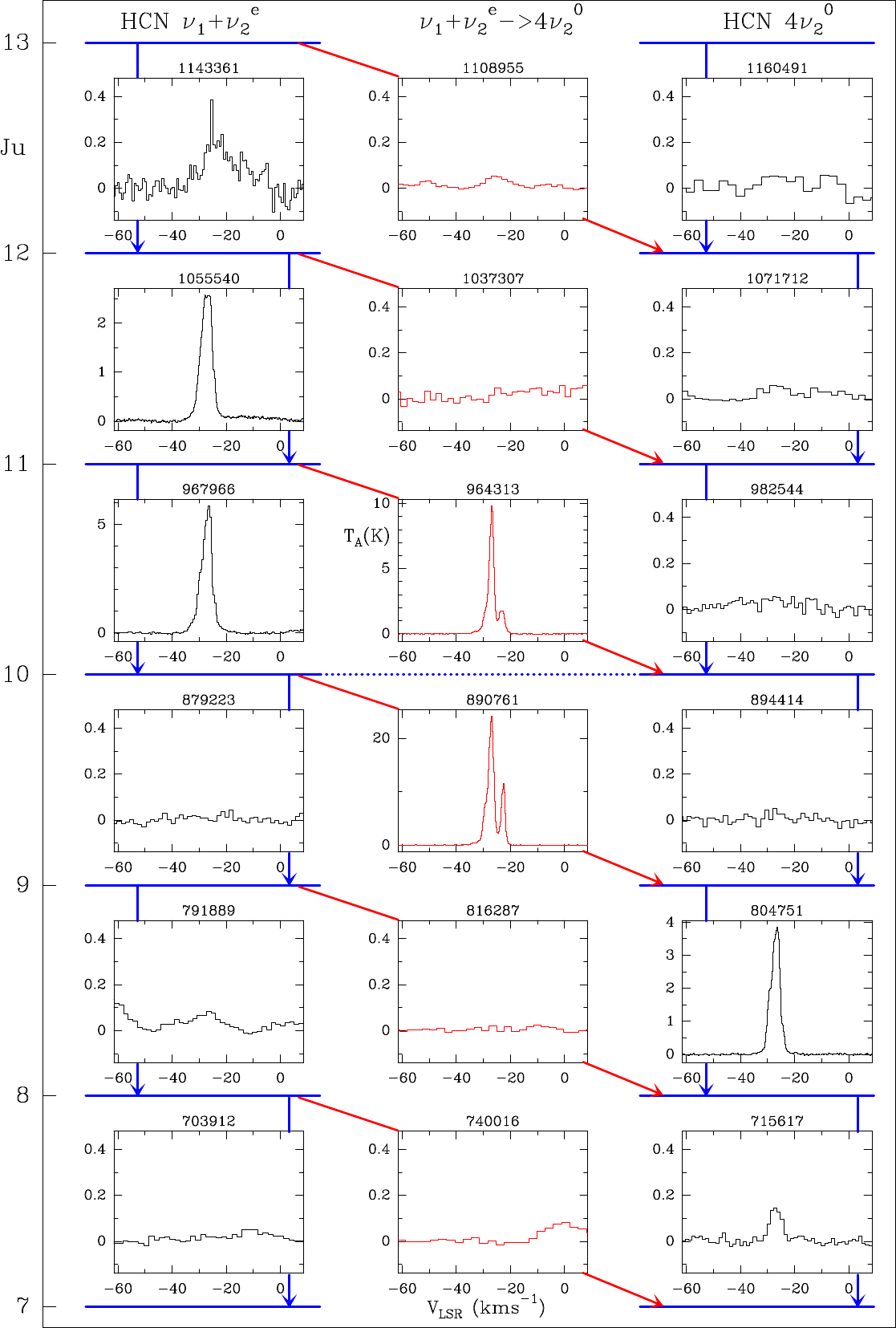}
\caption{HCN transitions between the (1,1$^{\rm 1e}$,0) and (0,4$^0$,0) vibrational states with the rotational levels from $J=7$ to $J=13$, observed towards IRC+10216 in 2010 May using {\it Herschel}/HIFI. 
Spectra are arranged from left to right: (left) the rotational lines in the (1,1$^{\rm 1e}$,0) vibrational state, (middle) the cross-ladder lines between two vibrational states, (right) the rotational lines in the (0,4$^{0}$,0) vibrational state. The intensity scale is antenna temperature in K. The frequency of each line is labelled at the top of the panel in MHz.
Blue solid horizontal lines represent the rotational levels, and the blue dotted line marks the $J$=10 rotational levels of the different vibrational states which are strongly coupled by Coriolis interaction.} \label{fig:complete_hcn}
\end{figure*}

Figure~\ref{fig:complete_hcn} shows a complete set of HCN spectra covering the rotational levels from $J=7$ to $J=13$ within and between the (1,1$^{\rm 1e}$,0) and (0,4$^0$,0) vibrational states towards IRC+10216. 
In this figure, we adopt another commonly used notation,  $\nu_1+\nu_2^{\rm e}$ and $4\nu_2^0$, to label the vibrational states of (1,1$^{\rm 1e}$,0) and (0,$4^{0}$,0), respectively.
The observations were carried out during 2010 May 11 -- 16 using {\it Herschel}/HIFI (Project id: GT1\_jcernich\_4, PI: Jos{\'e} Cernicharo).

Within the (1,1$^{\rm 1e}$,0) vibrational level, the strongest emission occurs at $J_{\rm up}=11$. The peak intensities of detected emissions decrease as $J$ increases, with no emission was detected for $J_{\rm up} \leq 10$. 
Within the (0,4$^0$,0) vibrational level, the emission at $J_{\rm up}=9$ is stronger than the emission at adjacent $J_{\rm up}=8$, and no emission was detected for $J_{\rm up} \geq 10$.
Between the two vibrational levels, cross-ladder lines at 891~GHz ($J_{\rm up}=10$) and 964~GHz ($J_{\rm up}=11$) are significantly stronger than the previously mentioned rotational lines and are the only detected cross-ladder transitions. 
The two strong cross-ladder emissions arise from strong coupling between the two vibrational states, with the coupling being stronger at $J=10$ and slightly weaker at $J=11$ \citep{1964JMoSp..12...45M,1967ApPhL..11...62L}, that further affect the population in the (1,1$^{\rm 1e}$,0) and (0,4$^0$,0) vibrational levels at certain $J$ (see detailed discussion in Sect.~\ref{Sec:exciatation}).


\end{appendix}

\end{document}